\documentclass[12pt]{article}
\pdfoutput=1 
\catcode`@=11
\@addtoreset{equation}{section}
\renewcommand{\theequation}{\arabic{section}.\arabic{equation}}

\usepackage{cite}

\usepackage[hyperfootnotes=false]{hyperref}
\usepackage{amsmath}
\usepackage{accents}
\usepackage{amssymb}
\usepackage{amsthm}
\usepackage{psfrag}
\usepackage{graphicx}
\usepackage{color}
\usepackage{upgreek}
\usepackage{cleveref}
\usepackage{bbm}

\usepackage[toc,page]{appendix}

\newcommand{\nocontentsline}[3]{}
\newcommand{\tocless}[2]{\bgroup\let\addcontentsline=\nocontentsline#1{#2}\egroup}

\allowdisplaybreaks[4]

\newcommand{\bz}{\bar{z}}

\newcommand{\be}{\begin{equation}}
\newcommand{\ee}{\end{equation}}
\newcommand{\beqa}{\begin{eqnarray}}
\newcommand{\eeqa}{\end{eqnarray}}

\renewcommand\l{\lambda}
\newcommand\m{\mu}

\newcommand\n{\nu}

\newcommand\vf{\varphi}
\renewcommand\a{\alpha}
\renewcommand{\b}{\beta}
\newcommand{\gc}{\text{g}}

\newcommand{\TT}{\mathcal{T}}
\newcommand{\N}{\mathcal{N}}
\renewcommand{\o}{\omega}
\renewcommand{\O}{\Omega}

\usepackage{slashbox}
\usepackage{setspace}

\def\e{{\rm e}}
\def\d{\partial}

\newcommand{\bseq}{\begin{subequations}}
	\newcommand{\eseq}{\end{subequations}}

\newcommand{\ch}{\mathop{\rm ch}\nolimits}
\newcommand{\sh}{\mathop{\rm sh}\nolimits}
\renewcommand{\ln}{\mathop{\rm ln}\nolimits}

\renewcommand{\Im}{\mathop{\rm Im}\nolimits}
\renewcommand{\Re}{\mathop{\rm Re}\nolimits}

\newcommand{\bra}[1]{\langle #1 |}
\newcommand{\ket}[1]{| #1 \rangle}
\newcommand{\braket}[2]{\langle #1 |#2 \rangle}

\newcommand{\vk}{\varkappa}
\newcommand{\g}{\gamma}

\newcommand{\U}{{\bar u}}
\newcommand{\V}{\bar v}

\newcommand{\GG}{\mathcal{G}}

\def\Res{\mathop{\mathrm{Res}}}
\newcommand{\arcch}{\mathop{\rm arcch}\nolimits}

\makeatletter
\newcommand\footnoteref[1]{\protected@xdef\@thefnmark{\ref{#1}}\@footnotemark}
\makeatother

\begin{document}
\begin{titlepage}
\clearpage

\title{\vspace{-2cm}
\begin{flushright}
{\normalsize FTPI-MINN-21-07, 
UMN-TH-4014/21,
INR-TH-2021-011}
\end{flushright}
\vspace{0.5cm}
{\bf Black hole induced false vacuum decay\\ from first principles}}

\author{Andrey Shkerin$^{a}$\footnote{ashkerin@umn.edu}~, 
Sergey Sibiryakov$^{b,c,d}$\footnote{ssibiryakov@perimeterinstitute.ca}\\[2mm]
{\small\it $^a$William I. Fine Theoretical Physics Institute, School
  of Physics and Astronomy,} \\ 
{\small\it University of Minnesota, Minneapolis, MN 55455, USA }\\
{\small\it $^b$Department of Physics \& Astronomy, McMaster
University,}\\
{\small\it Hamilton, Ontario, L8S 4M1, Canada}\\
{\small\it $^c$Perimeter Institute for Theoretical Physics, Waterloo,
 Ontario, N2L 2Y5, Canada}\\
{\small \it $^d$Institute for Nuclear Research of the Russian Academy
  of Sciences,}\\ 
{\small \it 60th October Anniversary Prospect, 7a, 117312 Moscow, Russia}
}

\date{}
\maketitle

\begin{abstract}
We provide a method to calculate the rate
  of false vacuum decay induced by a black hole. The method
  uses complex tunneling solutions and
  consistently takes into account the structure of different 
 quantum vacua in the
  black hole metric via boundary conditions. The latter are connected
 to the
  asymptotic behavior of the time-ordered Green's function in the
  corresponding vacua. We illustrate the technique on a
  two-dimensional toy model of a scalar field with inverted Liouville
  potential in an external background of a dilaton black hole. We
  analytically derive 
the exponential suppression of tunneling from the
  Boulware, 
  Hartle--Hawking and Unruh vacua and show that they are
  parametrically different. The Unruh vacuum decay rate is
  exponentially smaller than the decay rate of the
  Hartle--Hawking state, though both rates become unsuppressed 
  at high enough black hole temperature. We interpret the vanishing
  suppression of the Unruh vacuum decay at high temperature as an
  artifact of the two-dimensional model and discuss why this
  result can be modified in the realistic case of black holes in
  four dimensions. 
\end{abstract} 

\thispagestyle{empty}
\end{titlepage}

\newpage
\tableofcontents

\section{Introduction and summary}
\label{Sec:intro}

Description of false vacuum decay in the presence of a black hole (BH)
is a long-standing problem
\cite{Hiscock:1987hn,Berezin:1987ea,Arnold:1989cq,Berezin:1990qs}. The 
interest in it has been revived in recent years due to its
possible phenomenological relevance. The electroweak vacuum determined
by the Standard Model Higgs potential may not be absolutely stable
\cite{Flores:1982rv,Sher:1988mj,Isidori:2001bm,
1205.2893,1205.6497,1307.3536,Bednyakov:2015sca}.  
In the absence of excitations its decay rate is exponentially
suppressed and its lifetime exceeds the age of the Universe by many
orders of magnitude~\cite{1707.08124}. However, it has been argued
in~\cite{1401.0017,1501.04937,1503.07331,1601.02152}
that the decay can be strongly catalyzed if the Universe hosts at some
stages of its evolution light primordial BHs that later evaporate via 
Hawking radiation. Such BHs appear in a variety of 
early Universe models and can play important roles in cosmology,
including reheating of the Universe, production of baryon asymmetry
and dark matter, etc. 
\cite{GarciaBellido:1996qt,Fujita:2014hha,Allahverdi:2017sks,
Lennon:2017tqq,Morrison:2018xla,Hooper:2019gtx,Hooper:2020evu,DeLuca:2021oer} (see
also \cite{Carr:2020gox} for a review of primordial BH production
mechanisms and constraints). 
The results of Refs.~\cite{1401.0017,1501.04937,1503.07331,1601.02152}
would rule out the presence of any
evaporating BHs in our causal past and thereby put stringent
constraints on primordial BH models. Or, alternatively,
would imply that the Standard Model is completed in the way to prevent
the electroweak vacuum instability.

The intuitive reason behind the BH catalysis of vacuum decay is rather
simple. Due to the Hawking effect, a BH can be thought of as a body with
finite temperature. As the BH evaporates, it heats up scanning all
temperatures up to Planckian. On the other hand, it is known that the
false vacuum decay becomes unsuppressed at high enough temperatures,
comparable to the height of the energy barrier between the false and
the true vacuum. The latter height is given by the energy of the 
sphaleron configuration (also called critical bubble) 
--- static unstable solution of the equations
of motion separating the two vacua.\footnote{The term sphaleron
  was first introduced in \cite{Klinkhamer:1984di} in the context of
  fermion number violating transitions in the Standard Model. Here we
  are using it in a broader sense 
for the saddle-point solution on top of the potential
  barrier between different vacua.}  
Thus, one might expect a BH also to render the decay unsuppressed once
it becomes sufficiently hot.

However, the above reasoning has a caveat. A realistic BH is not in
thermal equilibrium with its environment. It radiates away a thermal
spectrum of particles, but does not receive anything
back.\footnote{For the sake of the argument, we neglect the grey-body
  factors and a possible effect of the medium surrounding the BH. Their
importance will be discussed in Sec.~\ref{Sec:disc}.} From the
technical viewpoint, this corresponds to the Unruh vacuum state 
\cite{Unruh:1976db}, as
opposed to the Hartle--Hawking vacuum \cite{Hartle:1976tp} 
describing a BH immersed in a
thermal bath with the same temperature. The deviation from equilibrium
is expected to reduce the catalyzing effect of BH and it is not clear
if it can overcome the exponential suppression of vacuum decay at any
BH temperature. The results in the literature addressing this issue
have been controversial
\cite{1606.04018,1704.05399,1706.04523,1708.02138,Hayashi:2020ocn}. 
Even if the exponential suppression persists for all BHs,
it is still important to know how much it is reduced compared to the
no-BH 
case. Indeed, for a given density of primordial BHs, each of them can
be a nucleation cite for the vacuum decay bubble. The small
probability of this event for a single BH will be multiplied by the
huge number of these BHs in the observable Universe 
\cite{1606.04018,1704.05399,1706.01364}. Thus the
condition that no vacuum decay occurs in our causal past can still
put relevant constraints on the primordial BH scenarios and/or
completion of the Standard Model.\footnote{Depending on the decay rate, bubbles of true vacuum seeded by black holes can percolate, completing the transition to the true vacuum phase \cite{Dai:2019eei}.}

Apart from the relevance for phenomenology, BH catalysis of
vacuum decay is of considerable theoretical interest. First, its study
is expected to give insight into nonperturbative quantum field theory
in curved spacetimes with nontrivial causal structure. Second, when
the dynamical metric is included, it can teach us about the properties
of semiclassical quantum gravity, similarly to 
Coleman--De Luccia~\cite{Coleman:1980aw} and Hawking--Moss~\cite{Hawking:1981fz}
instantons in de Sitter space. Third,
Ref.~\cite{1401.0017} pointed out an
intriguing connection between the probability of false vacuum decay and
BH entropy, which may shed a new light on the origin of the latter.  

All this calls for a self-consistent framework to calculate the effect
of BH on vacuum instability that takes into account the properties of
the Unruh vacuum. Developing such framework is the purpose of this
work. To clarify the analysis, we will consider the dynamical sector
consisting of a single scalar field $\varphi$ evolving in a fixed
background geometry. Further, for most of the paper we will
focus on a setup in two dimensions, commenting on its relation to
spherically-symmetric four-dimensional dynamics at the end. 

Even with these simplifications, our task is challenging. Being 
classically forbidden, the false vacuum decay represents a tunneling
process. In the semiclassical limit one expects it to be described by
a complex solution of the field equations representing the saddle
point of the Feynman path integral \cite{doi:10.1142/3768}. 
The first question that arises is:
\begin{itemize}
\item[i)]
	\emph{On which section of complexified spacetime coordinates
          the tunneling solution is defined?}
\end{itemize}
In equilibrium situations the answer to this question is well-known:
the tunneling solution lives in purely imaginary (Euclidean) time. The
standard way to arrive to these solutions is to work from the
beginning with the Euclidean partition function
\cite{Coleman:1977py,Callan:1977pt,Coleman:1978ae}. In the case of a
Schwarzschild BH this leads 
to the theory in the cigar-like geometry with compactified Euclidean
time coordinate playing the role of the angular variable and the
radial coordinate covering the region outside the horizon
\cite{Hartle:1976tp}. This picture corresponds to the partition
function in the Hartle--Hawking vacuum, i.e., an equilibrium thermal
state. It is not clear at all how it can be modified to accommodate
the Unruh state.  
 
Instead, we use an alternative approach that starts from the path
integral expression for the transition amplitude in real time from the
false vacuum at $t\to -\infty$ to the true vacuum at $t\to
+\infty$. To obtain the tunneling solution, the real time axis is
deformed into a contour in the complex time plane, on which the path
integral can be evaluated in the saddle-point
approximation~\cite{Miller,
Rubakov:1992ec,Bonini:1999kj,Bezrukov:2003tg,
Bramberger:2016yog}
(see
\cite{Turok:2013dfa,Cherman:2014sba,Andreassen:2016cff,Andreassen:2016cvx} 
for related
approaches). The contour consists of segments parallel to the real
axis that are connected by imaginary-time evolution and goes around
the singularities of the tunneling solution. This method is very
flexible and allows one to fix the initial and final quantum states
by an appropriate choice
of the boundary conditions at $t\to \mp\infty$. It has been 
employed to describe baryon number violating processes in the Standard
Model \cite{Bezrukov:2003er}, 
false vacuum decay in de Sitter space
\cite{Rubakov:1999ir}, tunneling induced by particle collisions
\cite{Kuznetsov:1997az,Levkov:2004ij,Demidov:2015bua}, creation of
solitons by highly 
energetic particles
\cite{Levkov:2004tf,Demidov:2011dk,Demidov:2015nea}, semiclassical
black hole S-matrix \cite{Bezrukov:2015ufa,Fitkevich:2020tcj}, as well
as a variety of transitions in quantum mechanics 
\cite{Levkov:2007ce,Levkov:2007zd,Levkov:2007yn,Levkov:2008csa}. In
this work we generalize this method to the case of mixed initial
states described by a density matrix and show that it naturally fits
into the in-in formalism of nonequilibrium quantum field theory. We
will see that for equilibrium initial states this method recovers the 
standard Euclidean results. 

It is still not clear at this point what time coordinate one shall
use. The nontrivial causal structure of BH spacetime provides several
inequivalent choices. First, one can work in Schwarzschild
coordinates, in which the metric is static. The latter property
appears desirable as it facilitates the analytic continuation into
complex time. This coordinate chart, however, is geodesically
incomplete covering only the region outside the BH horizon. A second
option is presented by Painlev\'e or Finkelstein coordinates which
preserve the stationarity of the metric while extending across the
future horizon. The third option is Kruskal coordinates covering the
whole maximally extended spacetime at the expense of rendering the
metric time-dependent. 

The Unruh vacuum is regular at the future BH horizon and is singular
at the past horizon. At first sight, this suggests to use the second
option above. However, one then encounters the following problem. If
one works in the coordinate chart covering the BH interior, it appears
that the analysis will depend on what happens inside the BH. Such
dependence would be unphysical: the vacuum decay rate measured by 
an observer outside the BH must be insensitive to the dynamics
shielded by the event horizon. Thus, we arrive to our second
question:
\begin{itemize}
\item[ii)]
	\emph{Is it possible to formulate the false vacuum decay
          problem referring only to the region outside the BH horizon?}
\end{itemize}
We answer this question in the affirmative. In fact, we will carry out the
whole analysis in the Schwarzschild coordinates and describe the 
Hartle--Hawking
and Unruh vacua as mixed states outside the BH.

In doing so, we will address the third and last question:
\begin{itemize}
\item[iii)]
	\emph{What are the boundary conditions on the tunneling
          solutions corresponding to different initial vacua?}
\end{itemize}
We derive these boundary conditions by performing a saddle-point
integration with the initial-state density matrix. This leads us to
linear relations between positive- and negative-frequency components
of the field in the asymptotic past. We show that the same relations
are obeyed by the mode decompositions of the time-ordered Green's
functions in the respective vacua. In other words, 
the boundary conditions at $t\to-\infty$ for the tunneling solutions
describing decay of a false vacuum are dictated by the time-ordered
Green's function in this vacuum. We argue that this result is general:
it is valid for arbitrary geometry and any state with a Gaussian density
matrix in the vicinity of the false vacuum. As for the final boundary
conditions, we will see that they do not need to be precisely
specified. It is enough to require that on the real axis the
tunneling solution ends up in the basin of attraction of the true
vacuum at $t\to+\infty$.  

We provide a detailed illustration of our method using a solvable toy
model of a scalar field with inverted Liouville potential and a mass
term in two dimensions.\footnote{Recently, the effect of black holes on vacuum decay in two dimensions has also been studied in \cite{Miyachi:2021bwd}.} We show that different boundary conditions
indeed discriminate between different vacuum states and lead to
manifestly different decay probabilities. In particular, we find the lifetime
of the Unruh vacuum to be exponentially longer than that of the
Hartle--Hawking state.

It is worth emphasizing that our method goes in an essential way
beyond the thin-wall approximation often used in the
literature. Indeed, the boundary conditions for the Unruh vacuum rely
on the properties of the solutions to the wave equation that are not
captured by the thin-wall Ansatz.  

The paper is organized as follows. In Sec.~\ref{Sec:gen} we develop
the general formalism for description of vacuum decay in the presence
of a BH. For concreteness, we consider a two-dimensional setup with a
scalar field $\varphi$ which we introduce in Sec.~\ref{Ssec:gen_geometry}.  In 
Sec.~\ref{Ssec:gen_vacuum} we discuss the mode decomposition and
various vacua, whereas in Sec.~\ref{Ssec:Green_gen} we present 
the corresponding Green's functions. In
Sec.~\ref{Ssec:gen_amplitude} we formulate the vacuum decay problem
using the in-in path integral and relate the boundary conditions for
the tunneling solution to the properties of the
time-ordered Green's function.

In Sec.~\ref{Sec:flat} we specify our toy model.
To get insight into its dynamics, we first study
it in flat geometry. In Sec.~\ref{Ssec:flat_sphaleron} we find the
sphaleron solution separating the false vacuum from the run-away
region $\varphi\to+\infty$, which in this model replaces the true
vacuum. In
Secs.~\ref{Ssec:flat_bounce} and \ref{Ssec:flat_thermal} we discuss
the tunneling solutions describing the false vacuum decay at zero and
finite temperature, respectively. We show how the standard results are
reproduced using our approach. 

In Sec.~\ref{Sec:R} we consider tunneling in Rindler metric which
describes the near-horizon region of a BH. This serves as a warm-up
before turning to the full BH case and allows us to develop the
necessary intuition. 
We revisit the
decay of Minkowski vacuum from the viewpoint of the Rindler
space where it corresponds to nontrivial boundary conditions,
analogous to the Hartle--Hawking state in the BH metric \cite{Ai:2018rnh}.

Sec.~\ref{Sec:BH} contains our key results for the toy model. In
Sec.~\ref{Ssec:BH_HH} we calculate the decay rate of the
Hartle--Hawking state as a function of BH temperature using our method
and show that it recovers the Euclidean result. As expected, the decay
rate increases with temperature and becomes unsuppressed when the
temperature gets high enough. In Sec.~\ref{Ssec:BH_U} we find the
tunneling solutions describing the decay of the Unruh vacuum and
evaluate their action. We consider both tunneling far away from the BH
and in the near-horizon region. In both cases the decay rate is
exponentially smaller that the decay rate of the Hartle--Hawking
state. Nevertheless, the suppression diminishes with temperature and
eventually disappears for sufficiently hot BHs. For completeness, we
also consider the Boulware vacuum \cite{Boulware:1974dm}
in Appendix~\ref{Ssec:BH_B} and show that its decay probability does not
essentially differ from that in flat space.

Sec.~\ref{Sec:disc} is devoted to discussion and outlook. In
particular, we point out that the vanishing suppression of the Unruh
vacuum decay at high BH temperature found in Sec.~\ref{Ssec:BH_U} is
likely to be a peculiarity of the two-dimensional theory. We highlight
the properties of realistic four-dimensional BHs that can alter this
behavior.

Several Appendices complement the analysis in the main text.

\section{The method}
\label{Sec:gen}

\subsection{Setup}
\label{Ssec:gen_geometry}

\begin{figure}[t]
	\centering 
	\includegraphics[width=0.4\textwidth]{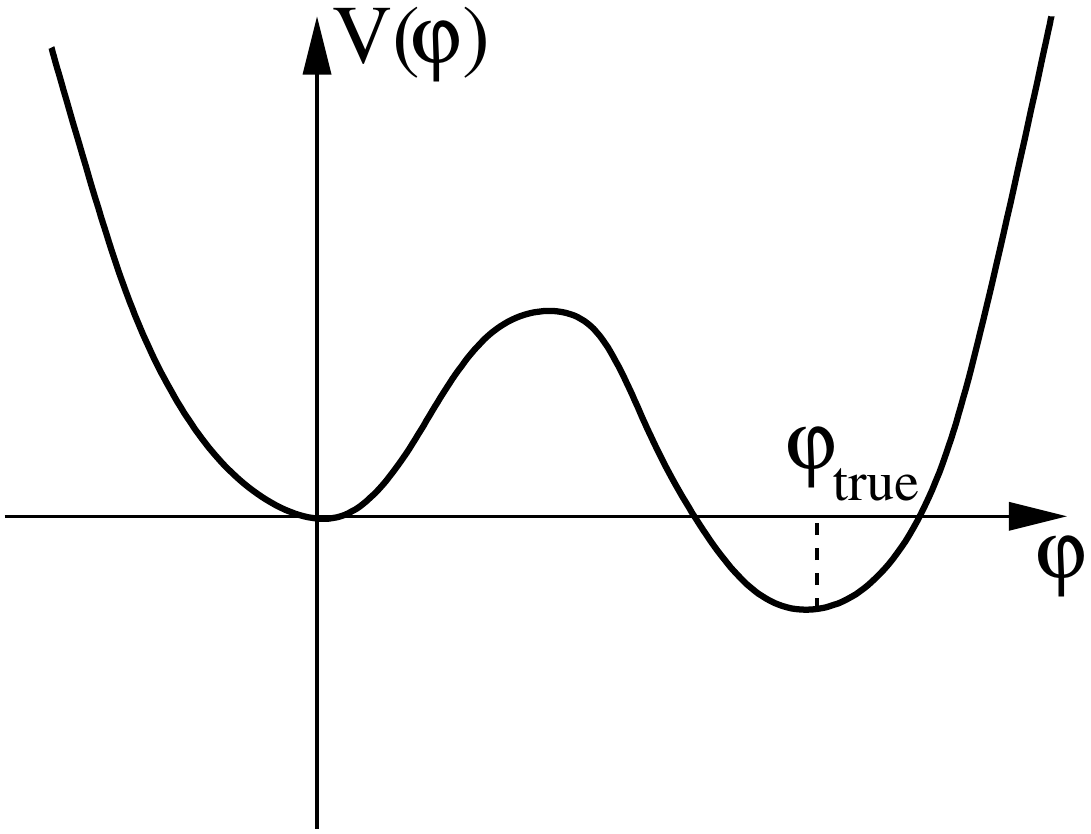}
\qquad\qquad\qquad
	\includegraphics[width=0.4\textwidth]{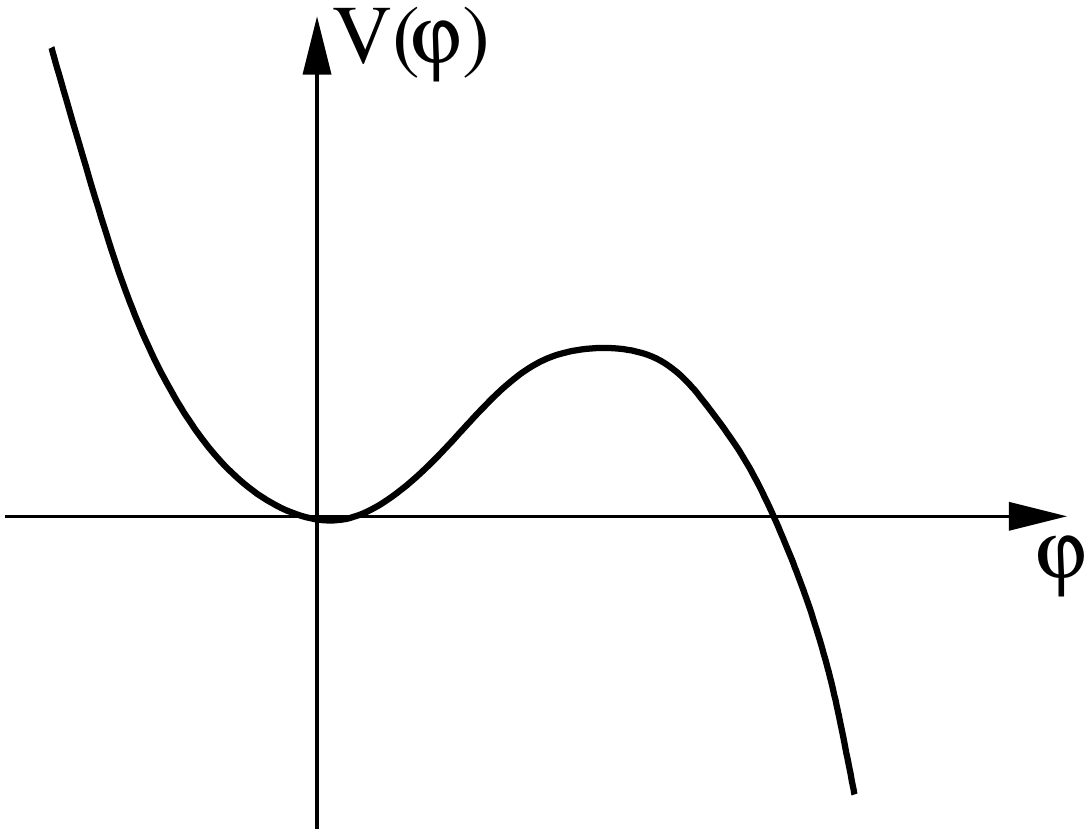}\\
a \qquad\qquad\qquad\qquad\qquad\qquad\qquad\qquad\qquad\qquad\qquad b
	\caption{Scalar potential with false vacuum at $\vf=0$. {\bf (a)}
          The true vacuum exists at a finite value of the field. {\bf (b)}
          The potential is unbounded from below and the false vacuum
          decay leads to the run-away $\vf\to +\infty$.}
	\label{fig:pot}
\end{figure}

We consider a scalar field $\vf$ in two spacetime dimensions with the
action\footnote{We adopt the metric signature $(-,+)$.} 
\begin{equation}\label{Action_main}
S=\dfrac{1}{\gc^2}\int d^2x\sqrt{-g}\left(-\dfrac{1}{2}g^{\mu\nu}\d_\mu\vf\d_\nu\vf-V(\vf)\right) \; .
\end{equation} 
Note that we have factored out the small coupling constant $\gc$ in
front of the action, which can always be achieved by a field
rescaling. This coupling will control the semiclassical expansion in
what follows. The potential $V(\vf)$ is assumed to have a local
minimum at $\vf=0$ where it vanishes, $V(0)=0$. This minimum
corresponds to a false vacuum separated from the region $V<0$ by a
potential barrier. Two situations are possible: (a) the potential is
bounded from below and a true vacuum exists at a finite value
$\vf_{\rm true}$; or (b) the potential is unbounded and the true
vacuum is replaced by the run-away $\vf\to +\infty$. These options are
depicted in Fig.~\ref{fig:pot}.

To model the BH geometry, we consider the static metric,
\begin{equation}\label{metral}
ds^2=-\Omega(r)\, dt^2 + \frac{dr^2}{\Omega(r)}\;,
\end{equation} 
where the function $\Omega$ approaches 1 at $r\to +\infty$ and has a
simple zero at $r=r_h$ corresponding to the horizon. Near the horizon
it is expanded as  
\be
\Omega\approx 2\l(r-r_h)
~~~~\text{at}~~r\approx r_h\;,
\ee
where the parameter $\lambda$ sets the horizon surface gravity and is
related to the BH temperature, $\l=2\pi T_{\rm BH}$ (see,
e.g., \cite{Birrell:1982ix}).
It is convenient to introduce the
``tortoise'' coordinate 
\begin{equation}\label{rx}
x=\int \frac{dr}{\Omega}\;,
\end{equation}
in which the metric becomes conformally-flat,
\begin{equation}\label{Line_element}
ds^2=\O(x)(-dt^2+dx^2) \; .
\end{equation}
The horizon is now located at $x\to-\infty$ and in the near-horizon
region the metric function has an exponential fall-off,
\be
\label{OnearH}
\O\approx \e^{2\l x}~~~~\text{at}~~x\to -\infty\;.
\ee
Explicitly, we will consider the metric of a two-dimensional dilaton
BH,\footnote{Some details of these solutions are given in
  Appendix~\ref{app:Dil_grav}.} 
\begin{equation}\label{O_BH}
\O=\big(1+\e^{-2\l x}\big)^{-1} \;,
\end{equation}
though most of our analysis will be insensitive to this precise form
of the function $\O(x)$. Note that while the coordinate size of the
near-horizon region in tortoise coordinates is infinite, its physical
size is finite and inversely proportional to $\l$,
\begin{equation}
\label{lphys}
l_h\sim \int_{-\infty}^0\sqrt{\O}\, dx\sim \frac{1}{\l} \;.
\end{equation}
With the choice of the metric (\ref{Line_element}) the scalar action
becomes 
\begin{equation}\label{ActionPhi}
S=\dfrac{1}{\gc^2}\int dt dx\left(
-\dfrac{1}{2}\eta^{\mu\nu}\d_\mu\vf\d_\nu\vf-\O(x)V(\vf)\right) \; ,
\end{equation}
where $\eta^{\mu\nu}=\text{diag}(-1,1)$ is the two-dimensional
Minkowski metric. We observe that the dependence on geometry has been
isolated into a position-dependent factor in front of the potential
term. 

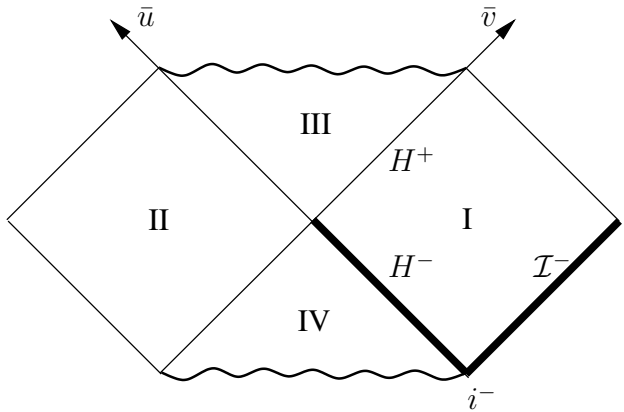
\begin{figure}[t]
	\centering 
\begin{picture}(230,140)
	\put(0,10){\includegraphics[width=3.25in]{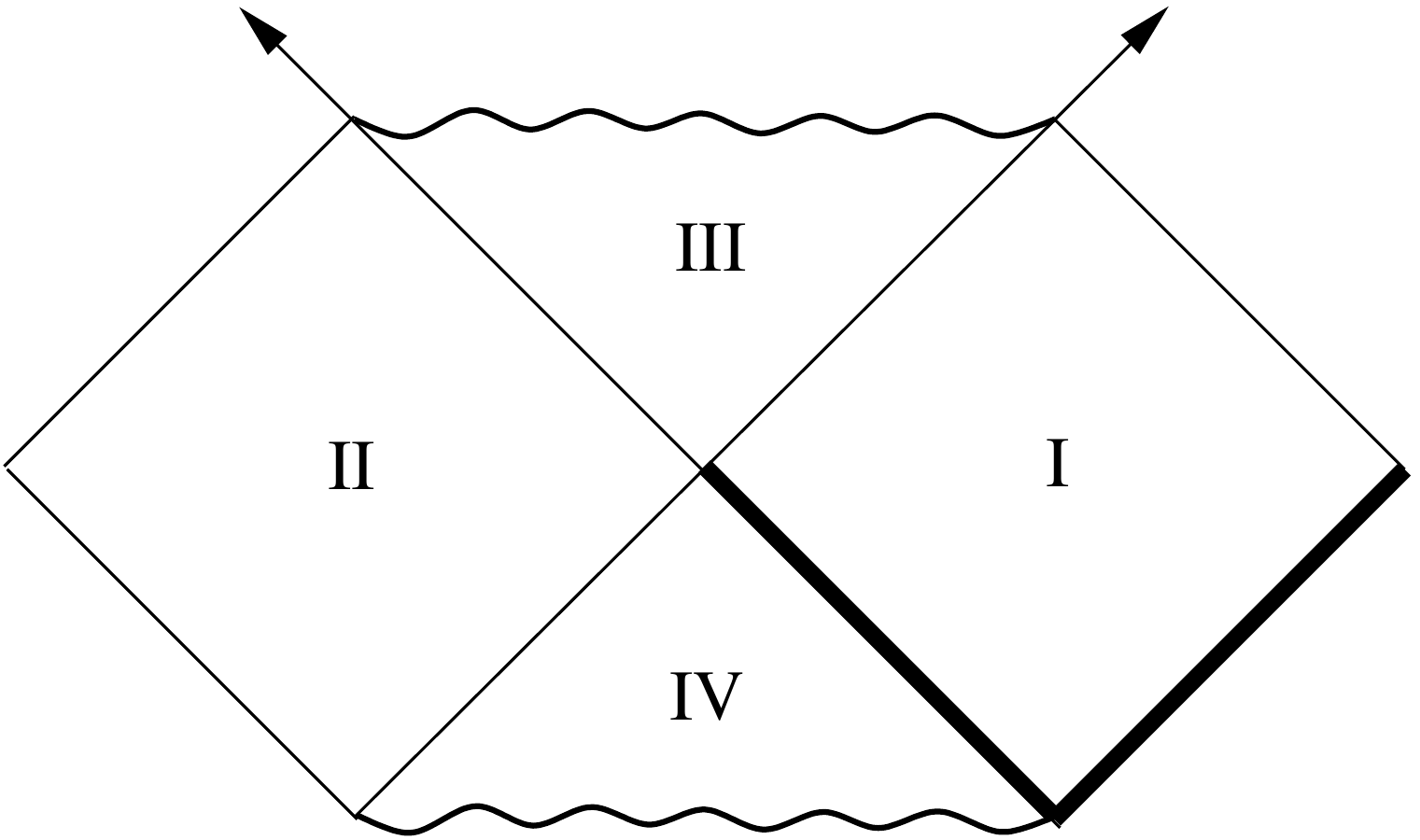}}
\put(50,145){$\bar u$}
\put(180,145){$\bar v$}
\put(175,0){$i^-$}
\put(145,90){$H^+$}
\put(145,50){$H^-$}
\put(200,50){${\cal I}^-$}
\end{picture}
	\caption{Penrose diagram of the maximally-extended BH
          spacetime. The tortoise coordinates $(t,x)$ cover the
          exterior region I. The conditions defining different vacua
          in the BH background will be imposed on the past
          boundary of this region consisting
          of the past horizon $H^-$, past 
          time-like infinity $i^-$ and past light-like infinity ${\cal
          I}^-$ (marked by the thick line).}
	\label{fig:Penrose}
\end{figure}

The coordinates $(t,x)$ cover the BH exterior. This corresponds to the
region I in the Penrose diagram of the maximally-extended BH
spacetime, see Fig.~\ref{fig:Penrose}. To obtain this maximal
extension, one first introduces the light-like coordinates
\begin{equation}\label{uv}
u=t-x \;, ~~~~ v=t+x\;,
\end{equation}
and then the Kruskal coordinates
\begin{equation}\label{Kruskal}
\bar u=-\l^{-1}\e^{-\l u} \;, ~~~~ \bar v=\l^{-1}\e^{\l v}\;.
\end{equation}
In the new coordinates the metric takes the form
\be
ds^2=-\frac{d\U d\V}{1-\l^2 \U\V}\;,
\ee
which is regular as long as $\U\V<1/\l^2$. The latter condition
defines the range of $(\U,\V)$ values covering the maximally-extended
spacetime. In region I we have $-\infty<\U<0$, $0<\V<+\infty$. The future
BH horizon $H^+$ corresponds to $\U=0$ and the past horizon $H^-$ to
$\V=0$. An important role in our analysis will be played by the past
boundary of the region I where we will impose the conditions defining
different vacua 
in the BH background.
It consists of the past horizon $H^-$, past
time-like infinity $i^-$ and past light-like infinity ${\cal I}^-$. 

Let us comment on the approximation of
static geometry. The metric of a realistic BH will evolve due to its
evaporation. Our approximation is valid as long as the evaporation
time is larger than the inverse of the energy scale characterizing
the vacuum decay. The latter should not be confused with
the vacuum decay rate. Rather, it is set by the size of the bubble of
the true vacuum inside the false one at the moment of nucleation. On
the other hand, the exponentially suppressed decay rate determines the
{\rm probability} of bubble nucleation in a unit time interval. If the
inverse decay rate exceeds the BH evaporation time, it just means that
the probability for a single BH to catalyze vacuum decay is small. As
with any probability, it acquires statistical significance when one
considers an ensemble of identical BHs, whose overall catalyzing
effect can become sizable due to their large number.

Our analysis does not capture the highly nonstationary stages of BH
formation and complete evaporation which may have
additional catalyzing effect on vacuum decay. The associated
enhancement of the decay rate is expected to depend strongly on the
details of these transient
events. By contrast, the catalyzing effect of a quasi-stationary BH
studied in this paper is universal and accumulates over the whole BH
lifetime. 

\subsection{Mode decomposition and vacua}
\label{Ssec:gen_vacuum}

\begin{figure}[t]
	\center{\includegraphics[width=0.5\linewidth]{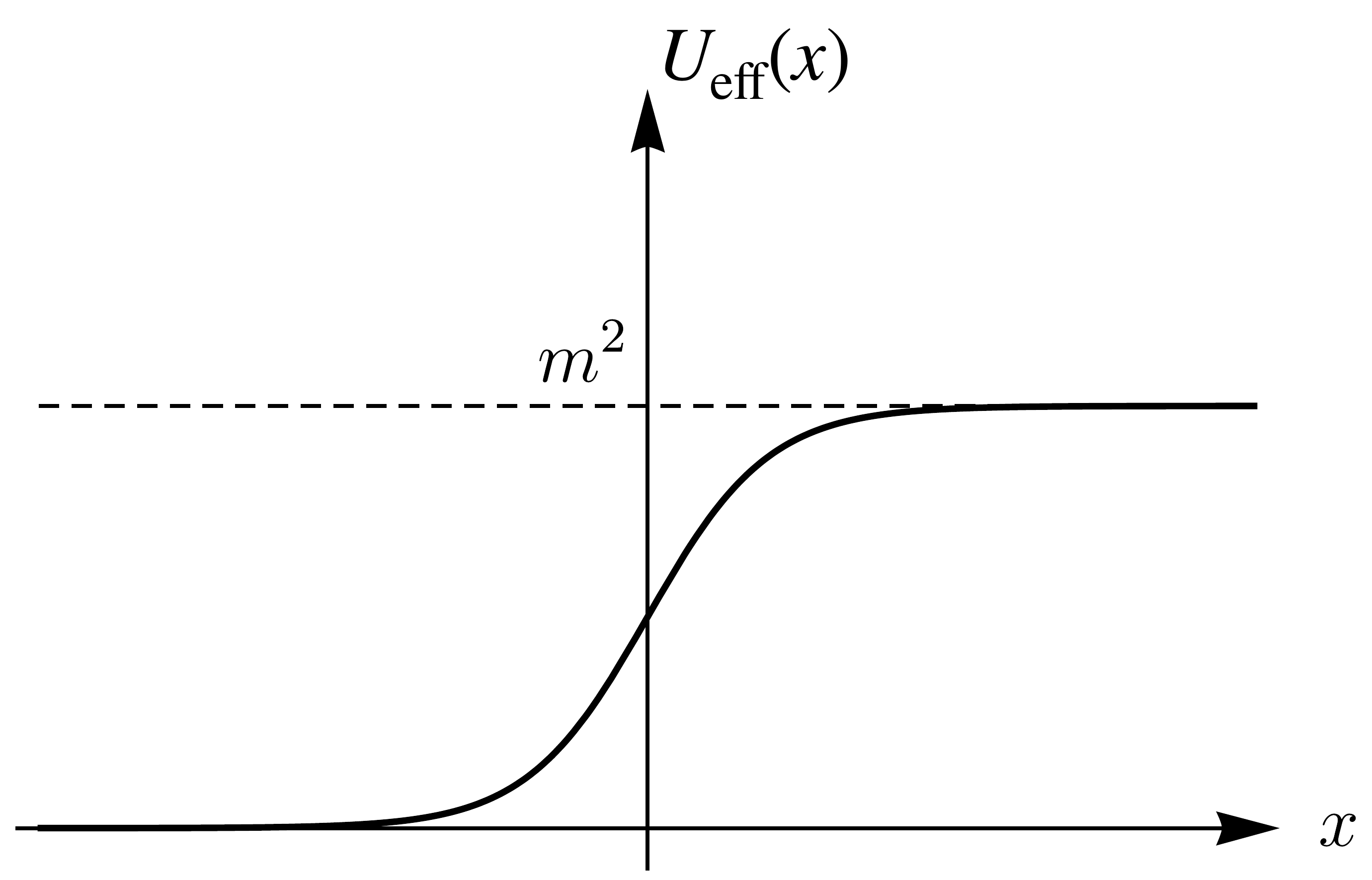} }
	\caption{Potential for massive scalar linear modes in the
          dilaton BH background 
          in two dimensions. The horizon is
          located at $x\to-\infty$.} 
	\label{fig:PotForModes1}
\end{figure}

In this section we study the dynamics of linear perturbations around
the false vacuum. Thus, we replace the potential term by the
free-field part,
\[
V(\vf)\mapsto m^2\vf^2/2\;,
\]
where $m$ is the mass of the field in the false vacuum. This leads to
the linearized field equation
\begin{equation}\label{EoMLin}
\Box \vf-m^2\O\vf=0 \;,
\end{equation}
where $\Box=\eta^{\m\n}\d_\m\d_\n$. The false vacuum is a quantum
state. To define it, we quantize the field $\vf$ using a complete set
of positive- and negative-frequency modes
\begin{equation}\label{Modes}
\vf^{+}_\o(t,x)=f_\o(x)e^{- i\o
  t}~,~~~~~\vf^{-}_\o(t,x)=f^*_\o(x)e^{i\o t}\; , ~~~ \o>0 \;,
\end{equation}
where the mode functions $f_{\o}(x)$ satisfy the eigenvalue equation
\begin{equation}\label{EqModes}
-\frac{d^2f_\o}{dx^2}+m^2\O f_\o=\o^2f_\o \;.
\end{equation}
This is a Schr{\"o}dinger equation with the potential $U_{\rm
  eff}(x)=m^2\O(x)$. The latter is shown in
Fig.~\ref{fig:PotForModes1} for the case of the dilaton BH.

At $\o>m$ equation (\ref{EqModes}) has two linearly-independent
solutions which we denote by $f_{R,\o}$ and $f_{L,\o}$. The first
solution $f_{R,\o}$ reduces to a right-moving plane wave at large
positive $x$: it describes radiation directed outward the
BH. In the near-horizon region $x\to-\infty$ it contains both left-
and right-moving waves. We have
\be
\label{fRoas}
f_{R,\o}=\begin{cases}
\alpha_\o\, \e^{i\o x}+\beta_\o\,\e^{-i\o x}\;, &x\to-\infty\\
\gamma_\o\,\e^{ikx}\;,  &x\to+\infty
\end{cases}
\ee
where 
\begin{equation}\label{k}
k=\sqrt{\o^2-m^2}\;.
\end{equation}
The second mode $f_{L,\o}$ becomes a pure left-moving wave at large
negative $x$: it describes radiation falling into BH. Far away from
the BH it is a sum of two plane waves,  
\be
\label{fLoas}
f_{L,\o}=\begin{cases}
\tilde\beta_\o\,\e^{-i\o x}\;, &x\to-\infty\\
\tilde\gamma_\o\,\e^{ikx}+\tilde\delta_\omega\,\e^{-ikx}\;,  &x\to+\infty
\end{cases}
\;,\qquad\o>m\;.
\ee
The modes $f_{R,\o}$, $f_{L,\o}$ are orthogonal to each other,
\begin{equation}\label{ModeOrth}
\int_{-\infty}^\infty dx\;f_{R,\o}(x)f^*_{L,\o'}(x)=0 \;,
\end{equation}
and are $\delta$-function normalizable. We fix their normalization as follows:
\begin{equation}\label{ModeNorm}
\begin{split}
&\int_{-\infty}^\infty dx\;f_{R,\o}(x)f^*_{R,\o'}(x)=2\pi\delta(\o-\o') \;, \\
&\int_{-\infty}^\infty
dx\;f_{L,\o}(x) f^*_{L,\o'}(x)=2\pi\delta(\o-\o') \;, ~~~ \o>m
\;. 
\end{split}
\end{equation}
As explained in Appendix~\ref{app:Green}, the coefficients of the
asymptotic expansions (\ref{fRoas}), (\ref{fLoas}) are not
independent. They can all be expressed through two parameters
$\beta_\o$ and $\gamma_\o$ which are the reflection and
transmission amplitudes through the potential barrier $U_{\rm
  eff}(x)$. Their absolute values are further related by
Eq.~(\ref{gammabeta}). 

For $\o<m$ only a single $\delta$-function normalizable mode exists,
which is a sum of two plane waves in the near-horizon region and falls
off exponentially at positive $x$. We keep for this mode the notation
$f_{R,\o}$ and still write its asymptotics in the form (\ref{fRoas}),
where $k$ now is purely imaginary, 
\be
\label{kappa}
k=i\sqrt{m^2-\o^2}\equiv i\vk\;.
\ee 
In this case
we clearly have 
\be
\label{modecoeff3}
\alpha_\o=1~,~~~~|\beta_\o|^2=1~,\qquad \o<m\;.
\ee
It is convenient to formally extend the definition of left-moving
modes to $\o<m$ by setting
\be
f_{L,\o}=0~,\qquad\o<m\;.
\ee 
With this convention the completeness condition of the mode basis
reads
\be
\label{completemodes}
\int_0^\infty\frac{d\o}{2\pi}\sum_{I=R,L} f_{I,\o}(x)f^*_{I,\o}(x')
=\delta(x-x')\;.
\ee 
For the concrete choice of the conformal factor (\ref{O_BH}) the modes
can be expressed in terms of the hypergeometric function (see
Eq.~(\ref{Modes_BH}) in Appendix~\ref{app:Green}). Note, however, that
the relations discussed above do not rely on this choice and apply to
modes in
any asymptotically-flat static metric with horizon.

Using the previously introduced modes, we write the quantum field as
\be
\label{phihat}
\hat\vf(t,x) ={\rm g}\int_0^\infty\!\!\frac{d\o}{\sqrt{4\pi\o}}
\sum_{I=R,L}\big[\hat a_{I,\o}\vf^+_{I,\o}(t,x)
+\hat a_{I,\o}^\dagger \vf^-_{I,\o}(t,x)\big]\;.
\ee 
Here $\hat a$, $\hat a^\dagger$ are the annihilation and creation
operators satisfying the usual commutation relations
\be
\label{acomm}
[\hat a_{R,\o},\hat a_{R,\o'}^\dagger]
=[\hat a_{L,\o},\hat a_{L,\o'}^\dagger]=\delta(\o-\o')\;,
\ee
with all other commutators vanishing.
The state annihilated by all $\hat a_{R,\o}$, $\hat a_{L,\o}$ is known as the
Boulware vacuum \cite{Boulware:1974dm},
\be
\label{aBoul}
\hat a_{R,\o}\ket{0}_B=\hat a_{L,\o}\ket{0}_B=0\qquad\qquad \text{(Boulware)}\;.
\ee 
This vacuum is a pure state and is empty from the viewpoint of a static
observer outside the BH. It is well-known, however, that it leads to a
divergent expectation value of the energy-momentum tensor at the
horizon and thus is not a regular state in BH geometry.

Regular states must include entanglement between modes inside and
outside the BH. In the part of spacetime outside BH they correspond to
mixed states. This is the case for the Hartle--Hawking
and Unruh vacua. The former is described by an exactly thermal density
matrix with the Hawking temperature~\cite{Birrell:1982ix}. This
implies that the occupation numbers of the modes follow
the Bose--Einstein distribution,
\be
\label{aHH}
\langle \hat a_{R,\o}^\dagger \hat a_{R,\o'}\rangle_{HH}
=\langle \hat a_{L,\o}^\dagger \hat a_{L,\o'}\rangle_{HH}
=\frac{\delta(\o-\o')}{\e^{2\pi\o/\l}-1}\qquad\qquad\text{(Hartle--Hawking)}\;.
\ee
This vacuum is regular both on the future and past BH horizons. It is
time-reversal invariant and describes a BH in
thermal equilibrium with the environment. It is not suitable to
describe an isolated BH formed by a gravitational collapse.

For the latter physical situation one uses the Unruh vacuum where only
the right-moving modes are thermally populated, whereas the
left-moving modes remain empty,
\be
\label{aUn}
\langle \hat a_{R,\o}^\dagger \hat a_{R,\o'}\rangle_{U}
=\frac{\delta(\o-\o')}{\e^{2\pi\o/\l}-1}\;,~~~~~
\langle \hat a_{L,\o}^\dagger \hat a_{L,\o'}\rangle_{U}=0
\qquad\qquad\text{(Unruh)}\;.
\ee 
The Unruh vacuum is regular at the future horizon and singular at the
past horizon. The latter fact is not a problem, since the past horizon
actually does not exist in the collapsing geometry, being shielded by
the collapsing matter.

\subsection{Time-ordered Green's functions}
\label{Ssec:Green_gen}

In what follows an important role will be played by the time-ordered
Green's functions of the field in various vacua. These are
defined as the time-ordered averages of the field operators in the
respective states,
\begin{equation}
\label{GX_op}
{\GG}(t,x;t',x')=\frac{1}{\gc^2}
\langle T\big(\hat{\vf}(t,x)\hat{\vf}(t',x')\big)\rangle \;.
\end{equation}
They satisfy the Klein--Gordon equation with a $\delta$-function
source
\begin{equation}
\label{KGG}
\big(\Box-m^2\O(x)\big)\GG(t,x;t',x')=i\delta(t-t')\delta(x-x') \; .
\end{equation}
Due to the commutativity of the field operators at coincident times,
the Green's functions are real if $t=t'$. It is straightforward to
express them using the mode decomposition of the field operator.

We start with the Boulware Green's function. An elementary
calculation yields
\begin{equation}\label{G_B}
\GG_B(t,x;t',x') = 
\int_0^\infty\dfrac{d\o}{4\pi \o}\sum_{I=R,L}f_{I,\o}(x) f^*_{I,\o}(x')\,
e^{-i\o |t-t'|}\; ,
\end{equation}
where we have used the relations between the modes and their complex
conjugate, Eqs.~(\ref{conjff}) from Appendix~\ref{app:Green}. Note
that, despite the appearance of an absolute value of the time
difference in Eq.~(\ref{G_B}), $\GG_B$ is an analytic function of $t-t'$
in the complex plane, regular everywhere except the light-cone
singularities on the real axis. To see this, one rewrites $\GG_B$ in
the form
\begin{equation}\label{G_Banalit}
\GG_B(t,x;t',x') =i\int_{-\infty}^\infty \frac{d\tilde{\o}}{2\pi}
\int_0^\infty \frac{d\o}{2\pi}\,
\frac{\sum_{I=R,L}f_{I,\o}(x) f^*_{I,\o}(x')}{\tilde{\o}^2-\o^2+i\epsilon}\e^{-i\tilde{\o}(t-t')}\;.   
\end{equation}  
Now one can rotate $(t-t')$ clockwise into the complex plane,
simultaneously counter-rotating the contour of integration in
$\tilde\o$ to keep the argument in the exponent real.

For the Hartle--Hawking state we use the averages (\ref{aHH}) and obtain
\begin{equation}\label{G_HH}
\begin{split}
\GG_{HH}(t,x;t',x')&=\int_0^\infty\dfrac{d\o}{4\pi\o}
\sum_{I=R,L}f_{I,\o}(x)f^*_{I,\o}(x')
\left[
  \dfrac{e^{-i\o|t-t'|}}{1-e^{-2\pi\o/\l}} +
  \dfrac{e^{i\o|t-t'|}}{e^{2\pi\o/\l}-1} \right]\\
&=\GG_{B}(t,x;t',x')+\int_0^\infty\dfrac{d\o}{4\pi\o}
\sum_{I=R,L}f_{I,\o}(x)f^*_{I,\o}(x')
\frac{\cos\o(t-t')}{e^{2\pi\o/\l}-1}.
\end{split}
\end{equation}
The second expression implies that $\GG_{HH}$ is regular in the strips 
$\{-\frac{2\pi}{\l}<\Im(t-t')<0\}$ and
$\{0<\Im(t-t')<\frac{2\pi}{\l}\}$. It has singularities on the lines
$\Im(t-t')=\pm 2\pi/\l$ that replicate the singularities on the
real axis. In fact, it happens to be periodic in the complex $(t-t')$
plane with the period $2\pi i/\l$ \cite{Hartle:1976tp}.

Finally, for the Unruh Green's function we use the averages
(\ref{aUn}) and after a straightforward calculation using
Eqs.~(\ref{conjff}) arrive at
\begin{equation}\label{G_U}
	\begin{split}
\GG_U(t,x;t',x') = \int_0^\infty &  \dfrac{d\o}{4\pi\o}\bigg\{
f_{R,\o}(x)f^*_{R,\o}(x')\left[
  \dfrac{e^{-i\o|t-t'|}}{1-e^{-2\pi\o/\l}} +
  \dfrac{e^{i\o|t-t'|}}{e^{2\pi\o/\l}-1} \right]\\
&+f_{L,\o}(x)f^*_{L,\o}(x')\e^{-i\o|t-t'|}\\
&+\left(|\b_\o|^2-1\right)
\left[f_{R,\o}(x) f^*_{R,\o}(x') -f_{L,\o}(x)f^*_{L,\o}(x')\right] 
\dfrac{e^{i\o (t-t')}}{e^{2\pi\o/\l}-1}\\
&+\sqrt{\dfrac{k}{\o}}
\left[\gamma_\o\b^*_\o f_{R,\o}(x) f^*_{L,\o}(x')
\!+\!\gamma^*_\o\beta_\o f_{L,\o}(x) f^*_{R,\o}(x')\right] 
\dfrac{e^{i\o (t-t')}}{e^{2\pi\o/\l}-1}\bigg\}.
	\end{split} 
\end{equation}
This expression is somewhat more complicated than in the previous
cases. In the first two lines we recognize the thermal
contributions for the right-moving modes and the vacuum term for
left-movers. In addition, there are terms explicitly depending on the
reflection and transmission amplitudes $\b_\o$, $\gamma_\o$ in the BH
effective potential. In particular, there is a term mixing the left
and right modes. Note that this mixing term disappears in the massless
limit $m=0$ since in that case $\beta_\o=0$.
Writing down $\GG_U$ as a sum of $\GG_B$ and a solution to the
homogeneous Klein--Gordon equation, we conclude that $\GG_U$ is an
analytic function of $(t-t')$ in the strip $|\Im(t-t')|<2\pi/\l$,
apart from the usual singularities on the real axis. 

Further properties of the Green's
functions are studied in Appendix~\ref{app:Green}.

\subsection{Bounce solution and tunneling rate}
\label{Ssec:gen_amplitude}

\begin{figure}[t]
	\centering 
	\includegraphics[width=0.47\textwidth]{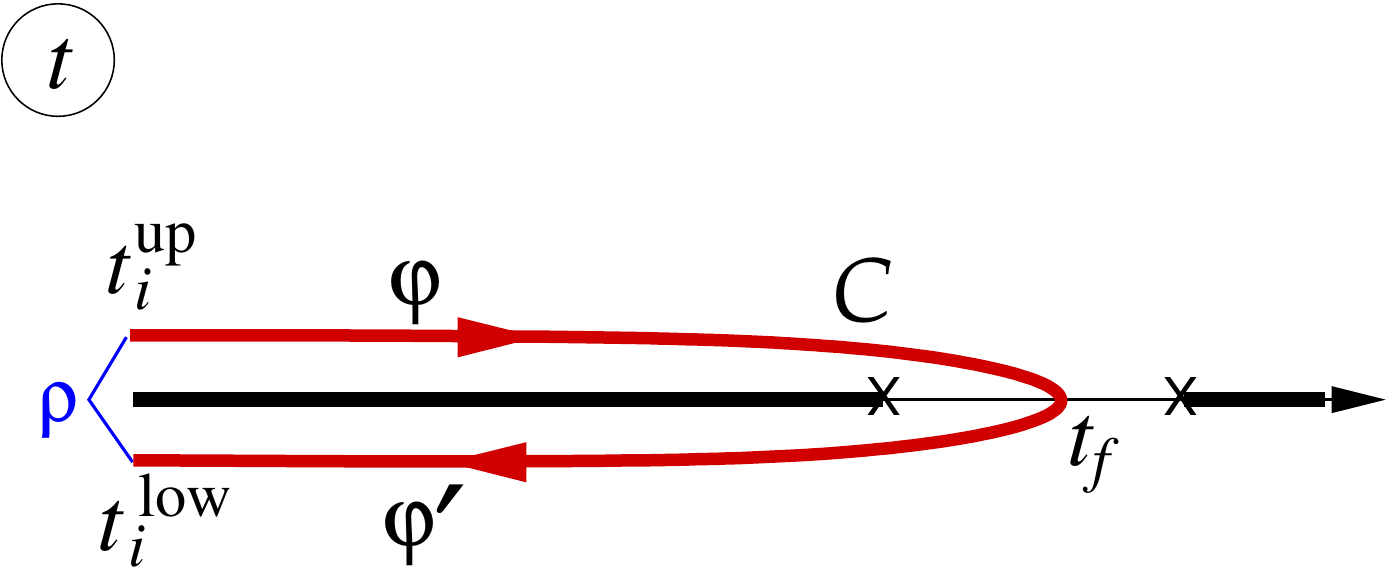}
\qquad\qquad
	\includegraphics[width=0.4\textwidth]{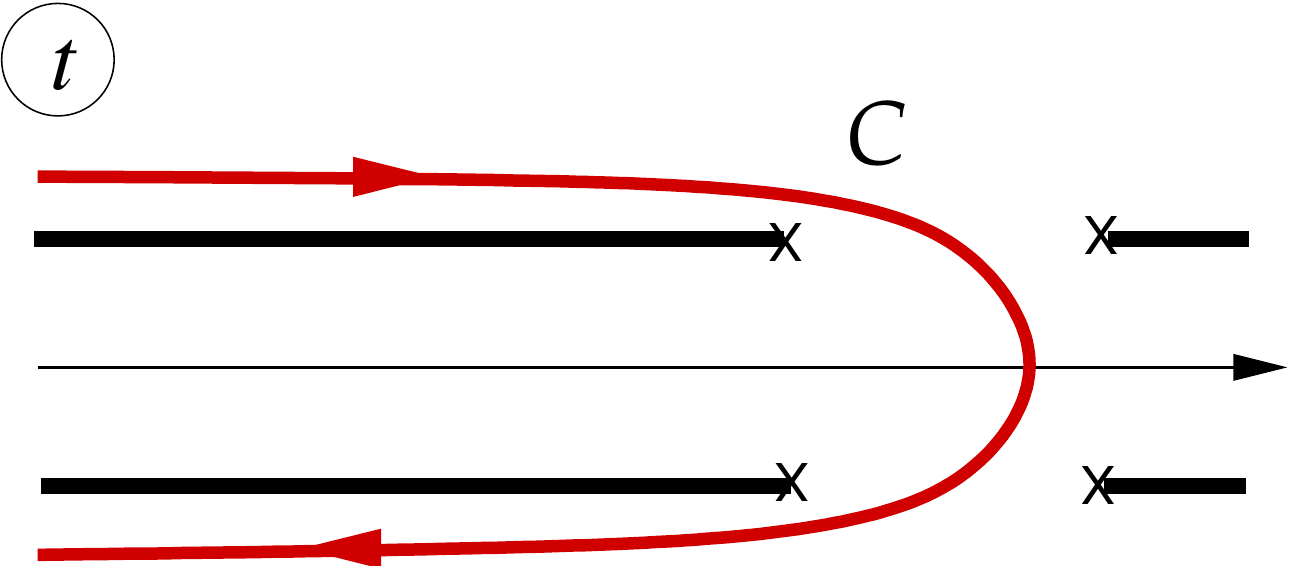}\\
~\\
a \qquad\qquad\qquad\qquad\qquad\qquad\qquad\qquad\qquad\qquad b
	\caption{{\bf (a)} Contour ${\cal C}$ in the complex time plane for the
          calculation of the false vacuum decay probability in the
          in-in formalism. It supports the bounce
          solution in theories with unbounded scalar
          potential. Crosses show the branch-point singularities of
          the bounce. {\bf (b)}~Singularities of the bounce
          in theories with scalar potential bounded
          from below. The contour ${\cal C}$ must be deformed to
          encircle a pair of branch points.} 
	\label{fig:Contour}
\end{figure}

Very generally, the quantum amplitude of transition between an
initial state $\ket{i}$ close to the false vacuum and a final state
$\ket{f}$ in the basin of attraction of the true vacuum is given by
the path integral 
\begin{equation}\label{A}
\braket{f}{i}=\int D[\vf_i(x)] D[\vf_f(x)] D[\vf(t,x)]
\,\braket{f}{\vf_f,t_f} \e^{iS[\vf]}\braket{\vf_i,t_i}{i}\;,
\end{equation}
where $\vf(t,x)$ are field configurations with boundary conditions
$\vf(t_{i,f},x)=\vf_{i,f}(x)$, and $\braket{\vf_i,t_i}{i}$, 
$\braket{f}{\vf_f,t_f}$ are wavefunctions of the initial and final
states in the con\-fi\-gu\-ra\-tion-space representation. Here we
introduced the eigenstates of the field operator,
\be
\label{eigenphi}
\hat\vf(t_i,x)\ket{\vf_i,t_i}=\vf_i(x)\ket{\vf_i,t_i}\;,
\ee
and similarly for $\ket{\vf_f,t_f}$. The initial state $\ket{i}$ is
assumed to belong to the Fock space of the linearized theory around
the false vacuum. The transition probability is obtained by squaring
the amplitude and summing over final states,
\begin{equation}\label{Pdecaypure}
{\cal P}_{\rm decay}=\sum_{f\in{\rm true}} \braket{i}{f}\braket{f}{i} 
\equiv \bra{i}P_{\rm true}\ket{i}\;,
\end{equation}
where $P_{\rm true}$ is a projector on states in the
basin of attraction of the true vacuum. We observe that the tunneling
probability is given by the average of this projector over the initial
state. This average can also be written as a path integral over two
sets of fields $\vf(t,x)$ and $\vf'(t,x)$, such that their values at
$t_f$ coincide, $\vf(t_f,x)=\vf'(t_f,x)=\vf_f(x)$. It is convenient to
think of them as a single field $\vf_{\cal C}$ on a doubly folded
time contour ${\cal C}$ depicted in Fig.~\ref{fig:Contour}a:
$\vf(t,x)$ is the value of the field on the upper side of the contour,
whereas $\vf'(t,x)$ is its value on the lower side. Of course, this is
just the usual representation of averages in the in-in
formalism. Thus, we can write
\be
\bra{i}P_{\rm true}\ket{i}=\int D[\vf_i]D[\vf_i'] D[\vf_{\cal C}]
\,\braket{i}{\vf'_i,t_i}\e^{iS[\vf_{\cal C}]}\braket{\vf_i,t_i}{i}\;,
\ee
where the configuration $\vf_{\cal C}$ is such that it is close to the
true vacuum at $t_f$. Note that we can freely shift the endpoints of
the contour, which we will denote by $t_i^{\rm up}$ and $t_i^{\rm
  low}$, to the upper and lower half-plane of complex time. We
choose them to be complex conjugate, $t_i^{\rm low}=(t_i^{\rm up})^*$.

It is now clear how to generalize this formula to an arbitrary mixed
state described by a density matrix $\varrho$. To compute the decay
probability, we have to average $P_{\rm true}$ with the density
matrix,
\be
\label{Pdecay}
{\cal P}_{\rm decay}=\langle P_{\rm true}\rangle_{\varrho}
=\int D[\vf_i]D[\vf_i']  D[\vf_{\cal C}]
\,\e^{iS[\vf_{\cal C}]}\bra{\vf_i,t_i^{\rm
    up}}\varrho\ket{\vf'_i,t_i^{\rm low}}\;.
\ee
In the semiclassical limit, ${\rm g}\ll 1$, the path integral can be
evaluated in the saddle-point approximation. The saddle point
corresponds to a solution of classical equations of motion on the
contour ${\cal C}$ which we will denote by $\vf_{\rm b}(t,x)$.
It starts from the
vicinity of the false vacuum at $t_i^{\rm up}$, 
evolves along the upper part of the contour to the basin of attraction of
the true vacuum at $t_f$, and then bounces back to the false vacuum
along the lower part of the contour. We refer to this solution as
``bounce''. As discussed below, it provides a generalization of
the Euclidean bounce describing the vacuum decay in flat spacetime
\cite{Coleman:1977py,Callan:1977pt,Coleman:1978ae}.

The boundary conditions for $\vf_{\rm b}$ at $t_i^{\rm up}$ and $t_i^{\rm
low}$ are set by the density matrix $\varrho$, upon taking the
saddle-point integrals in $\vf_i$, $\vf_i'$. We relegate the
derivation of these conditions to Appendix~\ref{app:Vac}. Here we
present the result. When $t_i^{\rm up}$, $t_i^{\rm low}$ have large
negative real part, the bounce solution
linearizes and we can decompose it into the eigenmodes (\ref{Modes}). At
the upper part of the contour ${\cal C}$ we have
\begin{align}
\label{vful}
\vf_{\rm b}\big|_{\Re t_i^{\rm up}\to-\infty}=
\int_0^\infty\frac{d\o}{\sqrt{4\pi\o}}\sum_{I=R,L}
\big[c_{I,\o}^{\rm up}\,\vf_{I,\o}^+(t_i^{\rm up},x)
+\bar c_{I,\o}^{\rm up}\,\vf_{I,\o}^-(t_i^{\rm up},x)\big]\;,
\end{align}
where $c_{I,\o}^{\rm up}$, $\bar c_{I,\o}^{\rm up}$ are constant
coefficients. Similar expansion holds at the lower part of the contour
for $\vf_{\rm b}\big|_{\Re t_i^{\rm low}\to-\infty}$ with the
coefficients $c_{I,\o}^{\rm low}$, $\bar c_{I,\o}^{\rm low}$.
The boundary conditions establish proportionality
between the components of the upper and lower parts, 
\be
\label{cpmrel}
c_{I,\o}^{\rm up}=r_I(\o)\,c_{I,\o}^{\rm low}~,~~~~~
r_I(\o)\,\bar c_{I,\o}^{\rm up}=\bar c_{I,\o}^{\rm low}\;,
\ee
where for different vacua we have
\bseq
\label{rBHHU}
\begin{align}
\label{rB}
&r_R(\o)=r_L(\o)=0& \text{(Boulware),}\\
\label{rHH}
&r_R(\o)=r_L(\o)=e^{-2\pi\o/\l}&\text{(Hartle--Hawking),}\\
\label{rU}
&r_R(\o)=e^{-2\pi\o/\l}~,~~~r_L(\o)=0&\text{(Unruh).}
\end{align}
\eseq
One can simplify these conditions by assuming that the bounce solution
is unique. Then its values on the upper and lower parts of the contour
must be complex conjugate,
\be
\label{vfbreal}
\vf_{\rm b}(t_i^{\rm low},x)=\vf_{\rm b}^*(t_i^{\rm up},x)\;,
\ee
otherwise the complex conjugate configuration $\vf_{\rm b}^*(t^*,x)$
would be a different solution. This implies the relations
between the frequency components, 
$c_{I,\o}^{\rm low}=(\bar c_{I,\o}^{\rm up})^*$, 
$\bar c_{I,\o}^{\rm low}=(c_{I,\o}^{\rm up})^*$, 
so
that Eqs.~(\ref{cpmrel}) reduce to a single condition
\be
\label{cpprel}
c_{I,\o}^{\rm up}=r_I(\o)\,(\bar c_{I,\o}^{\rm up})^*
\ee
imposed on the frequency components on the upper part of the contour. 

We now make the following observation. Consider, instead of
the tunneling probability, the generating functional for the
time-ordered Green's functions of the free theory,
\be
\label{ZJ}
Z[J]=\langle \e^{i(J\cdot\vf)}\rangle_\varrho\;,
\qquad\qquad (J\cdot\vf)\equiv\int
dtdx\sqrt{-g}\, J(t,x)\vf(t,x)\;. 
\ee 
This can also be written in the in-in formalism 
as a path integral along the contour ${\cal
  C}$ from Fig.~\ref{fig:Contour}a,
\be
\label{Zint}
Z[J]=\int D[\vf_i]D[\vf_i'] D[\vf_{\cal C}]\,
\e^{iS^{(2)}[\vf_{\cal C}]+i(J\cdot\vf_{\cal C})}
\bra{\vf_i,t_i^{\rm up}}\varrho\ket{\vf_i',t_i^{\rm low}}\;,
\ee
where $S^{(2)}$ is the quadratic action, and the interval $(t_i,t_f)$
includes the support of the external source $J$. 
Whenever the density matrix $\varrho$ is Gaussian, the integrals are
evaluated by the saddle point. The corresponding classical solution is
given by a convolution of the source with the Green's function, 
\be
\label{vfJ}
\vf_J(t,x)={\rm g}^2\int dt'dx' \GG(t,x;t',x') J(t',x')\;.
\ee
Here the value of $\vf_J$ on the lower part of the contour 
is obtained through the analytic continuation of the
Green's function into the lower half-plane of complex time.
The asymptotic behavior of this solution at $t\to-\infty$ is
determined by the saddle-point integrals 
over $\vf_i$, $\vf_i'$. These are exactly the same as in the derivation
of the boundary conditions for the bounce, implying that the
boundary conditions for the bounce and for the time-ordered
Green's function coincide. Indeed, it is straightforward to check that
the mode decomposition of the Green's functions (\ref{G_B}),
(\ref{G_HH}) and (\ref{G_U}) at $t\to-\infty+i\epsilon$ and $t'$ fixed
satisfies Eq.~(\ref{cpprel}). Being real at $t=t'$, they also satisfy
the relation $\GG(t^*,x;t'^*,x')=\GG^*(t,x;t',x')$, i.e., their values
on the upper and lower parts of the contour ${\cal C}$ are complex
conjugate to each other.

Turning the argument around, one can deduce the boundary conditions
for the bounce from the asymptotics of the time-ordered Green's
function. To this aim, one just needs to find the full set of linear
relations between the frequency components of the solution
(\ref{vfJ}), which hold independently of the choice of the external
source~$J$. 
The mode decomposition of the bounce solution in
the asymptotic past must then obey these relations. Note that this
method is general and can be applied to tunneling from arbitrary mixed
state described by a Gaussian density matrix.  
  
The relation between the properties of the Green's function and the
bounce solution opens the following way to search for the latter. Let
us split the scalar potential into the mass term $m^2\vf^2/2$ and the
interaction part $V_{\rm int}(\vf)$. The bounce satisfies the
classical field equations on the contour ${\cal C}$,
\begin{equation}\label{EoM}
\Box \vf_{\rm b}-m^2\O\vf_{\rm b}-\O V'_{\rm
    int}(\vf_{\rm b}) =0\;,
\end{equation}
where prime on the potential stands for its
derivative with respect to $\vf$.
This can be
recast into an integral equation using the Green's function,
\begin{equation}\label{SolInt}
\vf_{\rm b}(t,x)=-i\int_\mathcal{C}dt' dx'\,
\GG(t,x;t',x')\,\O(x') V'_{\rm int}\big(\vf_{\rm b}(t',x')\big) \; .
\end{equation}
Taking in this expression the time-ordered Green's function
corresponding to a specific vacuum state automatically ensures the
correct boundary condition for the bounce.

In general, the integral equation (\ref{SolInt}) is hard to solve,
perhaps even harder than the boundary value problem (\ref{cpmrel}) for the
differential equation (\ref{EoM}). However, there is a class of
theories where the task is greatly simplified. These are theories
where the nonlinear core of the bounce happens to be much smaller in
size than the inverse mass $m^{-1}$. Then the source in the integral
(\ref{SolInt}) is effectively pointlike and the solution outside
the core is simply proportional to the Green's function. On the other
hand, the core of the bounce can be found by neglecting the
mass. The full solution is obtained by matching the long-distance
asymptotics of the core with the short-distance behavior of the
Green's function. We will encounter precisely this situation in the
toy model studied later in this paper. 

A few comments are in order. First, the
condition (\ref{vfbreal}) implies that the bounce solution is real at
$t=t_f$. If $t_f$ is finite, the solution remains real when
continued 
from that point along the
real time axis. At $t>t_f$ it can be thought of as describing the
evolution of the field after tunneling. On the other hand, the
boundary conditions (\ref{cpprel}) are clearly incompatible with
$\vf_{\rm b}$, being real on the upper side of the contour ${\cal
  C}$. This implies that the bounce must have branch cuts in the
complex time plane which the contour ${\cal C}$ must encircle 
\cite{Rubakov:1992ec,Rubakov:1999ir,Bezrukov:2003tg,Bramberger:2016yog}. 
The details are somewhat different depending on whether the scalar
potential is bounded or not from below, see cases (a) and (b) in
Fig.~\ref{fig:pot}. If the potential is unbounded, the bounce
solution evolved from $t_f$ either forward or backward runs away to
$\vf_{\rm b}=+\infty$ in a finite time. This gives rise to singularities
on the real axis shown by crosses in Fig.~\ref{fig:Contour}a. These
singularities are also branch points and it suffices to draw 
the upper (lower) part of the
contour ${\cal C}$ slightly above (below) the left branch cut. 
We will see this situation realized in our toy model. 
On the other hand, if
the potential is bounded from below, the evolution of the scalar field
along the real axis is regular. The singularities of the bounce
are shifted into the complex plane. Due to the reality of
the solution on the real axis, they come in complex conjugate pairs,
see Fig.~\ref{fig:Contour}b.
The contour ${\cal C}$ should then be deformed to bypass them, 
as shown in the figure.   

Second, it may happen that the bounce solution does not exist for
finite $t_f$. This is the case when tunneling proceeds via formation
of the sphaleron, instead of a direct transition between the false and
true
vacua~\cite{Bezrukov:2003tg,Bezrukov:2003er,Takahashi_2003,
Levkov:2004tf,Levkov:2004ij, 
Takahashi_2005,Levkov:2007yn,Levkov:2008csa,Demidov:2015bua}. 
The tunneling solution can still be found if the
contour ${\cal C}$ is stretched to infinity, which corresponds to
$t_f=+\infty$. Then $\vf_{\rm b}$ must asymptotically approach the same
unstable configuration at $t\to +\infty$ along the upper and
lower parts of the contour.
However, because the
contour actually splits into two disjoint parts, the solutions 
$\vf_{\rm b}^{\rm up}$ and $\vf_{\rm b}^{\rm low}$ need not be
analytic continuations of each other. Still, their mode decompositions
at $\Re t\to -\infty$ must be related by Eqs.~(\ref{cpmrel}) and,
assuming uniqueness of the bounce solution, they must be complex
conjugate, $\vf_{\rm b}^{\rm low}(t,x)=(\vf_{\rm b}^{\rm
  up}(t^*,x))^*$. We will see that bounce solutions of this type
describe false vacuum decay at high temperatures, both in flat
spacetime and in the BH background. They correspond
to the transitions usually associated with thermal jumps onto the
sphaleron.  

Third, the boundary conditions (\ref{cpprel}), as well as the reality
condition (\ref{vfbreal}) are invariant under shifts of time by a real
constant. Therefore, the spectrum of perturbations around the bounce
contains a zero mode associated with time translations. As usual, the
presence of such mode implies that the probability (\ref{Pdecay})
linearly grows with time \cite{Coleman:1977py,Callan:1977pt,Coleman:1978ae}. 
Dividing out this growth,
one obtains the tunneling rate $\varGamma$.  

Last, but not least, we need to know how to calculate $\varGamma$ once
the bounce solution is found. In this paper we are interested only in
the exponential dependence
\begin{equation}\label{DecayRateLO}
\varGamma\sim \e^{-B} \;.
\end{equation}
From Eq.~(\ref{Pdecay}) we see that $B$ is essentially equal to the
imaginary part of the bounce action along the contour ${\cal C}$, plus
boundary terms coming from the initial-state
density matrix. It is shown in Appendix~\ref{app:Vac} that the latter
have the form
\be
\label{bterms}
\frac{i}{2{\rm g}^2}\int_{-\infty}^{\infty} dx\bigg[
\vf_{\rm b}\frac{\d\vf_{\rm b}}{\d t}\bigg|_{t_i^{\rm up}}
+\vf_{\rm b}\frac{\d\vf_{\rm b}}{\d t}\bigg|_{t_i^{\rm low}}\bigg].
\ee  
If we integrate by parts the kinetic term in the bounce action and use
the equation of motion (\ref{EoM}), the quadratic part of the action
and the boundary terms cancel out. We end up with
\begin{align}
B=-\frac{i}{{\rm \gc}^2}\int_{\mathcal{C}}
dt\int_{-\infty}^{\infty} dx\; \Omega(x)\bigg[\frac{1}{2}\vf_{\rm b}V'_{\rm int}(\vf_{\rm
b})-V_{\rm int}(\vf_{\rm b})\bigg],
\label{B}
\end{align}
where the time integral is taken along the contour ${\cal C}$. 
In this form it is manifest that only the region where the bounce
solution is nonlinear contributes to the suppression.

\section{Inverted Liouville potential with a mass term}
\label{Sec:flat}

\begin{figure}[t]
	\centering 
	\includegraphics[width=0.45\textwidth]{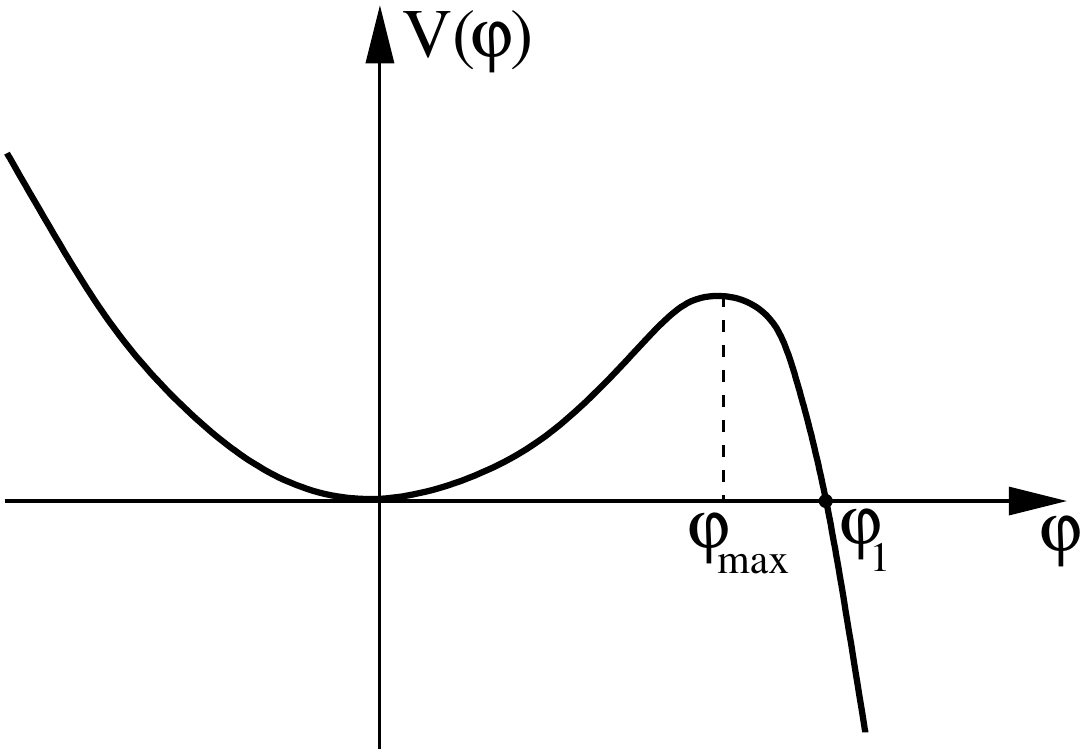}
	\caption{The toy model potential. }
	\label{fig:Potential}
\end{figure}

In the rest of the paper we illustrate the general formalism
of the previous section in a toy model with 
the scalar potential
\begin{equation}\label{Pot}
V(\vf)=\frac{m^2\vf^2}{2}-2\kappa (\e^\vf-1)\;,
\end{equation}
where $m^2,\kappa>0$.
The interaction 
term represents an inverted
Liouville potential and is unbounded from below. The mass term ensures
existence of a local minimum (false vacuum) at $\vf=0$. The constant
piece is chosen in such a way that $V(0)=0$. The potential is shown in
Fig.~\ref{fig:Potential}. 
We assume that the parameters $m$ and $\kappa$ obey the hierarchy
$m\gg\sqrt\kappa$ and, moreover, that the logarithm of their ratio is
large, 
\begin{equation}\label{mvkhier}
\ln\dfrac{m}{\sqrt{\kappa}}\gg 1 \;.
\end{equation}
This technical assumption will be crucial for analytic construction of
the relevant semiclassical solutions.

The potential has local maximum at 
\begin{equation}\label{Vmax}
\vf_{\rm max}\approx \ln{\frac{m^2}{2\kappa}}+\ln{\ln{\frac{m^2}{2\kappa}}}~,~~~~~~
V(\vf_{\rm max})\approx 2m^2\ln^2\frac{m}{\sqrt\kappa} \;,
\end{equation}
where we evaluated $\vf_{\rm max}$ up to doubly logarithmic
corrections, whereas $V(\vf_{\rm max})$ is calculated in the
leading-log approximation.
Above $\vf_{\rm max}$, the potential
quickly drops down and at
$\vf>\vf_1\approx\ln{\frac{m^2}{4\kappa}}+2\ln{\ln{\frac{m^2}{4\kappa}}}$ it
becomes negative. Note that $\vf_1$ differs from $\vf_{\rm max}$ only
by the doubly logarithmic terms. Thanks to the hierarchy
\eqref{mvkhier}, the theory possesses two intrinsic energy scales: the
mass scale $m$ and the scale associated with the barrier
$m\ln\frac{m}{\sqrt{\kappa}}$. Both will play an important role in the
studies of tunneling solutions in different environments. 

We start by studying the dynamics of the model
in flat spacetime. The equation of motion reads,
\begin{equation}\label{flateq}
\Box \vf-m^2\vf+2\kappa\,\e^{\vf}=0\;.
\end{equation}
For large $\vf\gtrsim \vf_{\rm max}$ one can neglect the mass term
and the equation reduces to the Liouville equation which has a general
solution 
\begin{equation}\label{Liouvgen}
\vf= \ln\bigg[\frac{4 F'(-u)G'(v)}{\big(1+\kappa F(-u)G(v)\big)^2}\bigg]\;,
\end{equation}
where $u$, $v$ are the advanced and retarded coordinates (\ref{uv}),
$F(-u)$, $G(v)$ are arbitrary functions, and primes stand for the
derivatives of these functions with respect to their arguments. On the
other hand, at $\vf\lesssim \vf_{\rm max}$ the mass term dominates and
the solution is the same as for the free massive theory. To find the
solution of the full Eq.~(\ref{flateq}), we adopt the strategy of
asymptotic expansion and matching. We will look for solutions in the
form (\ref{Liouvgen}) (in the form of a free massive field) in the
region where the second (third) term in (\ref{flateq}) can be
neglected. 
These two forms of solution will be patched together in
the overlapping region where they are both valid. The condition
(\ref{mvkhier}) will be instrumental to ensure that such overlap
region exists.

We now consider several solutions relevant for the false vacuum decay.

\subsection{Sphaleron}
\label{Ssec:flat_sphaleron}

Let us find the static unstable solution of Eq.~(\ref{flateq}) --- the
sphaleron $\vf_{\rm sph}$. This solution can decay either to the true
or to the false vacuum, so it can be thought of as sitting on the saddle
of the potential energy functional separating the two
vacua. The sphaleron energy gives the height of the
energy barrier between the vacua.

Without loss of generality, we can place the center of the sphaleron
at $x=0$. Then in the region $|x|\ll m^{-1}$ we can neglect the mass
term and the solution reads
\begin{equation}\label{sphaleronin}
\vf_{\rm sph}\Big|_{|x|\ll m^{-1}}=\ln\bigg[
\frac{\Lambda_0^2}{\kappa\ch^2(\Lambda_0 x)}\bigg]\;.
\end{equation}
Here $\Lambda_0$ is a constant which must be fixed from matching with
the long-distance solution. Assuming $\Lambda_0\gg m$, we expand
(\ref{sphaleronin}) at $\Lambda_0^{-1}\ll |x|\ll m^{-1}$ and obtain
\begin{equation}\label{sphaleroninas}
\vf_{\rm sph}\approx -2\Lambda_0 |x|+\ln(4\Lambda_0^2/\kappa)\;.
\end{equation}
On the other hand, in the outer region $|x|\gg \Lambda_0^{-1}$
the sphaleron is a solution to the free massive equation,
\begin{equation}\label{sphaleronout}
\vf_{\rm sph}\Big|_{|x|\gg \Lambda_0^{-1}}=A_{\rm sph}\e^{-mx}\;,
\end{equation}
where $A_{\rm sph}$ is another constant. At 
$\Lambda_0^{-1}\ll |x|\ll m^{-1}$ it becomes
$\vf_{\rm sph}=-A_{\rm sph}m |x|+ A_{\rm sph}$. Comparing this expression with 
(\ref{sphaleroninas}), we obtain $A_{\rm sph}=2\Lambda_0/m$ and an equation
determining $\Lambda_0$, 
\begin{equation}\label{Csol}
\frac{\Lambda_0}{\ln(2\Lambda_0/\sqrt\kappa)}=m~~~\Longrightarrow~~~
\Lambda_0=m\bigg(\ln{\frac{2m}{\sqrt{\kappa}}}
+\ln\ln{\frac{2m}{\sqrt{\kappa}}}+\ldots\bigg)\;. 
\end{equation}
We see that under the condition (\ref{mvkhier}) our assumption
$\Lambda_0\gg m$ is indeed justified. Note that the sphaleron has the
following structure: a narrow nonlinear core of the size
$\Lambda_0^{-1}$, where the field reaches $\vf_{\rm sph}\sim \vf_1$,
and a wide tail (\ref{sphaleronout}), where the field is
linear. This structure will be recurrent in the other
semiclassical solutions that we consider below.   

To calculate the sphaleron energy, it is convenient to integrate by
parts the gradient term in the standard expression for the energy and
use the equation of motion. This yields (up to a negligible
contribution of order $\mathcal{O}(\kappa/m)$)
\begin{equation}\label{SphEn}
E_{\rm sph}=\frac{1}{{\rm \gc}^2}\int_{-\infty}^\infty dx\,\kappa\,
(\vf_{\rm sph}-2)\, \e^{\vf_{\rm sph}}\;.
\end{equation}
The integral is saturated by the nonlinear core and substituting 
Eq.~(\ref{sphaleronin}), we obtain
\begin{equation}\label{Esphfin1}
E_{\rm sph}=\frac{4\Lambda_0}{{\rm
    \gc}^2}\bigg(\ln\frac{\Lambda_0}{\sqrt\kappa}
-2+\ln{2}\bigg)\approx 
\frac{4m}{{\rm \gc}^2}\bigg(\ln\frac{m}{\sqrt\kappa}\bigg)^2\;,
\end{equation}
where the last expression is written in the leading-log
approximation. Notice that the sphaleron energy is doubly enhanced: by
the inverse of the small coupling constant ${\rm \gc}$ and by the large
logarithm $\ln(m/\sqrt\kappa)$.

\subsection{Tunneling from Minkowski vacuum}
\label{Ssec:flat_bounce}

Next, we consider the bounce solution describing the false vacuum
decay in empty Minkowski spacetime. We first adopt the standard
Euclidean approach and then show how it is related to the in-in method
developed in Sec.~\ref{Sec:gen}.

In the standard approach, the bounce represents a saddle point of the
Euclidean partition function
\cite{Coleman:1977py,Callan:1977pt,Coleman:1978ae}. It 
is a solution of the field equations obtained upon Wick rotation of
the time variable to purely imaginary values, $t\mapsto -i\tau$. The
solution $\vf_{\rm b}$ is assumed to be real for real $\tau$, vanish
at infinity, and have zero time derivative at $\tau=0$. The latter
property ensures that the analytic continuation of the
bounce onto the real time axis is real and describes the evolution
of the field after tunneling. It is customary to assume that the
bounce with the smallest Euclidean action, and hence giving the least
suppressed channel for the vacuum decay, is spherically-symmetric in
the Euclidean spacetime.\footnote{This assertion has been widely
  discussed in the literature and proven under various
  assumptions. See \cite{Coleman:1977th,Blum:2016ipp} for the
  proof in $d>2$ spacetime dimensions and
  \cite{2008arXiv0806.0299B} 
for the
  proof including the $d=2$ case.} 
This means that the bounce depends only on $\rho=\sqrt{x^2+\tau^2}$
and obeys the equation 
\begin{equation}
\d^2_\rho\vf_{\rm b}+\frac{1}{\rho}\d_\rho\vf_{\rm b}-m^2\vf_{\rm
  b}+2\vk\, \e^{\vf_{\rm b}} =0 \;.
\end{equation}

We again use the strategy of splitting the solution into a core and a
tail and matching them in the overlap. At $\rho\ll m^{-1}$ we neglect
the mass term and obtain
\begin{equation}\label{bouncein}
\vf_{\rm b}\Big|_{\rho\ll m^{-1}}=
\ln\left[\frac{4C^2_{M}}{(1+\kappa C_M^2\rho^2)^2}\right] \;,
\end{equation}
where $C_M$ is a constant. This corresponds to the following choice of linear
functions $F$, $G$ in the general solution (\ref{Liouvgen}):
\begin{equation}\label{fg_flat}
F(z)=C_M\, z \;, ~~~ G(\bz)=C_M\,\bz \;,
\end{equation}
where we have adapted the notations to the Euclidean signature,
\begin{equation}\label{zzbar}
-u\mapsto z=x+i\tau \;,~~~~v\mapsto \bz=x-i\tau \;.
\end{equation}
At $(C_M\sqrt{\kappa})^{-1}\ll\rho\ll m^{-1}$ the core solution becomes 
\begin{equation}
\label{bouncein1}
\vf_{\rm b} \approx -4\ln (\sqrt\kappa \rho)-2\ln C_M+2\ln 2\;.
\end{equation}
On the other hand, the tail is given by the solution of the free
massive equation,
\begin{equation}\label{bounceout}
\vf_{\rm b}\Big|_{\rho\gg (C_M\sqrt\kappa)^{-1}} =A_M K_0\left(m\rho\right)\;, 
\end{equation}
where $K_0$ is the modified Bessel function of the second
kind and $A_M$ is another constant. Expansion at small $\rho$ gives
$\vf_{\rm b}\approx 
-A_M\ln (m\rho) +A_M(\ln 2-\gamma_E)$, where $\gamma_E$ is the Euler
constant. Comparing with Eq.~(\ref{bouncein1}), we obtain $A_M=4$ and
\begin{equation}\label{Aabounce}
C_M=\frac{m^2}{2\kappa}\e^{2\gamma_E}\;.
\end{equation}
We can now verify a posteriori that the matching region exists. The
condition is $C_M\sqrt\kappa\gg m$, which is indeed implied by our
assumption (\ref{mvkhier}). 

\begin{figure}[t]
	\centering 
	\includegraphics[width=0.45\textwidth]{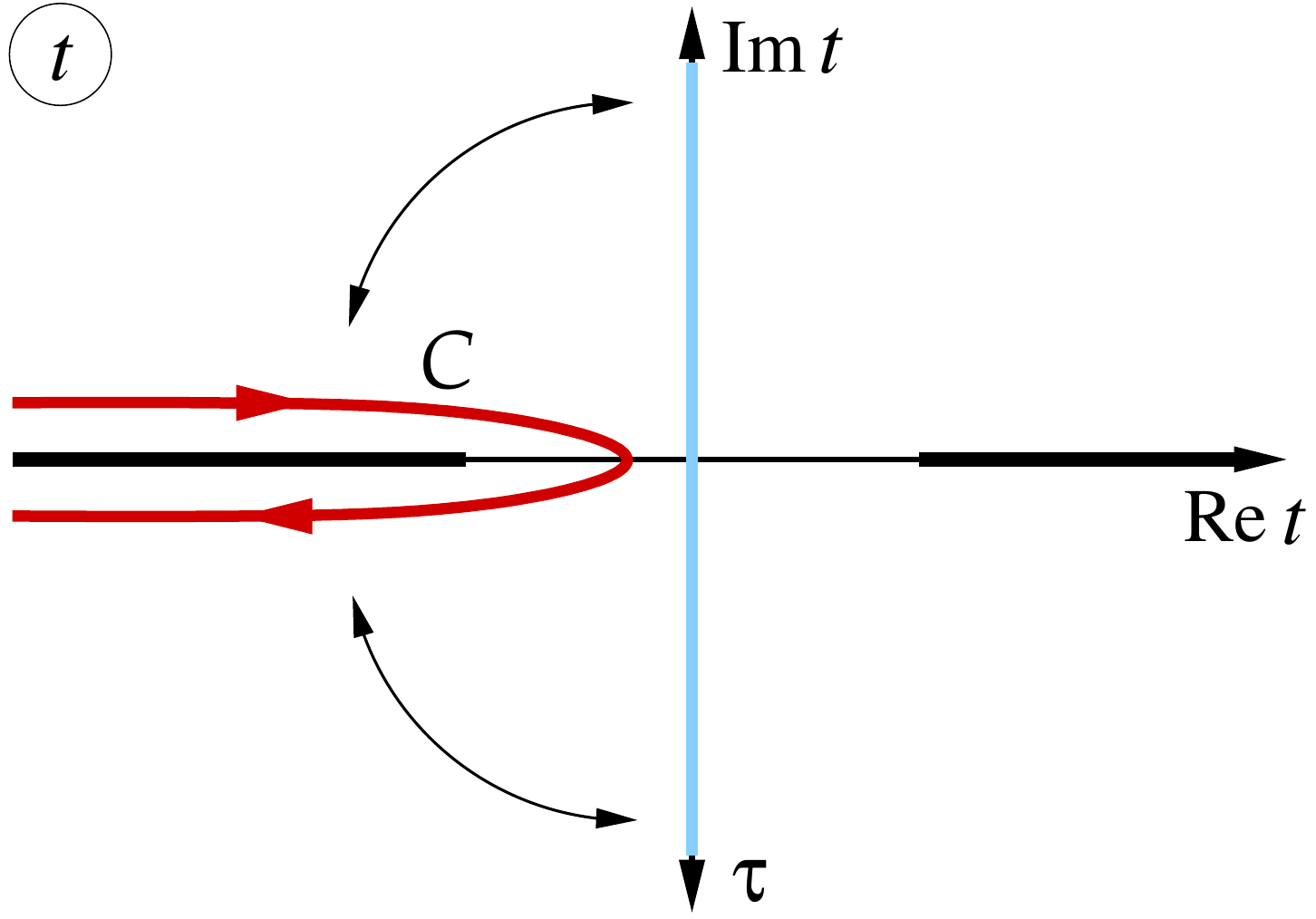}
	\caption{Structure of Minkowski bounce in the complex time
          plane. The standard Euclidean bounce is defined on the
          imaginary time axis (blue). Its analytic continuation to the
        contour ${\cal C}$ (red) satisfies Feynman boundary conditions
      at $\Re t\to -\infty$. Thick black lines show the branch cuts.} 
	\label{fig:bounceM}
\end{figure}

Let us see how the above results are reproduced by the method of
Sec.~\ref{Sec:gen}. We notice that the core of the solution
(\ref{bouncein}) is an analytic function of complex time with branch-cut singularities on the real axis at
\be
\label{bouncesing}
t=\pm t_{M,s}(x)~,~~~~ t_{M,s}(x)=\sqrt{x^2+(C_M\sqrt\kappa)^{-1}}\;.
\ee
Further, the tail of the solution (\ref{bounceout}) is proportional
to the analytic continuation to the Euclidean time of the Feynman
Green's function\footnote{This can be obtained from
  Eq.~(\ref{G_B}) by substituting the plane-wave mode
  functions,
$$
f_{R,\o}=\sqrt{\o/k}\,\e^{ikx} \;, ~~~
f_{L,\o}=\sqrt{\o/k}\,\e^{-ikx} \;, ~~~ \o>m \;. 
$$
}
\begin{equation}\label{G_F_flat}
\GG_{ F}(t,x;0,0) = \frac{1}{2\pi}
K_0\left(m\sqrt{x^2-t^2+i\epsilon}\right) \; .
\end{equation}
This implies that the bounce solution can be analytically continued to
the whole complex plane of $t$ with only singularities at
(\ref{bouncesing}), see Fig.~\ref{fig:bounceM}. In particular, it is
defined on the contour ${\cal C}$ introduced in
Sec.~\ref{Ssec:gen_amplitude}. Moreover, at the endpoints of this
contour it obeys the Feynman boundary conditions, as required for the
tunneling from vacuum. We conclude that for the problem at hand the
tunneling solution given by the method of
Sec.~\ref{Ssec:gen_amplitude} and the standard Euclidean bounce are
just different representations of the same analytic function ---
simply stated, they coincide.

The singularity of the bounce solution at $t>0$ has a natural physical
interpretation. It corresponds to the run-away of the field towards
$\vf=+\infty$ after tunneling. We observe that it is mirrored by a
twin singularity at $t<0$. The latter does not appear to have any
transparent physical meaning. However, as discussed in
Sec.~\ref{Ssec:gen_amplitude}, its presence is necessary for
existence of a nontrivial tunneling solution on the contour~${\cal
  C}$.  

To compute the tunneling suppression, we can either integrate the bounce
action in the Euclidean time, as in the standard approach, or use the
integral (\ref{B}) along the contour ${\cal C}$. The two results will
coincide, because we can continuously deform the contour ${\cal C}$
into the imaginary time axis, and vice versa,\footnote{To get exactly
  the same integral, one has to integrate the Euclidean action by
  parts and use the field equations, as it was done in the derivation
  of Eq.~(\ref{B}). These manipulations do not alter the value of the
action as the corresponding boundary terms vanish.} 
see
Fig.~\ref{fig:bounceM}. Notice that the integrals along the arcs
at infinity vanish. Indeed, at $|t|\to \infty$, $\Im t\neq 0$ the
field linearizes and does not contribute into the tunneling
suppression, as is clear from the expression (\ref{B}). The result reads
\begin{equation}\label{B0}
B_M=\frac{16\pi}{{\rm
\gc}^2}\bigg(\ln\frac{m}{\sqrt\kappa} +\gamma_E-1\bigg)\;. 
\end{equation}
We observe that the suppression is enhanced by the large logarithm
(\ref{mvkhier}). This contribution can be traced back to the large
value of the field at the core of the bounce 
$\vf_{\rm b}(0)=4\big(\ln(m/\sqrt\kappa)+\gamma_E\big)$. It is easily
computed by replacing the field in the integral for $B$, when it
appears outside the exponent, by its value at the core and taking the
resulting integral with residues,
\be
\begin{split}
B_M\Big|_{\rm leading-log}&=\frac{i}{{\rm g}^2}
\int_{\cal C}dt\int_{-\infty}^\infty dx\,\kappa\,\vf_{\rm b}(0)\,
\e^{\vf_{\rm b}(t,x)}\\
&=\frac{2\pi \vf_{\rm b}(0)}{{\rm g}^2}
\int_{-\infty}^\infty
dx\,\kappa\,\Res_{t=-t_{M,s}(x)} 
\e^{\vf_{\rm b}(t,x)}=\frac{4\pi}{{\rm g}^2}\vf_{\rm b}(0)\;.
\label{B0ll}
\end{split}
\ee
One can use this replacement to quickly get the leading-log contribution
to the suppression in the cases when the full calculation may be 
complicated.

\subsection{Thermal transitions in flat spacetime}
\label{Ssec:flat_thermal}

Here we study false vacuum decay in flat spacetime at finite
temperature. 
The results of this section will be important in what follows for the
analysis of instanton solutions in the 
Hartle--Hawking and Unruh vacua. To make contact with those cases, we
denote the temperature by $\l/(2\pi)$ and assume $\l$ to be much
larger than~$m$.

We again begin with the standard approach which prescribes to look for
a real solution of the Euclidean field equation periodic in
Euclidean time $\tau$ with the period $2\pi/\l$. This periodic instanton
is the saddle point of the thermal partition function. We make an
educated guess for the functions $F$ and $G$ describing the core of
the instanton,   
\begin{equation}\label{fg_th}
F(z)=\dfrac{C_{th}}{\l}\left(\e^{\l z}-d_{th}\right) \;, ~~~~~
G(\bz)=\dfrac{C_{th}}{\l}\left(\e^{\l\bz}-d_{th}\right) \;, 
\end{equation}
with real constants $C_{th}$, $d_{th}$. Substituting into the
expression for the field (\ref{Liouvgen}), after some elementary
manipulations we obtain
\begin{equation}\label{PIin2}
\vf_{\rm b}\Big|_{\rm core}=\ln\left[
\frac{\l^2b_{th}}{\kappa\left(\ch \l x-\sqrt{1-b_{th}}\cos\l\tau\right)^2}\right] \;,
\end{equation}
where we have denoted 
\be
\label{bth}
b_{th}=\frac{\l^2}{\kappa C^2_{th}}
\ee
and have placed the center of the instanton at $x=0$ by setting 
$d_{th}=\sqrt{1-b_{th}}$. We have provisionally denoted the solution
as $\vf_{\rm b}$, and we will see shortly that it indeed describes
the bounce in the sense of Sec.~\ref{Ssec:gen_amplitude}.
The solution is real as long as $b_{th}<1$. 

Let us first assume that $b_{th}\ll 1$. Then at $|x|,|\tau|\gtrsim
1/\l$ the solution becomes 
\begin{equation}\label{PIin}
\vf_{\rm b}\approx -2\ln \left[4\sh\bigg(\frac{\l z}{2}\bigg) \sh\bigg(\frac{\l \bar
    z}{2}\bigg)\right]
+\ln\left[\frac{4\l^2 b_{th}}{\kappa}\right] \;.
\end{equation}
This has the same form as the Wick rotated
thermal Green's function when its two arguments are separated by less
than $1/m$ (``close separation''), see Eq.~(\ref{G_th_as}). Thus,
the tail of the instanton is given by this Green's
function,
\be
\label{PIout}
\vf_{\rm b}\Big|_{\rm tail}=8\pi \GG_{th} (-i\tau,x;0,0)\;,
\ee 
where the proportionality coefficient has been fixed by matching the
singular part of $\GG_{th}$. Comparing the constant pieces in
(\ref{PIin}) and (\ref{G_th_as}), we fix
\begin{equation}\label{aTh}
b_{th}=\frac{\kappa}{4\l^2} \e^{\frac{2\l}{m}} \;,
\end{equation}
which is indeed small if the temperature does not exceed a certain
critical value. It is easy to see that the latter coincides with
$\Lambda_0$ determined by Eq.~(\ref{Csol}).

This is not an accident: when $\l$ approaches $\Lambda_0$ and $b_{th}$
approaches $1$, the periodic instanton becomes $\tau$-independent and
degenerates into the sphaleron. The matching procedure used above
becomes problematic in this limit and fails to describe this
transition since the region where the instanton can be written in the
form (\ref{PIin}) seizes to exist in Euclidean time (we will see shortly how
to remedy this problem). Still, the transition of periodic
instantons into the sphaleron is expected on general grounds. It is
well-known that in field theory at high temperature there are no
nontrivial periodic instantons and the false vacuum decay proceeds by
thermal jumps over the barrier separating it from the true vacuum
\cite{Linde:1981zj}. The
probability of the latter process is suppressed by the Boltzmann
exponent involving the height of the barrier, i.e., the sphaleron
energy, divided by the temperature, 
$\Gamma\sim\exp(-2\pi E_{\rm sph}/\l)$. Alternatively, the suppression
can be obtained as the sphaleron action over the Euclidean time
interval~$2\pi/\l$.  

\begin{figure}[t]
	\centering 
	\includegraphics[width=0.3\textwidth]{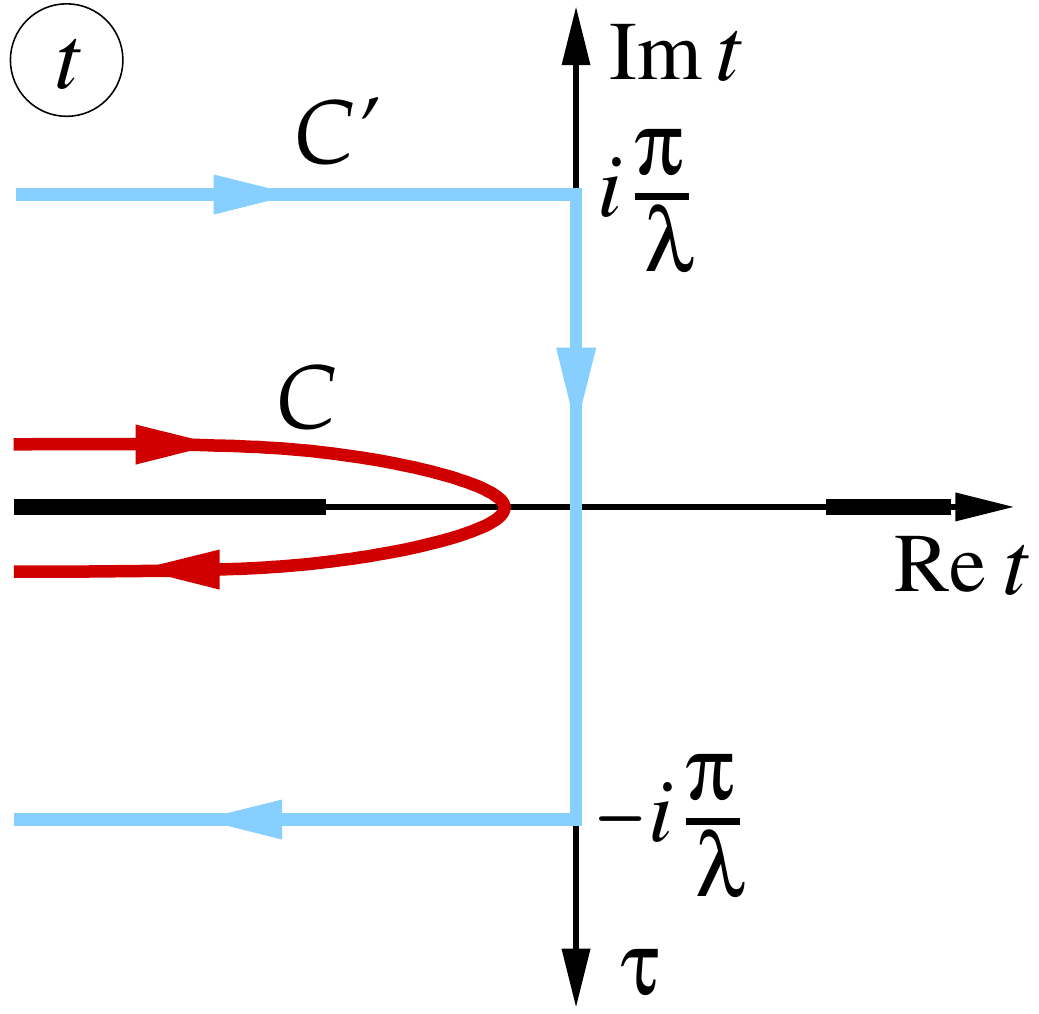}~~~~
\includegraphics[width=0.3\textwidth]{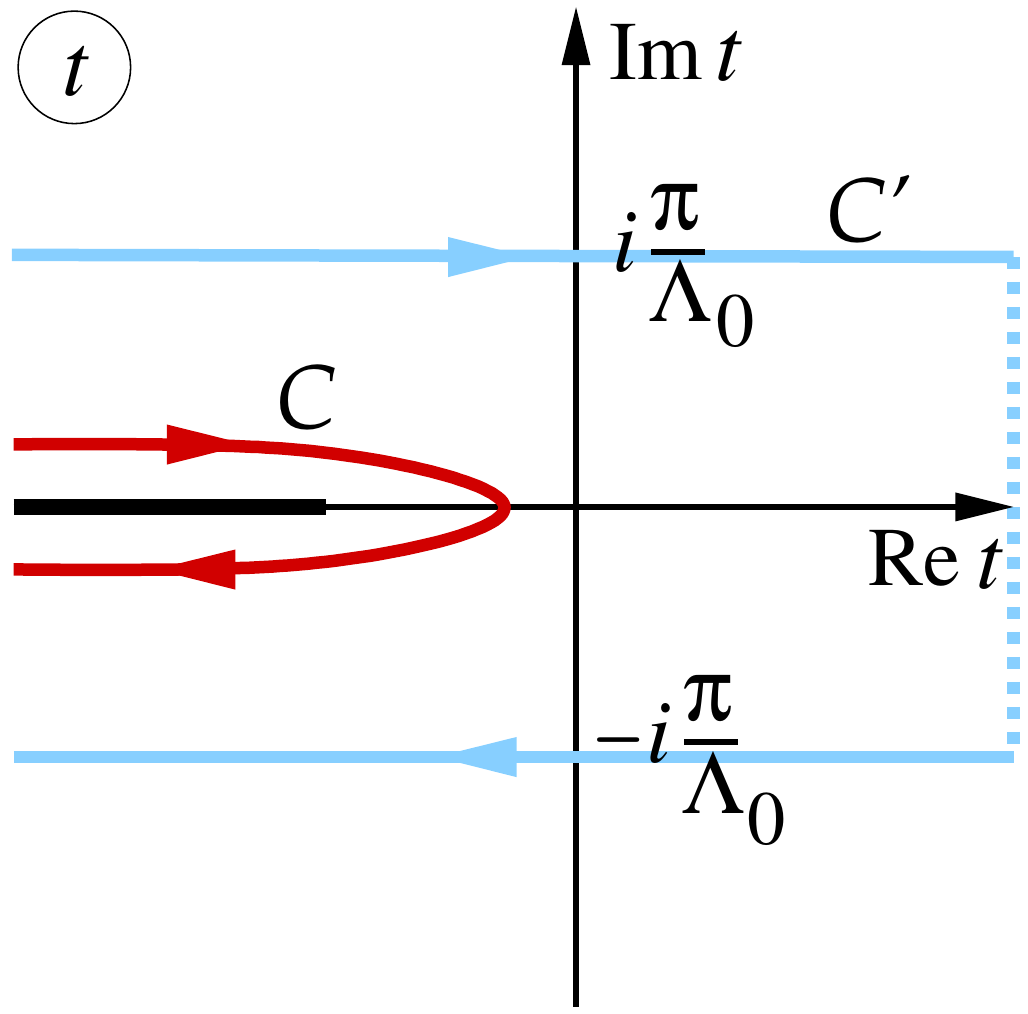}~~~~
\includegraphics[width=0.3\textwidth]{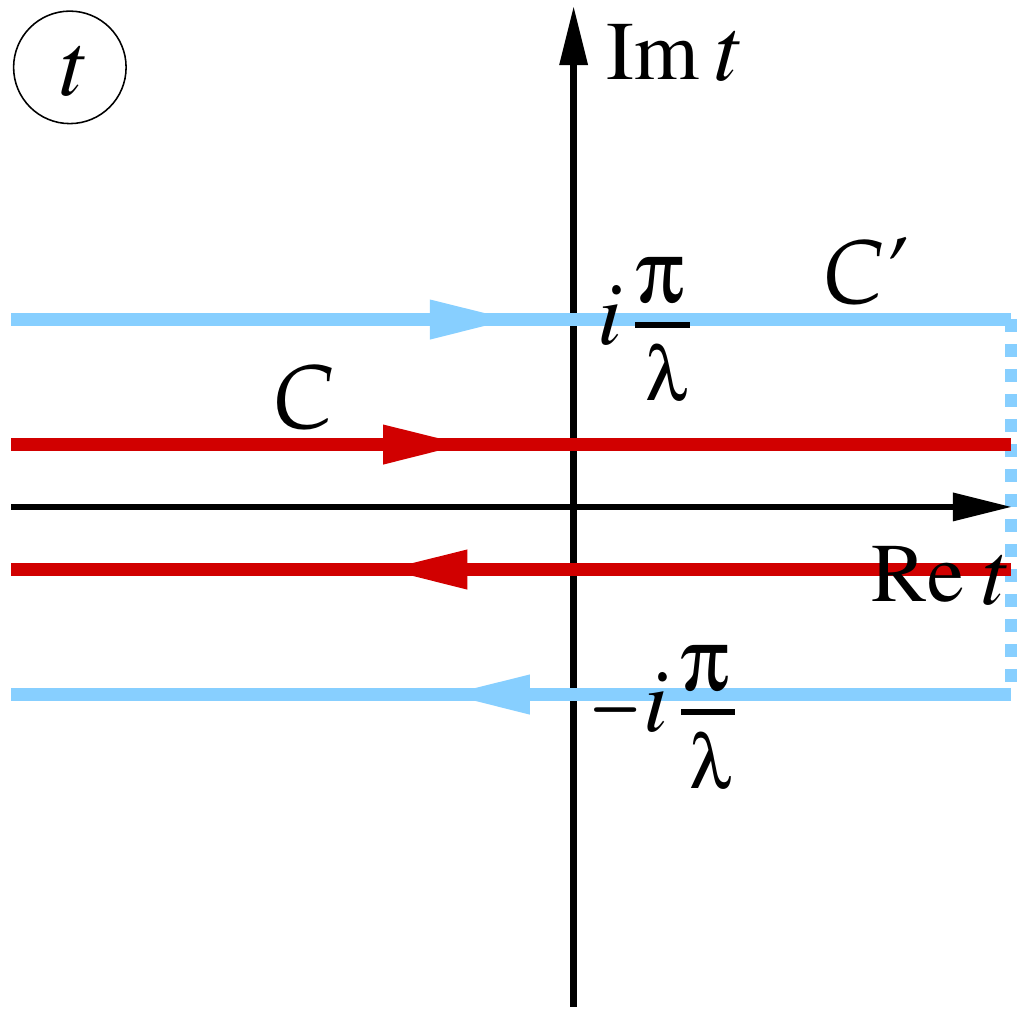}\\
~\\
\quad~ a\qquad\qquad\qquad\qquad\qquad\qquad\quad\, b \qquad\qquad\qquad\qquad\qquad\quad\quad~~ c\quad~
	\caption{Structure of the thermal bounce in the complex time
          plane. {\bf (a)} Low temperature solutions at $\l<\Lambda_0$
          correspond to periodic instantons in Euclidean time. At
          real $t>0$ they describe run-away of the field $\vf$ to the true
          vacuum at $\vf=+\infty$. {\bf (b)} The bounce solution at
          $\l=\Lambda_0$ tends to the sphaleron at $\Re
          t\to+\infty$. {\bf (c)} Solutions at $\l>\Lambda_0$ describe
          tunneling onto the sphaleron. They are given by different
          analytic functions on the upper and lower parts of the
          contour ${\cal C}$ that runs from $t=-\infty$ to $t=+\infty$
          and backward. See the text for more details.} 
	\label{fig:bounceth}
\end{figure}

Let us now reinterpret the above results along the lines of
Sec.~\ref{Sec:gen}. Periodic instanton (\ref{PIin2}) is an analytic
function of complex time with branch cuts on the real
axis starting at\footnote{These cuts are periodically replicated at $\Im t=2\pi
  n/\l$ with integer $n$, but only those with $n=0$ are relevant for
  our discussion.}
\be
\label{PIcuts}
t=\pm t_{th,s}(x)~,~~~~
t_{th,s}(x)=\frac{1}{\l}\arcch\left(\frac{\ch\l x}{\sqrt{1-b_{th}}}\right)\;,
\ee
see Fig.~\ref{fig:bounceth}a. This structure is similar to that of the
vacuum Minkowski bounce (cf. Fig.~\ref{fig:bounceM}). The singularity
at $t>0$ corresponds to the run-away of the field after tunneling. On
the other hand, the branch cut at $t<0$ ensures the correct
asymptotics 
of the solution along the contour ${\cal C}$.
Indeed, in the far
past the solution linearizes and coincides with the thermal Green's
function, see Eq.~(\ref{PIout}). Therefore, its mode decomposition on
the upper part of the contour 
(\ref{vful}) satisfies the relations (\ref{cpmrel}) with the thermal
coefficient (\ref{rHH}). Thus, the periodic
instanton, analytically continued onto the contour ${\cal C}$, 
satisfies all the requirements on the bounce solution formulated in
Sec.~\ref{Ssec:gen_amplitude}.

The corresponding tunneling suppression can be computed along the
contour ${\cal C}$. It is convenient, however, to deform the latter
into the contour ${\cal C}'$ shown with blue in
Fig.~\ref{fig:bounceth}a. Due to the periodicity of $\vf_{\rm b}$ in
complex time, the integrals over the parts of this contour at $\Im
t=\pm \pi/\l$ cancel each other and we are left with the contribution
along the portion of the imaginary time axis from $t=i\pi/\l$ to
$t=-i\pi/\l$. This is nothing but the Euclidean action of the periodic
instanton over a single period. We evaluate it in Appendix~\ref{app:Bcalc} with
the result
\begin{equation}\label{SPI}
B_{th}=\dfrac{16\pi}{\gc^2}\left(
  \ln\dfrac{\l}{\sqrt{\kappa}}-\dfrac{\l}{2m}+\ln 2-1\right)~,
~~~~~ \l< \Lambda_0 \;. 
\end{equation}
Notice that for $\l\sim m$ we recover the vacuum suppression
(\ref{B0}) in the leading-log approximation. The $\mathcal{O}(1)$-terms are
different, because in deriving (\ref{SPI}) we used the assumption
$\l\gg m$.  

When $b_{th}\to 1$, i.e., when we approach the sphaleron regime, the
singularities (\ref{PIcuts}) move away from the origin and at
$b_{th}=1$ run to infinity. One may be puzzled how one can obtain a
bounce solution with correct asymptotics in this case, given that the
sphaleron is time-independent and thus never linearizes. The answer is
simple: one just needs to slightly modify the limit by shifting the
periodic instanton in time in such a way that the left singularity is
kept at a finite distance. Namely, one makes a replacement
\be
t\mapsto t+\frac{1}{2\l}\ln(1-b_{th})\;,
\ee 
so that the core solution (\ref{PIin2}) becomes
\be
\label{PIin3}
\vf_{\rm b}\Big|_{\rm core}=
\ln\bigg[\frac{\l^2 b_{th}}{\kappa\left(\ch\l x-\tfrac{1}{2}\e^{-\l t}
-\tfrac{1}{2}(1-b_{th})\e^{\l t}\right)^2}\bigg]\;.
\ee
This can now be matched to the tail (\ref{PIout}) at $\Re t<0$ for any
values of $b_{th}$, not necessarily small. The matching is performed
in the region $\Re t<0$; $\l^{-1}\ll |\Re t|$;  $|t|,|x|\ll
m^{-1}$. In this region the coefficient of the term $\e^{\l t}$ in
Eq.~(\ref{PIin3}) is irrelevant. If we set it to $1/2$, we recover the
form of the Green's function at close separation,
Eq.~(\ref{G_th_as}). Matching the constant parts of $\vf_{\rm
  b}\big|_{\rm core}$ and $\vf_{\rm
  b}\big|_{\rm tail}$ reproduces Eq.~(\ref{aTh}) for $b_{th}$, which
is now valid for any $b_{th}\leq 1$.  

Importantly, when $b_{th}=1$ the solution (\ref{PIin3}) still
linearizes at $\Re t\to-\infty$ and obeys the boundary conditions
appropriate for tunneling from a thermal state with temperature
$\Lambda_0$. The solution does not have singularities at $\Re t>0$,
thus it does not directly interpolate to the true vacuum (see
Fig.~\ref{fig:bounceth}b). Instead, at $\Re t\to+\infty$ it
asymptotically tends to the sphaleron, approaching it along the
unstable direction. This phenomenon can be called ``tunneling onto the
sphaleron'' and has been previously observed in the context of
semiclassical transitions induced by particle collisions 
\cite{Bezrukov:2003er,Levkov:2004ij,Levkov:2004tf,Demidov:2015bua} and
in quantum mechanics with multiple degrees of freedom
\cite{Bezrukov:2003tg,Takahashi_2003,Takahashi_2005,
Levkov:2007yn,Levkov:2008csa}. 
The 
sphaleron formed in this way later decays into the true vacuum with
order-one probability, therefore, all exponential suppression comes
from the first stage of the process --- formation of the sphaleron ---
which is captured by the semiclassical solution.

What is the structure of the bounce at $\l>\Lambda_0$? Naively, one
could think that it is given by the continuation of the expression
(\ref{PIin3}) to $b_{th}>1$. However, this does not work: it is
straightforward to see that the resulting configurations decay back to
the false vacuum at $t>0$ and thus do not describe appropriate
transitions. The true bounce solution is still expected to tunnel on
top of the sphaleron. However, at this temperature 
there is no single analytic function
that would be a solution of the field equations, 
satisfy the boundary conditions (\ref{cpmrel}), (\ref{rHH}) at $\Re
t\to -\infty$, and approach the sphaleron at $\Re
t\to +\infty$. We are in the regime discussed at the end of
Sec.~\ref{Ssec:gen_amplitude} when the bounce cannot be found on a
contour ${\cal C}$ with a finite turn-around point $t_f$. However, we
can still construct the solution if we pull the turn-around point to
infinity as in Fig.~\ref{fig:bounceth}c. In this case the solution on
the upper and lower halves of the contour need not be the same
analytic function, the only requirement being that they have the same
limit at $\Re t\to+\infty$. It is straightforward to see that the
following Ansatz will do the job:
\be
\label{bouncehighT}
\begin{split}
&\vf_{\rm b}^{\rm up}(t,x)=\vf_{\rm b}^{(\Lambda_0)}
\left(t+i\pi\big(\tfrac{1}{\Lambda_0}-\tfrac{1}{\l}\big),x\right)\;,\\
&\vf_{\rm b}^{\rm low}(t,x)=\vf_{\rm b}^{(\Lambda_0)}
\left(t-i\pi\big(\tfrac{1}{\Lambda_0}-\tfrac{1}{\l}\big),x\right)\;,
\end{split}
\ee
where $\vf_{\rm b}^{(\Lambda_0)}$ is the bounce solution for the
critical temperature $\l=\Lambda_0$.

The corresponding tunneling suppression can be evaluated along the
contour ${\cal C}$. It is simpler, however, to deform it into the
contour ${\cal C}'$ as shown in Figs.~\ref{fig:bounceth}b,c. The
integrals over the upper and lower halves of the contour cancel due
to the periodicity of $\vf_{\rm b}^{(\Lambda_0)}$, and the only
remaining contribution comes from the piece at $\Re
t=+\infty$ (shown with dashed lines in the figure). The
solution there simply coincides with the static sphaleron and 
the suppression is
given by its energy times the difference in the imaginary time
between the upper and lower parts of the contour,
\begin{equation}\label{SPIhigh}
B_{th}=\frac{2\pi E_{\rm sph}}{\l}~,~~~~~
\l\geq \Lambda_0\;.
\end{equation}
Thus, we have recovered with the in-in formalism of Sec.~\ref{Sec:gen}
the standard high-energy transition rate associated with jumps over
the potential barrier.

\begin{figure}[t]
	\centering 
	\includegraphics[width=0.5\textwidth]{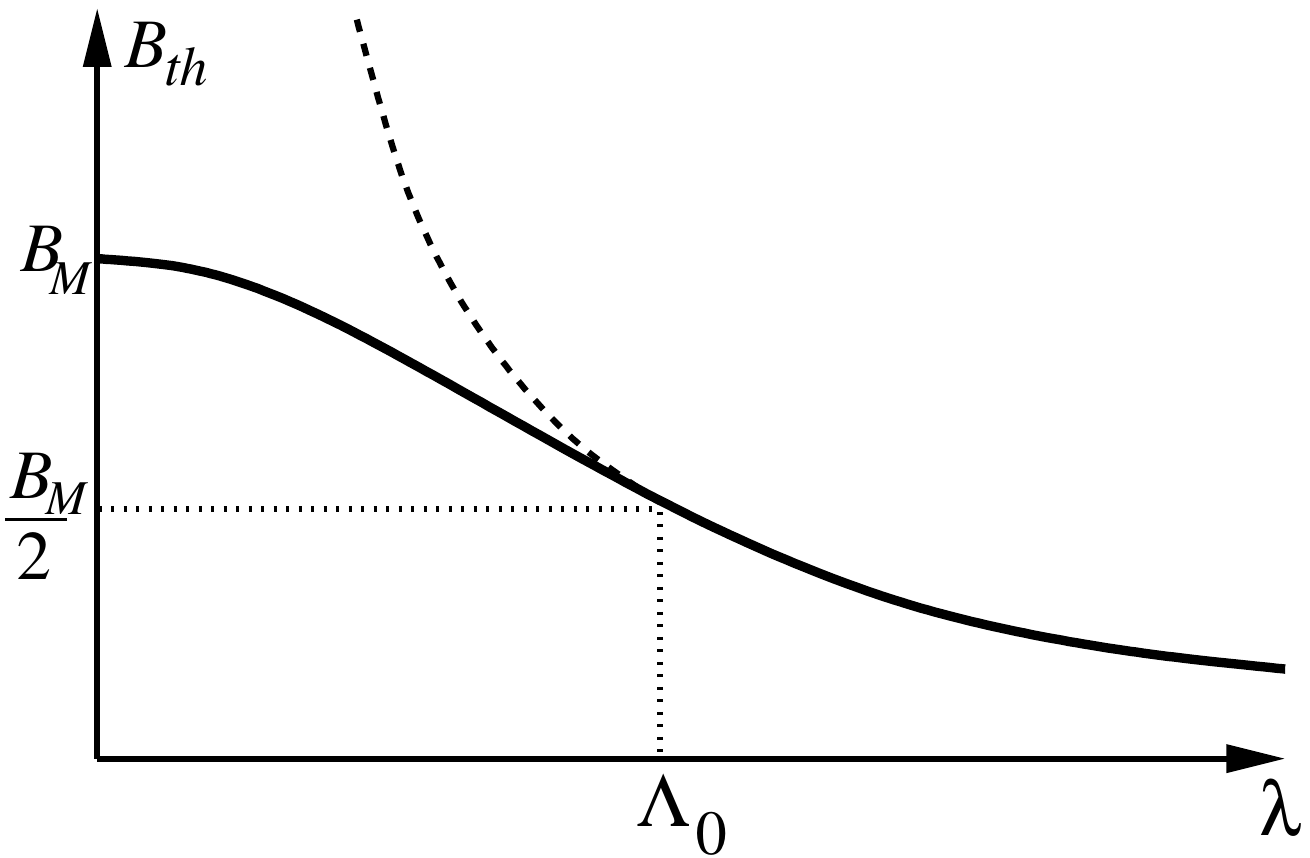}
	\caption{Tunneling suppression $B_{th}$ in flat spacetime at
          different temperatures $\l/(2\pi)$. Dashed line represents the decay
          channel via the sphaleron. The change 
         to sphaleron dominated transitions occurs at 
          $\Lambda_0$ given by Eq.~(\ref{Csol}). $B_M$ represents
          suppression in empty space, Eq.~(\ref{B0}).} 
	\label{fig:MainPlotFlat}
\end{figure}

Recalling the formula for the sphaleron energy (\ref{Esphfin1}), we
see that at $\l=\Lambda_0$ the two expressions (\ref{SPI}),
(\ref{SPIhigh}) smoothly match, up to the first derivative with
respect to $\l$, whereas the second derivative is discontinuous. At
the matching point the suppression is roughly equal to half the
suppression of the vacuum tunneling (\ref{B0}). These findings are
summarized in Fig.~\ref{fig:MainPlotFlat}.

Finally, let us make an observation which will be useful later,
when studying decay of the Unruh vacuum. The leading
term in the suppression of transitions at high temperature can be
found with a different method. When $\l/(2\pi)\gg m$, the occupation
numbers of modes with $\omega\sim m$ are large. Hence, $\vf$ can be
viewed as a classical stochastic field. The low-frequency modes
dominate thermal field fluctuations. Their variance is found from the thermal
Green's function at coincident points, upon renormalizing it by
subtraction of the Green's function in empty space,
\begin{equation}
\label{thermfluc}
\delta\vf_{th}^2=\gc^2\lim_{t,x\to
  0}\big[\GG_{th}(t,x;0,0)-\GG_{F}(t,x;0,0)\big]
\approx\frac{{\rm g}^2\l}{4\pi m} \;,
\end{equation}
where we have used Eqs.~(\ref{G_F_close}), (\ref{G_th_as}). Now we can
estimate the transition probability as 
the probability of the field fluctuation reaching beyond 
the maximum of the potential barrier $\vf_{\rm
  max}$. Since the interaction quickly dies
out at $\vf<\vf_{\rm max}$, we can take the fluctuations to be
Gaussian, so that
\begin{equation}\label{Stoch_Gamma1}
\varGamma_{\text{high-}\l}\sim \exp\left(-\frac{\vf_{\rm
      max}^2}{2\,\delta\vf_{th}^2}\right)\sim
\exp\left[-\frac{8\pi m}{{\rm g}^2\l}
\left(\ln\frac{m}{\sqrt\kappa}\right)^2\right] 
 \;,  
\end{equation}
where we used the leading term in the expression (\ref{Vmax}) for
$\vf_{\rm max}$. This coincides with the leading-log part of the exact
high-temperature suppression~(\ref{SPIhigh}).

It is worth stressing that the possibility to make the simple
estimate (\ref{Stoch_Gamma1}) hinges on two peculiar properties of our
model. The first is the dominance of the field fluctuations by long
modes with wavelengths of order $1/m$, which is due to the
two-dimensional nature of the model. Thanks to this property, the
field changes coherently in large regions of space, comparable to the
size of the sphaleron. 
The second property is the abrupt variation of the scalar potential
around $\vf_{\rm max}$, which implies that
the field is
essentially linear at $\vf<\vf_{\rm max}$, whereas almost any
fluctuation towards $\vf>\vf_{\rm max}$
leads to a roll-over of the field into the true vacuum. In principle,
the stochastic approach can also work in more general situations, but will
require full-fledged simulations of the classical field dynamics to
determine the vacuum decay 
rate~\cite{Grigoriev:1988bd,Grigoriev:1989je,Grigoriev:1989ub,Khlebnikov:1998sz}.

\section{Minkowski bounce as periodic
  instanton in Rindler space}
\label{Sec:R}

Rindler spacetime presents the simplest example of a nontrivial
metric to test our approach. It corresponds to the line element
(\ref{Line_element}) with 
\begin{equation}\label{O_R}
\O=\e^{2\l x} \;.
\end{equation}
The curvature of spacetime is still zero,\footnote{Recall that we
  define $\Box=\eta^{\m\n}\d_\m\d_\n$.}
\be
\label{OmegaInt}
R=-\Omega^{-1}\Box\ln\O=0\;,
\ee 
and the change of variables
\begin{equation}\label{R_flat}
T=\l^{-1} \e^{\l x}\sh \l t \; , ~~~~~ X=\l^{-1} \e^{\l x}\ch \l t 
\end{equation}
brings the line element to the Minkowski form
$ds^2=-dT^2+dX^2$. The original coordinates $(t,x)$ cover the right
wedge of Minkowski space $X>|T|$. The lines of constant $x$
represent trajectories of uniformly accelerated observers with
the acceleration $\l\e^{-\l x}$. Note that the acceleration decreases at
large $x$. The time variable $t$ is the proper time of the observer at
$x=0$. 
While Rindler space is interesting on its own right, for us it
has an additional value since it describes the near-horizon
region of a BH, as it is clear from Eq.~(\ref{OnearH}). Thus,
understanding the bounce solutions in the Rindler geometry will give
us insight about tunneling in BH background. 

The field equation now reads
\be
\label{phieqOmega}
\Box\vf-m^2\,\O\,\vf+2\kappa\,\O\,\e^{\vf}=0\;.
\ee
If we neglect the mass term, it is still exactly solvable due to the
property (\ref{OmegaInt}) with the general solution
\begin{equation}\label{GenSolCurv}
\vf=\ln\left[\frac{4F'(-u)G'(v)}{\O(u,v)\big(1+\kappa
    F(-u)G(v)\big)^2}\right] \; . 
\end{equation}
This is, of course, a consequence of the solvability of the Liouville
equation in flat spacetime.

The complete set of Rindler mode functions is given by
Eq.~(\ref{Modes_R}) from the Appendix. Due to the unbounded growth of
the effective potential in the mode equation (\ref{EqModes}), 
all modes quickly vanish at $x\to+\infty$. At $x\to-\infty$ they
represent the sum of right- and left-moving waves with equal
amplitudes. This means that, unlike flat or BH background, there is no
separation into left- and right-moving modes. In particular, in
Rindler space 
there is
no analog of the Unruh vacuum which requires
different occupation of left and right modes.

On the other hand, an analog of the Hartle--Hawking state does exist
and is given by the Minkowski vacuum. We focus on tunneling from this state.
In principle, one can find the bounce directly in the coordinates
$(t,x)$ by using an appropriate Ansatz for the core and matching it to
the Green's function at the tail. We do not need to do it, however,
because we already know the form of the bounce in flat spacetime,
Eqs.~(\ref{bouncein}), (\ref{bounceout}).
We consider a bounce centered at a point $(\TT=0,X=X_0)$ in flat
Euclidean space, with $\TT=iT$. Then replacing $\rho$ in
Eq.~(\ref{bouncein}) by $\sqrt{\TT^2+(X-X_0)^2}$ and performing the
Euclidean version of the coordinate change (\ref{R_flat}),
\be
\TT=\l^{-1}\e^{\l x}\sin{\l\tau}~,~~~~~X=\l^{-1}\e^{\l x}\cos{\l\tau}\;,
\ee
we obtain in the Rindler frame
\begin{equation}\label{R_SolM}
\vf_{\rm b}\Big|_{\text{core}}=\ln\left[
  \dfrac{\l^2b_R}{\kappa\left(\ch\l(x-x_0)\mp \sqrt{1-b_R}\cos\l\tau\right)^2}
\right] -2\l x\; .
\end{equation}
Here
\begin{equation}\label{R_FlatVsTh}
x_0=\frac{1}{2\l}\ln\left[\frac{\l^2}{\kappa C^2_M}+\l^2X_0^2\right]~,~~~~~~
b_R=\frac{\l^2}{\kappa C^2_M}\e^{-2\l x_0} \;, 
\end{equation}
and the minus (plus) sign corresponds to $X_0>0$ ($X_0<0$). Similar
transformations with Eq.~(\ref{bounceout}) give
\be
\label{R_SolMout}
\vf_{\rm b}\Big|_{\text{tail}}=4K_0\left(\frac{m}{\l}
\sqrt{\e^{2\l x}\mp 2\sqrt{1-b_R}\,
  \e^{\l(x+x_0)}\cos\l\tau+(1-b_R)\e^{2\l x_0}}\right)\;.
\ee

We first focus on the solutions with the minus sign in
Eqs.~(\ref{R_SolM}), (\ref{R_SolMout}). We observe that the core
solution (\ref{R_SolM}) is the same as the core of the periodic
instanton in flat space, Eq.~(\ref{PIin2}), up to a linear term that
comes from the
factor $\O^{-1}(x)$ inside the logarithm in the
general solution (\ref{GenSolCurv}). Thus, the flat-space vacuum
bounce in Cartesian coordinates becomes a periodic instanton in the
Rindler frame. This is what one expects because the
Minkowski vacuum corresponds to a thermal state from
the viewpoint of an accelerated observer \cite{Ai:2018rnh}.

The physical temperature seen by observers at different positions $x$
is, however, different due to the redshift introduced by the
space-dependent metric. The Green's function probes the field
nonlocally
and is sensitive to this deviation from equithermality. As a
consequence, the tail of the Rindler bounce (\ref{R_SolMout}) {\em is
  not} the same as in the flat-space periodic instanton. To see this
explicitly, let us expand the tail (\ref{R_SolMout}) in the
region where the argument of the Bessel function is small. Assuming
for simplicity $b_R\ll 1$, we obtain
\be
\label{R_SolMapprox}
\vf_{\rm b}\approx -2\ln\left[
4\sh\left(\frac{\l z}{2}\right)\sh\left(\frac{\l \bar
    z}{2}\right)\right]
-2\l x-2\l x_0 +4\ln\frac{\l}{m}+4\ln 2-4\gamma_E\;.
\ee
This must be contrasted with the flat-space thermal Green's function at
close separation, Eq.~(\ref{G_th_as}). We see that while the singular
parts of the two expressions are proportional to each other, the
constants have different dependence on $\l/m$. Furthermore, the expression
(\ref{R_SolMapprox}) 
contains a linear-in-$x$ piece, which is absent from
(\ref{G_th_as}). It matches a similar linear piece in
the core solution (\ref{R_SolM}). Different constant in
Eqs.~(\ref{R_SolMapprox}), (\ref{G_th_as}) lead to different
expressions for the parameter $b$, cf. Eqs.~(\ref{aTh}),
(\ref{R_FlatVsTh}). This, in turn, translates into different tunneling
suppressions, see below.

\begin{figure}[t]
	\centering 
	\includegraphics[width=0.7\textwidth]{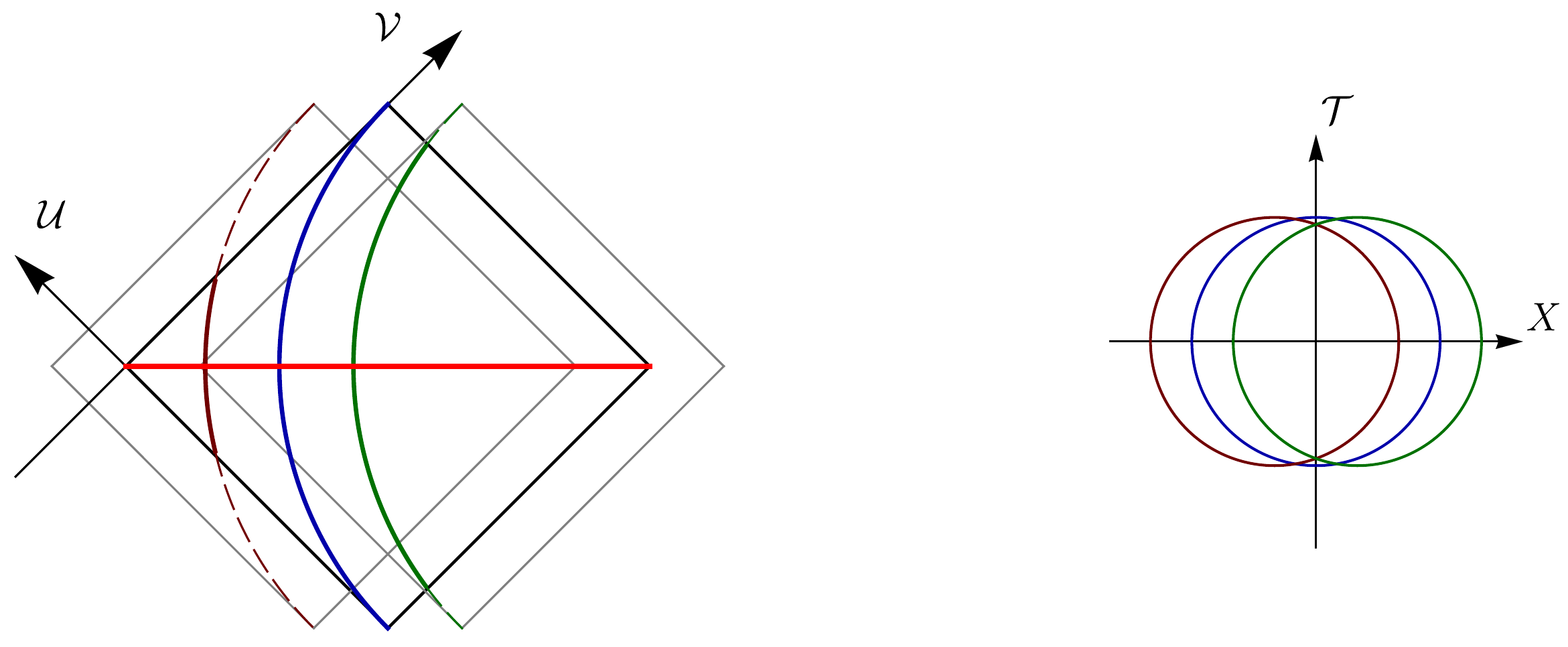}
	\caption{Schematic plot of the family of periodic instantons
          in Rindler spacetime ({\bf left}) and their form in Euclidean
          time in Cartesian coordinates ({\bf right}). Red line marks
          the surface at which the instanton is continued to real
          time. From left to right: the collapsing bubble (brown), the
          sphaleron (blue), the expanding bubble (green).} 
	\label{fig:RindlerInstanton}
\end{figure}

Let us now discuss the choice of sign in Eqs.~(\ref{R_SolM}), (\ref{R_SolMout}). We notice that
the evident freedom in choosing
the center of the instanton at $X=X_0$ in Minkowski coordinates becomes
somewhat nontrivial when expressed in terms of $(\tau,x)$. For
different values of $X_0$, the solutions \eqref{R_SolM} on 
the real positive time axis describe different dynamics of a true
vacuum region (see Fig.~\ref{fig:RindlerInstanton}):
\begin{itemize}

\item
$X_0>0$ (negative sign in Eqs.~(\ref{R_SolM}), (\ref{R_SolMout})). 
In Cartesian coordinates, $\vf_{\rm b}$ is a vacuum
bounce shifted to the right with respect to the origin.
On the real positive time axis $t$,
$\vf_{b}$ describes a bubble of true vacuum expanding outwards
the horizon. 

\item
$X_0=0$. Then $b_R=1$ and $\vf_{\rm b}$ becomes $\tau$-independent. 
It represents the sphaleron of Rindler observers. 
In Cartesian coordinates
this sphaleron is a flat vacuum bounce sitting symmetrically around the
origin $X=0$. 

\item
$X_0<0$  (positive sign in Eqs.~(\ref{R_SolM}), (\ref{R_SolMout})).
In Cartesian coordinates,
$\vf_{\rm b}$ is a vacuum bounce shifted to the left with respect to
the origin. Its part in the Rindler wedge at real positive $t$
describes a bubble of true vacuum collapsing towards the horizon. This
leaves the false vacuum in the Rindler wedge intact. We conclude that
this branch of solutions is irrelevant for the false vacuum decay in
Rindler space and
should be discarded.

\end{itemize}

We draw one more lesson from the above discussion. 
Although the Rindler metric is not homogeneous, there is still a
freedom in choosing the center of the instanton $x_0$ in
Eq.~\eqref{R_SolM}, corresponding to the choice of $X_0>0$. This is a
nontrivial observation. Shifts in $X$ do not preserve the position of
the horizon. Hence, they are not an isometry of the Rindler
spacetime. Nevertheless, $x_0$ represents a zero mode of the
solution. By varying $x_0$, the branch of periodic instantons is
continuously connected to the sphaleron.

Finally, we compute the tunneling action. As usual, we take the
general formula \eqref{B}, substitute the solution in the core
\eqref{R_SolM} and integrate over one period of oscillation in
Euclidean time, $0<\tau<2\pi/\l$. We obtain, as we should, that the
action does not depend on $x_0$ or $\l$ and coincides
with the action of the flat vacuum bounce \eqref{B0}. 

Two comments are in order. First, the independence of the action of the
position of the instanton and of temperature might seem
counter-intuitive from the viewpoint of a Rindler
observer. However, it follows inevitably from the invariance of the
tunneling probability under changes of the reference frame. Second, note
that the Rindler observer does not have access to the portion of the
Minkowski vacuum bubble hidden by the horizon. In particular, the
Rindler sphaleron is only half of the Minkowski bubble on the slice
$T=0$. It is this half that we use to compute the sphaleron Rindler
energy and the corresponding suppression.  
However puzzling it might seem at first,
the result we obtain coincides with the full flat-space
integration. This supports the conclusion that parts of
spacetime outside the physical wedge are not relevant for tunneling
and one can exclude them completely from consideration.

\section{Tunneling in black hole background}
\label{Sec:BH}

We are now ready to address tunneling in the BH background. The field
equation has the form (\ref{phieqOmega}) with the metric function
given by Eq.~(\ref{O_BH}). Even if we neglect the mass term, this
equation is not in general exactly solvable. One could try to maintain
solvability by adding a coupling between the field $\vf$ and curvature
to ensure the conformal invariance of the Liouville part of the
action. This path is, however, not suitable for our purposes. As
discussed in Appendix~\ref{app:Dil_nonmin}, the nonminimal coupling
leads to a deformation of the classical vacuum in the presence of BH
and increases the barrier between the false and true vacua. This leads
to an artificial suppression of the tunneling rate, which is not
present in realistic situations.

Therefore, we stick to the minimal coupling case and notice that, in
the absence of mass, Eq.~(\ref{phieqOmega}) can still be solved in two
regions: near horizon, and far away from BH. In both these cases we
have $|x|\gg 1/\l$ and $(\ln \O)''=0$, hence the general solution is
given by Eq.~(\ref{GenSolCurv}). This will suffice to find the bounce
solutions whose cores are contained entirely in one of those
regions. Notice that this does not impose any restrictions on the
tails of the solutions described by the Green's functions of the free
massive theory, which can extend across the boundary between the two
regions. We are going to see that the majority of bounce solutions
satisfy this requirement.

In the main text we focus on the physically relevant cases of
tunneling from the Hartle--Hawking and Unruh states. For completeness
we also consider the Boulware vacuum in Appendix~\ref{Ssec:BH_B}. We
find that the suppression in the latter case is essentially the same as 
in flat space, with only a minor
enhancement due to the vacuum polarization by the gravitational field. 
On the other hand, the thermal excitations
present in 
the
Hartle--Hawking and Unruh vacua have a
dramatic effect on the decay rate, as we presently show. Throughout
this section we assume $\l\gg m$.

\subsection{Hartle--Hawking vacuum}
\label{Ssec:BH_HH}

\subsubsection{Moderate temperature: Tunneling near horizon}
\label{Sssec:BH_HH_PI}

Let us make an assumption that tunneling is dominated by periodic
instantons with the core in the near-horizon region $x<0$, $|x|\gg
1/\l$. We will see that this is indeed the case as long as the BH
temperature does not exceed a certain critical value\footnote{Which is
parametrically larger than $m$, so $\l\lesssim \Lambda_{HH}$ is
compatible with $\l\gg m$.} $\Lambda_{HH}$. We will work in the Euclidean
signature to make contact with previous studies and refer the reader
to Sec.~\ref{Ssec:flat_thermal} for the discussion of the relation to the in-in
formalism. We make the thermal Ansatz for the functions
parameterizing the solution in the core (cf. Eq.~(\ref{fg_th})),
\begin{equation}\label{fg_HH}
F(z)=\frac{C_{HH}}{\l}\left(\e^{\l z}-d_{HH}\right) \;, ~~~~~
G(\bz)=\frac{C_{HH}}{\l}\left(\e^{\l\bz}-d_{HH}\right) \;. 
\end{equation} 
This yields
\begin{equation}\label{HHcore}
\vf_{\rm b}\Big|_{\text{core}}=
\ln \left[\frac{\l^2 b_{HH}}{\kappa\big(\ch\l(x-x_{HH})
-\sqrt{1-b_{HH}}\cos\l\tau\big)^2}\right]-2\l x\;, 
\end{equation}
where
\be
\label{xHHbHH}
x_{HH}=\frac{1}{2\l}\ln\left[\frac{\l^2}{\kappa
    C_{HH}^2}+d_{HH}^2\right]\;,~~~~~
b_{HH}=\frac{\l^2}{\kappa C_{HH}^2}\e^{-2\l x_{HH}}\;.
\ee
Notice the similarity of these expressions with Eqs.~(\ref{R_SolM}),
(\ref{R_FlatVsTh}) in Rindler space. It is not surprising, since
the physics in the near-horizon region of a BH is the same as in Rindler space.

We now have to match the core to the tail of the solution given by the
Green's function in the Hartle--Hawking vacuum which in the
near-horizon region has the form (\ref{G_HH_near}). The matching is
most easily performed if $b_{HH}\ll 1$. In this case 
\be
\label{HHtail}
\vf_{\rm b}\Big|_{\text{tail}}=8\pi \GG_{HH}(-i\tau,x;0,x_{HH})
\ee
and the matching region exists in the Euclidean strip
$-\pi/\l<\tau<\pi/\l$. Notice that the last linear-in-$x$ term in the
core solution (\ref{HHcore}) has an exact counterpart
in the Green's function (\ref{G_HH_near}). One obtains the
following relation between the parameters:
\begin{equation}\label{HH_a}
b_{HH}=\frac{\kappa}{4\l^2}\e^{\frac{4\l}{m}-2\l x_{HH}} \;.
\end{equation}
By extending the matching region into the complex time plane,
as in the case of periodic instantons in flat
space (see Sec.~\ref{Ssec:flat_thermal}), one can show that this
relation remains valid even if $b_{HH}$ is order-one.

Thus, we have obtained a family of solutions labeled by a single
parameter --- the position of the bounce core $x_{HH}$. By
construction, this parameter is restricted to negative values in order
for the bounce to fit into the near-horizon region,
$x_{HH}<0$. Besides, we have the requirement that $b_{HH}$ cannot
exceed unity, $b_{HH}\leq 1$. Together these two conditions restrict
from above the range of temperatures for the existence of periodic
instantons, $\l<\Lambda_{HH}$, where  
\begin{equation}\label{TempCritHH}
\frac{\Lambda_{\rm HH}}{\ln(2\Lambda_{HH}/\sqrt\kappa)} =
\frac{m}{2}~~~ \Longrightarrow ~~~
\Lambda_{HH}=\frac{m}{2}\bigg(\ln{\frac{m}{\sqrt{\kappa}}}
+\ln\ln{\frac{m}{\sqrt{\kappa}}}+\ldots\bigg)\ \;.
\end{equation}
This can be compared to the similar situation with thermal transitions
in flat spacetime where periodic instantons exist only at temperatures
below the critical value $\Lambda_0$ given by
Eq.~\eqref{Csol}. Notice that $\Lambda_{HH}$ is approximately half of
$\Lambda_0$. 

The calculation of the tunneling suppression corresponding to the
periodic instantons parallels the calculation in flat space. It is
outlined in Appendix~\ref{app:Bcalc}. The result is independent of the
instanton position and reads
\begin{equation}\label{B_BH_HH}
B_{HH}=\dfrac{16\pi}{\gc^2}\left( \ln\dfrac{\l}{\sqrt{\kappa}}
-\dfrac{\l}{m}+\ln 2-1\right)~, ~~~~~ \l<\Lambda_{HH} \;.
\end{equation}
It starts from the flat vacuum suppression $B_M$ (see Eq.~(\ref{B0}))
at $\l\sim m$ and decreases down to $B_M/2$ at $\l\sim \Lambda_{HH}$.

When $x_{HH}$ takes the value
\begin{equation}
\label{xHHsph}
x_{HH,{\rm sph}}=\frac{2}{m}-\frac{1}{\l}\ln\frac{2\l}{\sqrt{\kappa}} \;,
\end{equation}
the parameter $b_{HH}$ becomes equal to $1$ and the bounce reduces to
a static sphaleron with the core
\begin{equation}\label{HH_sph}
\vf_{\rm sph}\Big|_{\text{core}}=\ln\left[
\frac{\l^2}{\kappa\ch^2\big(\l(x-x_{HH,{\rm sph}})\big)}\right]-2\l x \;,
\end{equation} 
which at $(x-x_{HH,{\rm sph}})\gg 1/\l$ matches to a static solution
of the free massive equation. Note that this sphaleron differs from
its flat-space counterpart in two respects. First, 
its width depends on the BH temperature, and second, it gives the same
suppression as the periodic instantons and thus provides a valid
tunneling channel at low temperatures. In Appendix~\ref{app:Sph} we
show that the family of sphalerons extends to $\l\lesssim m$, and
at $\l\to 0$ their suppression coincides precisely with the vacuum
suppression (\ref{B0}), including the subleading terms.

Existence of a one-parameter family of periodic instantons with
identical suppression continuously connected to a sphaleron reproduces
the situation in Rindler space discussed in Sec.~\ref{Sec:R}. This is
natural, since the latter describes the BH near-horizon region. In
Rindler space this was a consequence of the exact translation
invariance of the underlying Minkowski geometry. However, no such
invariance exists for a BH. Thus, one does not expect the flat
direction corresponding to the parameter $x_{HH}$ to be exact. It will
be tilted by the terms of order $\e^{4\l x}$ in the expansion of the
function $\O(x)$ at $x<0$ distinguishing the BH from the Rindler metric. As
a result, one expects to get a unique tunneling solution with the
least suppression. The most likely candidate for this unique solution
is the sphaleron that lies at the endpoint of the flat
direction. Unlike other periodic instantons, it has a monotonic field
profile with the maximum achieved at the horizon, see the left panel
of Fig.~\ref{fig:HH_sph}. This appears to be the most natural
morphology for a tunneling solution `seeded' by the BH.\footnote{
In two-dimensional models obtained as spherical
  reduction from four dimensions, the static solution will correspond
  to formation of a true vacuum bubble encompassing the BH. Whereas
  the four-dimensional analog of a periodic instanton is closer to a
spherical shell of true vacuum.} We
plan to address the relation between Hartle--Hawking periodic
instantons and sphalerons in more detail elsewhere.    

The procedure of finding the tunneling solutions presented above breaks
down when $\l$ exceeds $\Lambda_{HH}$. So, what are the solutions at
higher BH temperatures? To answer this question, let us focus on the
sphaleron and understand what happens with it when $\l$ approaches
$\Lambda_{HH}$ from below. It is instructive to estimate the physical
size of its core,  
\begin{equation}\label{HH_rc}
l_{HH,{\rm sph}}\sim\int_{-\infty}^{x_{HH,{\rm sph}}} dx\;\e^{\l x} 
=\frac{\sqrt{\kappa}}{2\l^2}\:\e^{\frac{2\l}{m}} \;.
\end{equation}
We see that the size grows with temperature and at $\l=\Lambda_{HH}$
reaches the physical size of the near-horizon
region\footnote{Note that at the same temperature the periodic
  instantons in the near-horizon region cease to exist.}  
$l_h\sim 1/\l$. At higher temperatures the core of the
sphaleron simply does 
not fit inside. The study of thermal tunneling in flat space teaches
us that at high temperature the transition must still proceed through
jumps onto the sphaleron, just now the sphaleron core will extend
outside the near-horizon region. We presently study this case. 

\subsubsection{High-temperature sphaleron}
\label{Sssec:BH_HH_sph}

\begin{figure}[t]
\includegraphics[width=0.45\linewidth]{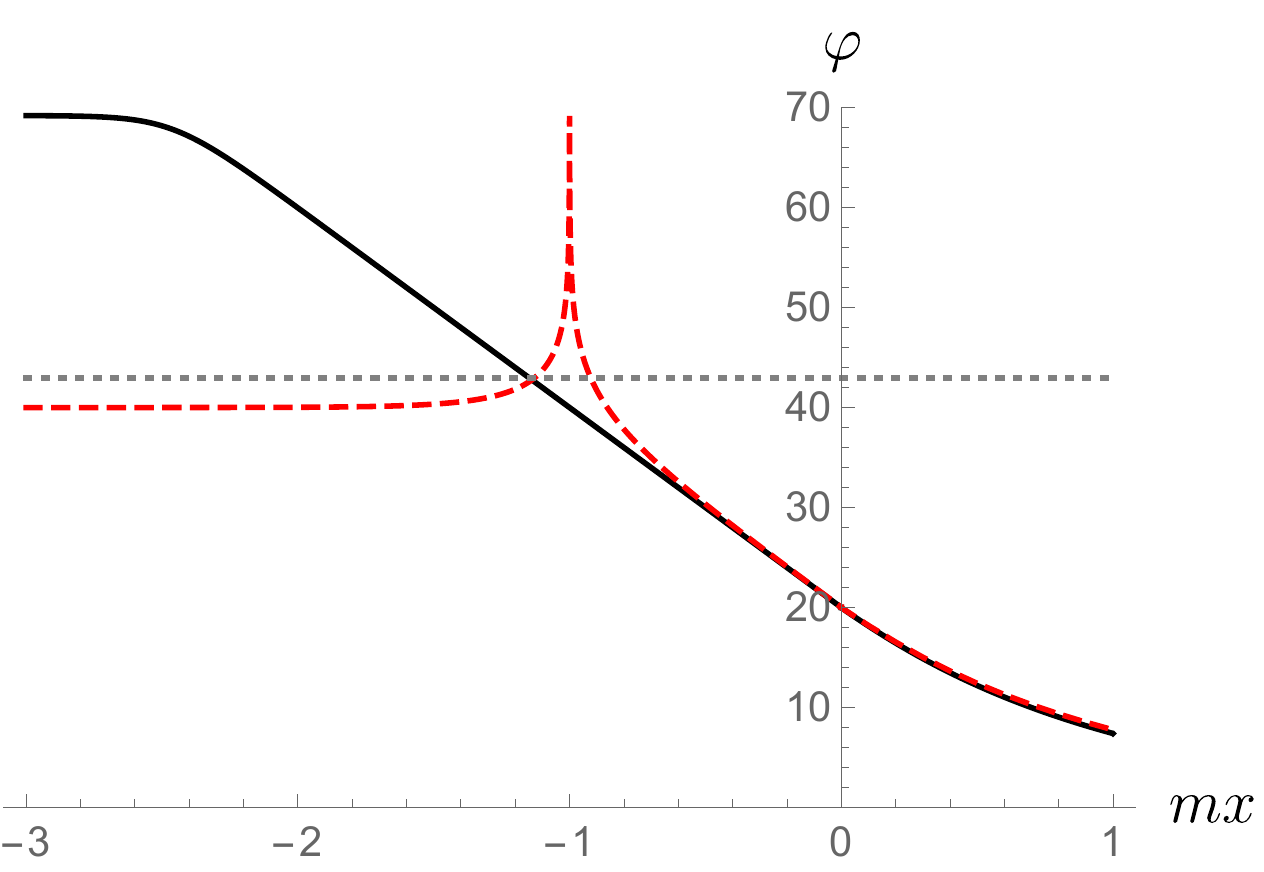}
	\hfill
\includegraphics[width=0.45\linewidth]{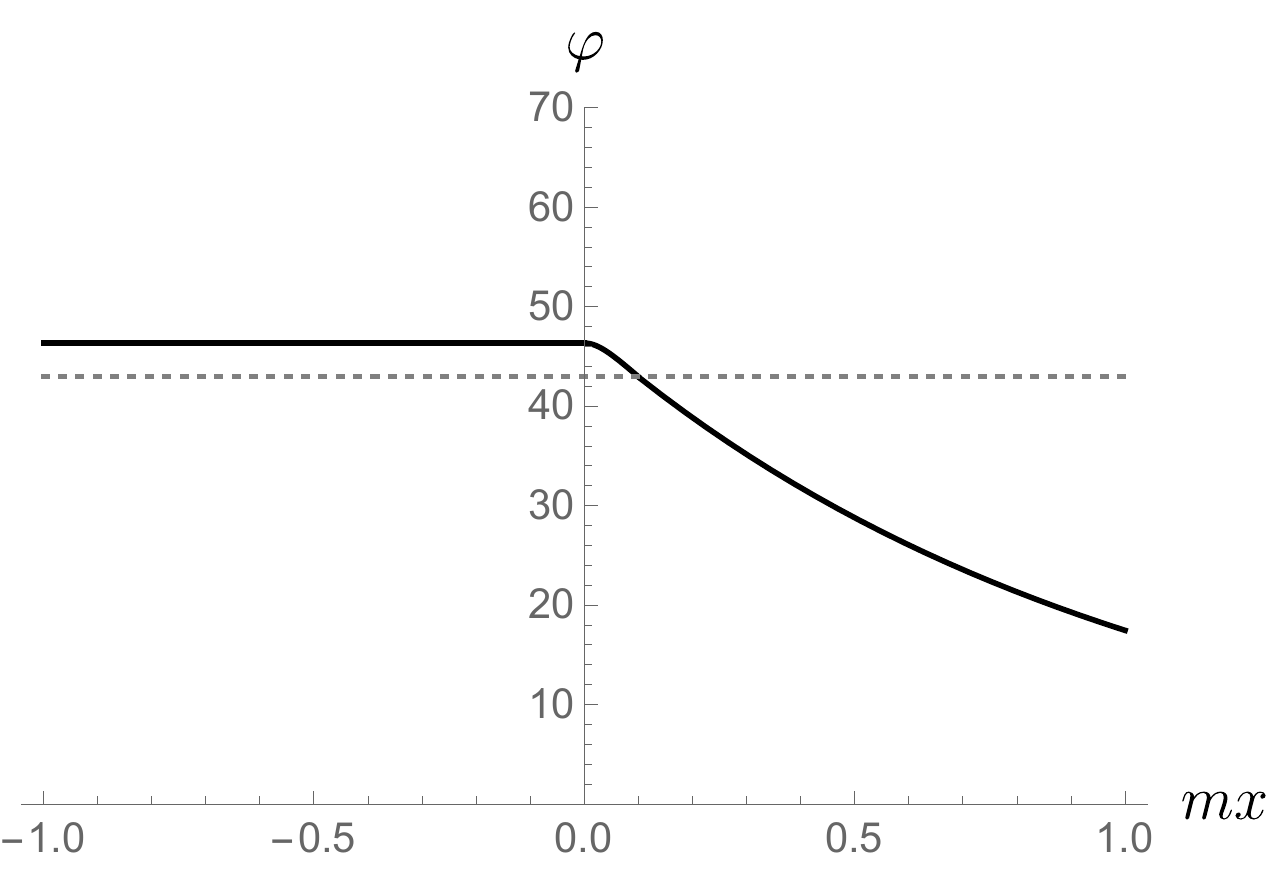}
	\caption{{\bf Left:} Profiles of the Hartle--Hawking
          sphaleron (black solid) and a slice of periodic instanton at
          $\tau=0$ (red dashed) for 
          $\l=0.5\Lambda_{HH}$. The center of the periodic instanton
          is taken at $x_{HH}=-1/m$. The grey dotted line marks the field
          value $\vf_{\rm max}$ corresponding to the top of the
          potential barrier, see Fig.~\ref{fig:Potential}. We take
          $\ln(m/\sqrt\kappa)=20$. {\bf Right:} 
High-temperature sphaleron for the same model parameters.
} 
	\label{fig:HH_sph}
\end{figure}

For the static sphaleron configuration the general equation
(\ref{phieqOmega}) reduces to 
\begin{equation}\label{HH_sph_eom}
\d_x^2\vf_{\rm sph}- m^2\,\O\,\vf_{\rm sph}+2\kappa\,\O\,\e^{\vf_{\rm sph}}=0 \;.
\end{equation}
Its solution can be easily found numerically for any given values of
parameters. In the previous subsection we have found the solution
analytically at $\l<\Lambda_{HH}$. Now we construct it in the opposite
limit $\l\gg\Lambda_{HH}$.

We notice that at high BH temperatures, $\O$ can be approximated by a
step-function: $\O=0$ at $x<0$ and $\O=1$ and $x>0$. Therefore,
Eq.~\eqref{HH_sph_eom} can be solved separately at negative and
positive $x$, with matching at $x=0$. In the inner region,
$x<0$, the equation is simply $\d_x^2\vf_{\rm sph}=0$. Requiring
regularity at the horizon leads to a constant solution, $\vf_{\rm
  sph}(x<0)=$~const. In the outer region, $x>0$, the equation
coincides with the flat-space one. Hence, one can employ the same
strategy as with the flat-space sphaleron studied in
Sec.~\ref{Ssec:flat_sphaleron}: find the nonlinear core centered at $x=0$ (as
required for the smooth matching with the inner region) and
glue it with the massive linear tail. Overall, we obtain 
\begin{equation}\label{HH_sph3}
\vf_{\rm sph}=\begin{cases}
\ln\frac{\Lambda_0^2}{\kappa} \;, & x<0 \\
\ln\left[ \frac{\Lambda_0^2}{\kappa\ch^2(\Lambda_0 x)}\right] \;, &
x>0\;,~~x\ll 1/m\\
\frac{2\Lambda_0}{m} \e^{-mx}\;,& x>0\;,~~x\gg 1/\Lambda_0
\end{cases}
\end{equation}
where $\Lambda_0$ is given in Eq.~\eqref{Csol}. The solution is shown
in the right panel of Fig.~\ref{fig:HH_sph}.  

As $\l\to\infty$, the physical size of the near-horizon region shrinks
to zero. Hence, the high-temperature Hartle--Hawking sphaleron is just
a half of the flat-space spha\-le\-ron. Correspondingly, its energy is
one half of the energy of the flat-space sphaleron
(cf. Eqs.~(\ref{SphEn}), (\ref{Esphfin1})),
\begin{equation}\label{SphEnBH}
E_{\text{sph, high-}\l}=\frac{1}{{\rm \gc}^2}\int_{-\infty}^\infty
dx\,
\O(x)\kappa (\vf_{\rm sph}-2)\e^{\vf_{\rm sph}}
\approx \frac{2\Lambda_0}{{\rm
    g}^2}\left(\ln\frac{\Lambda_0}{\sqrt\kappa}
-2+\ln 2\right)\;.
\end{equation}
This reduction of the sphaleron energy by a factor 2 can be viewed as
the purely geometric effect of the BH on the height of the energy
barrier between the false and true vacua. Its analog in
four-dimensional Schwarzschild metric was studied in
\cite{1606.04018,1706.01364}.   
Finally, the tunneling suppression due to the Hartle--Hawking sphaleron
at high temperatures is 
\begin{equation}\label{SBHsph1}
B_{HH}=\frac{4\pi \Lambda_0}{{\gc}^2\l}
\left(\ln\frac{\Lambda_0}{\sqrt{\kappa}}-2+\ln 2\right)~,~~~~~
\l\gg \Lambda_{HH} \;.
\end{equation}
At $\l\simeq \Lambda_{HH}$ this expression matches with
\eqref{B_BH_HH} to the leading-log approximation, 
providing a smooth transition between the low- and high-temperature
regimes. 

It is worth mentioning that, similarly to the flat-space case, the
tunneling rate at high BH temperature can be estimated using the
stochastic picture. From the expressions for the Green's function at close
separation --- the upper line in Eq.~(\ref{GHHcloseright}) and
Eq.~(\ref{G_HH_near}) --- one reads out the variance of the thermal
fluctuations of the field in the
neighborhood of the BH,
\be
\label{HHvar}
\delta\vf_{HH}^2\approx \frac{{\rm g}^2\l}{2\pi m}\;.
\ee
Note that it is twice bigger than in flat space at the same
temperature, Eq.~(\ref{thermfluc}), due to the contribution of modes
localized on the BH. This gives the vacuum decay rate
\be
\label{GammaHHstoch}
\varGamma_{HH,\,\text{high-}\l}\sim\exp\left(-\frac{\vf_{\rm
      max}^2}{2\delta\vf_{HH}^2}\right)
\sim \exp\left[-\frac{4\pi m}{{\rm
      g}^2\l}\left(\ln\frac{m}{\sqrt\kappa}\right)^2\right]\;. 
\ee
It coincides with the suppression (\ref{SBHsph1}) in the
leading-log approximation.\footnote{We stress again that the
  applicability of the estimate (\ref{GammaHHstoch}) relies on
  specific properties of our model, such as its two-dimensional nature
and the form of the interaction, see the comment at the end of
Sec.~\ref{Ssec:flat_thermal}. A similar estimate in the case of the
four-dimensional Schwarzschild BH \cite{1708.02138} where the field
fluctuations are dominated by modes with $\omega\sim T_{BH}$ appears
unjustified.}


\begin{figure}[t]
	\centering 
	\includegraphics[width=0.5\textwidth]{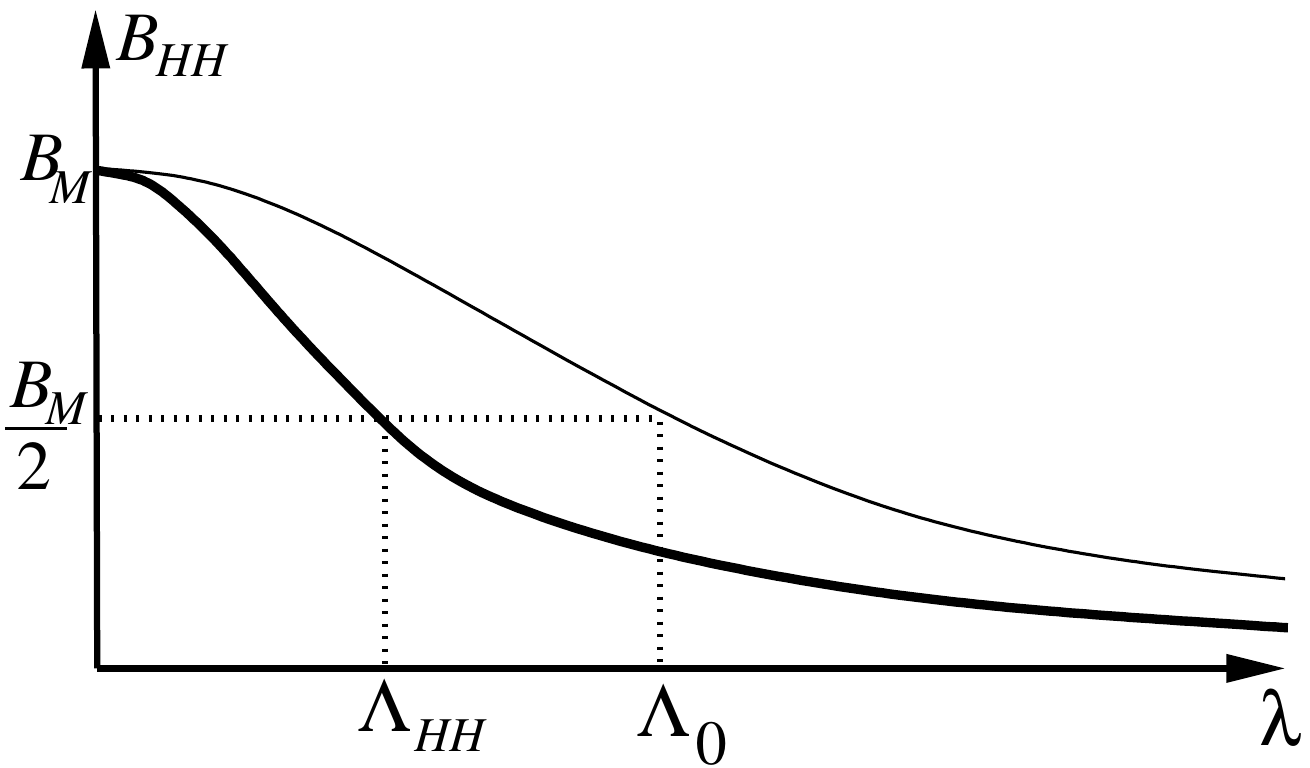}
	\caption{Suppression of the decay rate of the
          Hartle--Hawking vacuum as a function of the BH
          temperature $T_{BH}=\l/(2\pi)$ (thick line). Transition
          between the low- and high-temperature regimes occurs at
          $\Lambda_{HH}$ given by Eq.~(\ref{TempCritHH}). 
The tunneling
          suppression in flat spacetime at the same temperature is
          shown with the thin line
          (cf. Fig.~\ref{fig:MainPlotFlat}). For $\Lambda_0$ and $B_M$
see Eqs.~(\ref{Csol}),
          (\ref{B0}).} 
	\label{fig:MainPlotHH}
\end{figure} 

Let us summarize. At low temperatures, the decay of the Hartle--Hawking
vacuum in the vicinity of a BH proceeds via periodic
configurations. One of these configurations is static, and it is
plausible that it is actually preferred when the subleading
corrections to the metric are taken into account. The suppression is
given by Eq.~(\ref{B_BH_HH}). At the critical
temperature $\Lambda_{HH}$ the nonlinear core of the tunneling solution stops
fitting the near-horizon region. At higher temperatures, the tunneling
proceeds via the sphaleron that extends outside the near-horizon
region, 
and the tunneling suppression is half of
that in flat space. The summary of our findings is shown in
Fig.~\ref{fig:MainPlotHH}.

\subsection{Unruh vacuum}
\label{Ssec:BH_U}

\subsubsection{Tunneling far from the black hole}
\label{Sssec:BH_U_far}

The Unruh state corresponds to a flux of thermal radiation emitted by
the BH. In one spatial dimension the flux propagates without spreading
and leads to an enhancement of vacuum decay rate at an arbitrary
distance from the BH. It is instructive to first consider this case,
where tunneling proceeds in flat geometry, with the difference from
the Minkowski vacuum entirely due to the presence of
(out-of-equilibrium) excitations. This will serve us as a benchmark
for the subsequent study of tunneling near horizon where both effects
of the geometry and excitations are present. 

Specifically, we look for a bounce centered at $x_1\gg 1/m$. The
Euclidean formalism is no longer useful, so we work with the
Lorentzian time $t$ and construct the solution on the contour ${\cal
  C}$ of Sec.~\ref{Ssec:gen_amplitude}. As before, we assume that
outside the nonlinear core, the solution is proportional to the
time-ordered Green's function,
\be
\label{U1tail}
\vf_{\rm b}\Big|_{\rm tail}=8\pi\GG_U(t,x;0,x_1)\;,
\ee
where we take the same proportionality coefficient as in the cases
studied above. 
When $(t,x)$ gets close to the center $(0,x_1)$, the tail
must be matched to the solution (\ref{Liouvgen}) of the nonlinear
Liouville equation. The Green's function at close separation for the
case at hand is given by the lower line in Eq.~(\ref{G_U_far}) from
the Appendix. Its singular part is a mixture of a thermal contribution
for the right-moving modes and a vacuum contribution for
left-movers. This suggests to take the thermal (vacuum) Ansatz
for the function $F$ ($G$) of the general Liouville solution. Namely,
we write
\begin{equation}
\label{fg_U}
F(-u)=\frac{C_{U1}}{\l}\left(\e^{-\l(u-u_1)}-1\right) \;, ~~~~~ G(v)=C_{U1}(v-v_1) \;,
\end{equation}
where $v_1=-u_1=x_1$ and $C_{U1}$ is an unknown
constant.\footnote{Note that we have not reduced generality by
  choosing the same constant in $F$ and $G$, as only the product of 
these functions 
  enters the solution.}
Substituting this into Eq.~(\ref{Liouvgen}), we obtain
\begin{equation}
\label{U_core_far}
\vf_{\rm b}\Big|_{\rm
  core}=\ln\left[\frac{4\l^2 b_{U1}}{\kappa
    \left(-2\l(v-v_1)\sh\left(\frac{\l}{2}(u-u_1)\right)
+b_{U1}\e^{\frac{\l}{2}(u-u_1)}\right)^2}
\right] 
\end{equation}
with
\be
b_{U1}=\frac{\l^2}{\kappa C_{U1}^2}\;.
\ee
This indeed matches to Eq.~(\ref{G_U_far}) (lower line) describing the
Green's function at close separation when the first term in the
denominator wins over the second,
\be
\label{U1cond}
\left|2\l(v-v_1)\sh\left(\tfrac{\l}{2}(u-u_1)\right)\right|\gg 
\left|b_{U1}\e^{\frac{\l}{2}(u-u_1)}\right|\;.
\ee
Equating the constant parts in $\vf_{\rm b}\big|_{\rm core}$ and
$\vf_{\rm b}\big|_{\rm tail}$ fixes
\begin{equation}
\label{U_a_far}
b_{U1}=\frac{\kappa}{m^2}\:\e^{\frac{8\l}{3\pi m}-2\gamma_{E}+\frac{1}{2}} \; .
\end{equation}
Note that $b_{U1}\ll 1$ for not-so-large $\l\gtrsim m$, but grows
exponentially with $\l$.

\begin{figure}[t]
\centering 
	\includegraphics[width=0.47\textwidth]{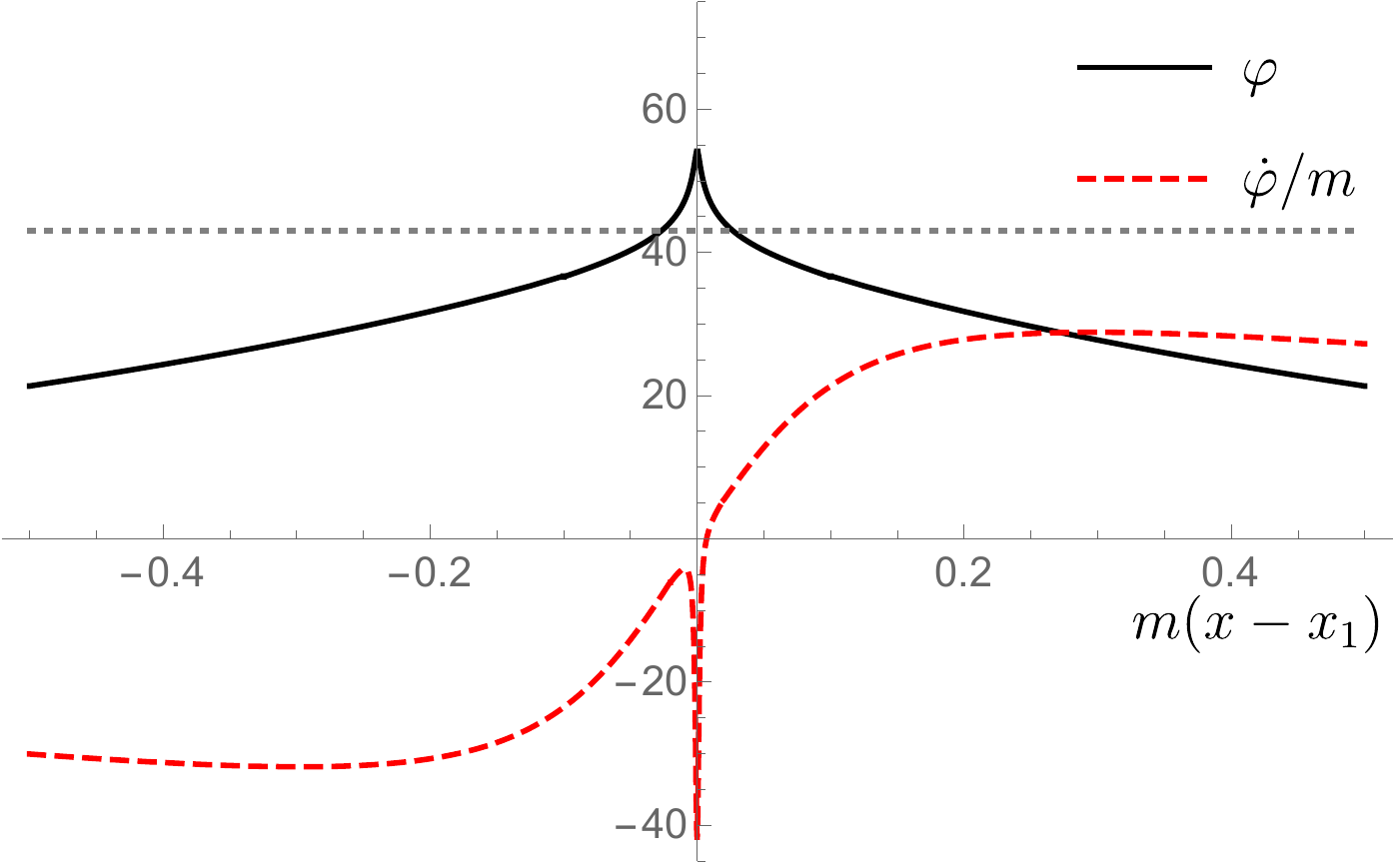}
\hfill
	\includegraphics[width=0.47\textwidth]{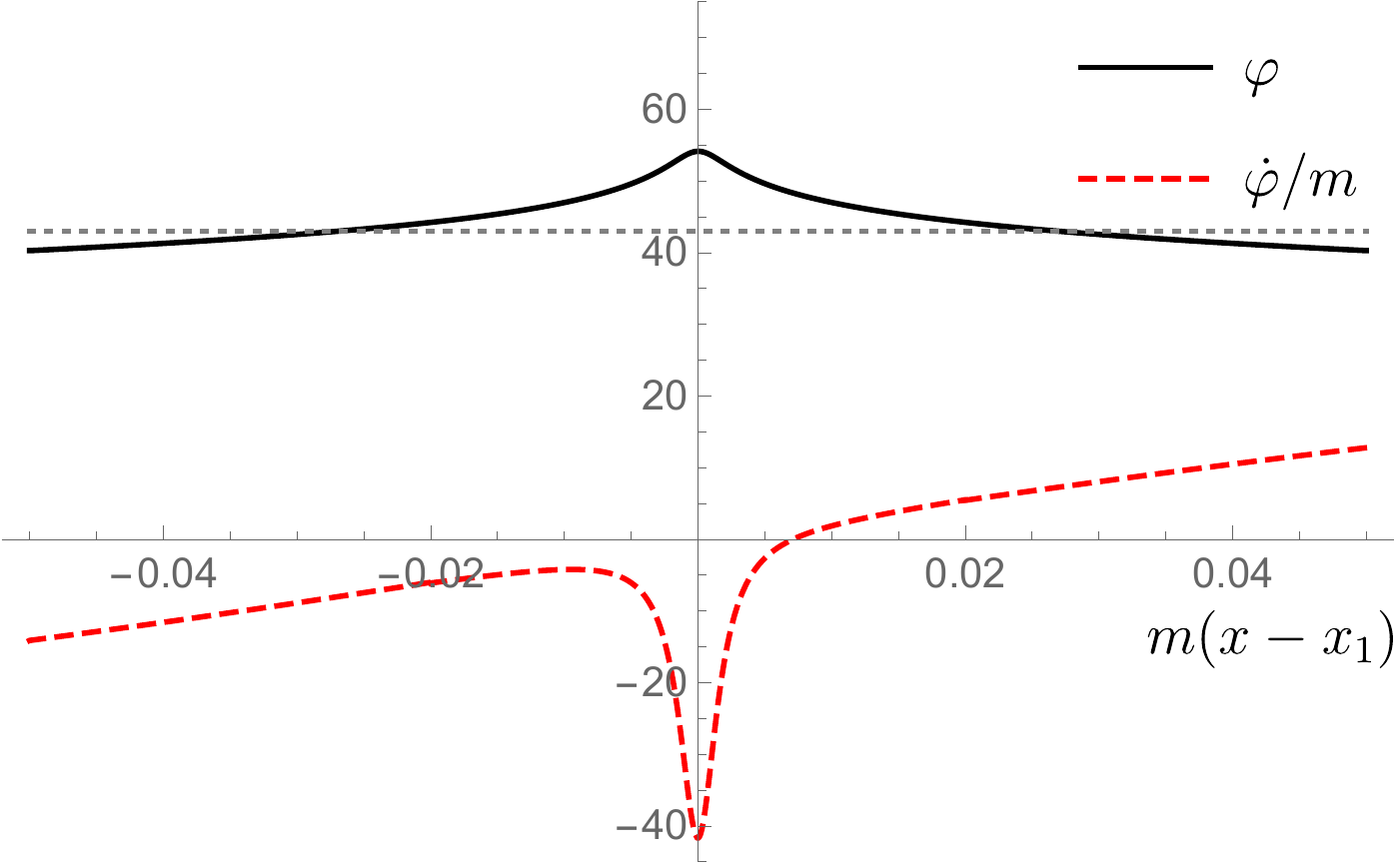}
\caption{Bounce solution describing tunneling from the Unruh vacuum
  far away from the BH. {\bf Left:} Profiles of the
  bounce (black solid) and its time derivative (red dashed) 
at $t=0$ for $\l=0.87\Lambda_{U1}$, where $\Lambda_{U1}$ is defined in
  Eq.~(\ref{TempCritU0}). {\bf Right:} Zoom-in on the central region
  of the left plot. 
We take $\ln(m/\sqrt\kappa)=20$. The
 grey dotted line marks the field value $\vf_{\rm max}$
  at the maximum of the potential barrier, see
  Fig.~\ref{fig:Potential}.
}
\label{fig:UI}
\end{figure}

Clearly, the solution (\ref{U1tail}), (\ref{U_core_far}) is real on
the real time axis and describes run-away towards $\vf\to+\infty$ at
positive time. What distinguishes it from vacuum or thermal
bounces, is the absence of a constant-time slice on which the
solution would have 
zero time derivative $\dot\vf_{\rm b}$. 
The profiles of $\vf_{\rm b}$ and $\dot\vf_{\rm b}$ at $t=0$
are shown in Fig.~\ref{fig:UI}. 

Let us scrutinize the matching procedure. For this purpose, we deform
the contour ${\cal C}$ on which the bounce is 
defined into ${\cal C}'$ consisting of semi-infinite parts at $\Im
t=\pm\pi/\l$, $\Re t<0$ and a Euclidean portion at $-\pi/\l<\Im
t<\pi/\l$, $\Re t=0$ (see Fig.~\ref{fig:bounceth}a). If
\be
\label{bU1cond}
b_{U1}\ll 1\;,
\ee
the core of the bounce fits entirely inside the Euclidean part of the
contour. In other words, the matching region where (\ref{U1cond}) is
satisfied  surrounds the core in Euclidean time.
This region also comfortably overlaps with the domain of
validity of the expression for the Green's function at close
separation, which is bounded by (see Appendix~\ref{Ssec:Green_as})
\be
\label{U1cond1}
|x-x_1|,~|t|\ll 1/\sqrt{\l m}\;.
\ee
On the other hand, when $b_{U1}>1$, the matching procedure in
Euclidean time breaks down. It is unclear if it can be extended to
higher values
of $b_{U1}$ by matching on the parts of the contour
parallel to the real axis.\footnote{In any case, these values are
  bounded from above by $b_{U1}\ll\sqrt{\l/m}$, as required for the
  compatibility of inequalities (\ref{U1cond}), (\ref{U1cond1}).}  
A careful analysis of this issue would
require studying corrections to the core and tail of the bounce which
is beyond the scope of this paper. Thus, we take (\ref{bU1cond}) as a
conservative condition for the validity of the bounce solution
constructed above. 
In view of the formula (\ref{U_a_far}), it translates
into an upper bound on the BH temperature, $\l\lesssim \Lambda_{U1}$,
where
\begin{equation}
\label{TempCritU0}
\Lambda_{U1}=\dfrac{3\pi
  m}{4}\bigg(\ln\frac{m}{\sqrt{\kappa}}+\g_E-\frac{1}{4}\bigg)  \; .
\end{equation}
We will
discuss what happens at higher BH temperatures shortly.

Turning to the tunneling suppression, we need to compute the integral
(\ref{B}). Unlike Minkowski or thermal cases, we cannot deform the
contour to cast this integral into the form of an Euclidean
action. Therefore, we work directly with the contour ${\cal C}$. The
computation requires some care and is relegated to
Appendix~\ref{app:Bcalc}. The result reads 
\begin{equation}
\label{B_U_far}
B_{U1}=\frac{16\pi}{\gc^2}\bigg(\ln\sqrt{\frac{\l m}{\kappa}}
-\frac{2\l}{3\pi m}+\frac{\ln 2+\gamma_E}{2}-\frac{9}{8}\bigg)\;,~~~~~
\l\lesssim\Lambda_{U1} \;.
\end{equation}
Notice that the leading logarithmic part of the suppression can be
easily found by the method outlined at the end of
Sec.~\ref{Ssec:flat_bounce} which relates it to the
field value at the core of the bounce. Substituting 
\be
\vf_{\rm b}(t=0,x=x_1)\approx\ln\bigg(\frac{4\l^2}{\kappa
  b_{U1}}\bigg)
\ee  
into Eq.~(\ref{B0ll}) and using Eq.~(\ref{U_a_far}), we indeed recover
Eq.~(\ref{B_U_far}) up to order-one corrections in the brackets.

We observe that the suppression decreases with the BH temperature and
reaches down to 
approximately half of the vacuum suppression at $\l\approx
\Lambda_{U1}$. 
Comparing it with the suppression of the flat thermal bounce \eqref{SPI}, we
see that the main difference is in the linear-in-$\l/m$ term. The latter is
smaller in the case of Unruh vacuum, hence tunneling from it is more
suppressed. This is, of course, expected due to the deviation of the Unruh
state from thermality.

We presently discuss transitions at BH temperature higher than
$\Lambda_{U1}$. The lessons learned from thermal and Hartle--Hawking
cases suggest that this will be driven by jumps onto sphaleron. An
indication in favour of this guess is that the value of the field in
the core of the bounce (\ref{U_core_far}) 
decreases from $\sim 4\ln(m/\sqrt\kappa)$ at
$\l\approx m$ (the value in the center of the Minkowski bounce) down
to $\sim 2\ln(m/\sqrt\kappa)\approx \vf_{\rm max}$ at
$\l\approx\Lambda_{U1}$, which is the same as the value in the center
of the sphaleron. Of course, due to asymmetry of the Unruh particle
flux, the sphaleron will not be produced at rest. 
Rather, one expects the created sphaleron to move in the
direction of the flux and be accompanied by particle excitations (see
\cite{Levkov:2004tf,Bezrukov:2003tg,Levkov:2007yn,Levkov:2008csa,Demidov:2015bua} 
for the formation of similar ``excited sphalerons'' in
quantum mechanical and field theoretical models). Our analytic
cut-and-match procedure does not allow us to capture this type of
solutions, which do not have a well-localized core in time and space. 

Still, we can try to estimate the corresponding suppression using the
stochastic approach, which we saw to give the right leading-log
results for the thermal and Hartle--Hawking vacuum decays. We expect it to work
in the Unruh case as well, because the Unruh flux in our model is
dominated by soft modes with low frequencies $\omega\sim m\ll \l$ and
large occupation numbers. In other words, the fluctuations of the
field in this flux are essentially semiclassical. This will lead to
classical ``jumps on the barrier'', for which the stochastic picture
provides a fair description. 

Similarly to Eq.~(\ref{thermfluc}), we
estimate the field fluctuations in the Unruh flux as
\begin{equation}
\label{U1fluc}
\delta\vf_{U1}^2=\gc^2\lim_{\begin{smallmatrix}
t\to 0\\x\to  x_1\end{smallmatrix}}\big[\GG_{U}(t,x;0,x_1)-\GG_{F}(t,x;0,x_1)\big]
\approx\frac{{\rm g}^2\l}{3\pi^2 m} \;,
\end{equation}
where we have used the lower line of Eq.~(\ref{G_U_far}) and retained
only the leading term. Then, the vacuum decay rate is
\be
\label{GammaU1stoch}
\varGamma_{U1,\,\text{high-}\l}\sim\exp\left(-\frac{\vf_{\rm
      max}^2}{2\delta\vf_{U1}^2}\right)
\sim \exp\left[-\frac{6\pi^2 m}{{\rm
      g}^2\l}\left(\ln\frac{m}{\sqrt\kappa}\right)^2\right]\;. 
\ee
At $\l\approx\Lambda_{U1}$ the
stochastic estimate coincides with the bounce suppression (\ref{B_U_far})
in the leading-log approximation. It 
is exponentially smaller than the thermal rate (\ref{Stoch_Gamma1}). 
Still, it reaches order-one values at $\l\to\infty$. In other words, the
exponential suppression disappears at high enough BH temperatures. As
we will explain in Sec.~\ref{Sec:disc}, we believe this property to be
special for our model and do not expect it to be
generic. 

\begin{figure}[t]
	\centering 
	\includegraphics[width=0.8\textwidth]{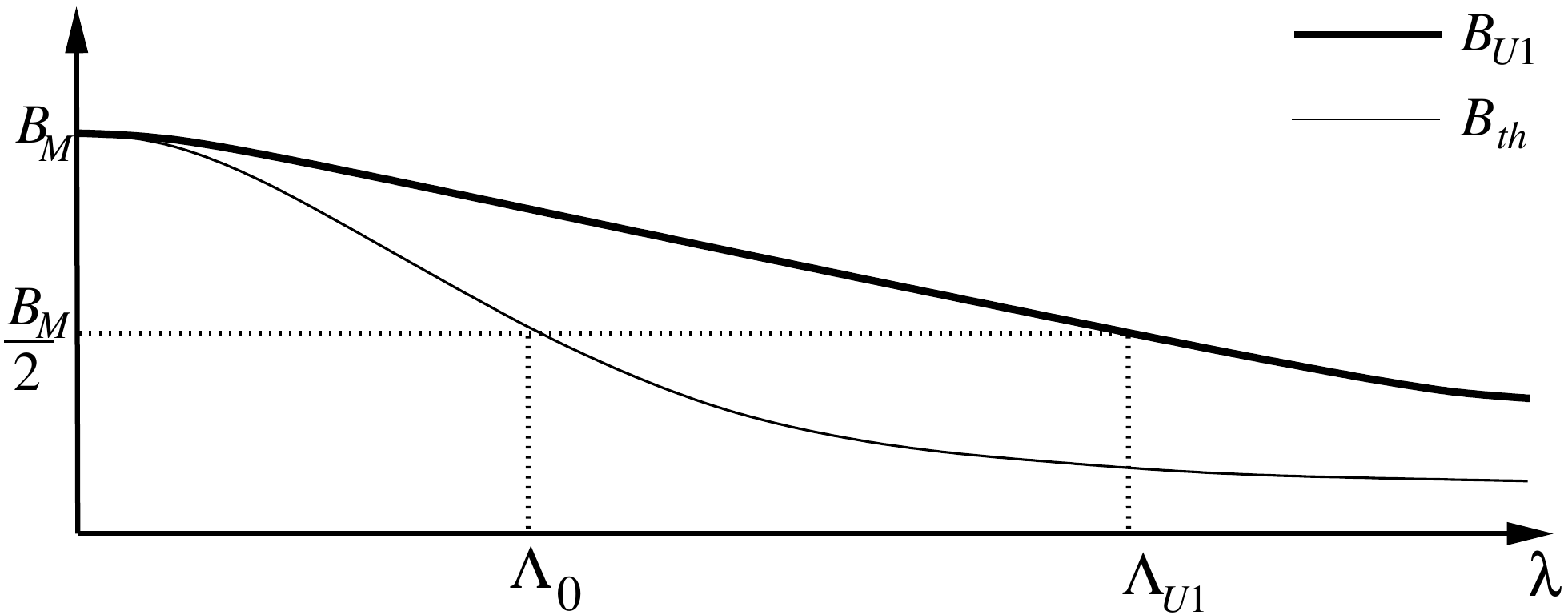}
	\caption{Suppression of the Unruh vacuum decay far from the
          BH as a function of the BH temperature
          $T_{BH}=\l/(2\pi)$ (thick line). The transition between the
          low-temperature bounces and the high-temperature stochastic
          jumps happens at $\Lambda_{U1}$, see
          Eq.~(\ref{TempCritU0}). Thin line shows for comparison the
          suppression of false vacuum decay in a thermal bath with the
          same temperature. Expressions for $\Lambda_0$ and $B_M$ are
          given in Eqs.~(\ref{Csol}),
          (\ref{B0}).} 
	\label{fig:MainPlotU1}
\end{figure}

The results of this subsection for the suppression of the Unruh
vacuum decay far from the BH are summarized in Fig.~\ref{fig:MainPlotU1}.

\subsubsection{Tunneling near horizon}
\label{Sssec:BH_U_near}

In the BH vicinity the vacuum decay is affected both by excitations
and by the nontrivial geometry. The latter significantly contributes to the
enhancement of the decay rate, as we are now going to see.

With the insight from the previous subsection, we can immediately
write down the Ansatz for the bounce centered at a point
$(t_2,x_2)$ in the near-horizon region,
\bseq 
\label{Inst_U}
\begin{align}
\label{Inst_U_core2}
&\vf_{\rm b}\Big|_{\rm core}=\ln\left[ \frac{4\l^2 b_{U2}}{\kappa\left(
    -2\l (v-v_2)\sh\left(\frac{\l}{2}(u-u_2)\right)
+b_{U2}\,\e^{\frac{\l}{2}(u-u_2)}\right)^2} \right] -2\l x\;, \\
\label{Inst_U_tail}
&\vf_{\rm b}\Big|_{\rm tail}=8\pi\GG_{U}(t,x;t_2,x_2)\;,
\end{align}
\eseq
where $u_2=t_2-x_2$, $v_2=t_2+x_2$. The term $-2\l x$ in the first
expression comes from the metric factor $\O(x)$ in the general
solution (\ref{GenSolCurv}) of the Liouville equation in the
near-horizon geometry. The Green's function at close separation near
horizon is given by Eq.~(\ref{G_U_near}) from
Appendix~\ref{app:Green}. Matching it to the asymptotics of the expression
(\ref{Inst_U_core2}), 
we find
\begin{equation}\label{U_a}
b_{U2}=\bar b_{U2}\e^{-2\l x_2}~,~~~~~\bar
b_{U2}=\frac{\kappa}{m^2}\e^{\frac{32\l}{3\pi 
    m}-2\gamma_E-\frac{3}{2}} \;.
\end{equation}

Let us discuss the conditions for the validity of the matching
procedure, which are quite subtle. It is convenient
to introduce the advance Kruskal coordinate $\bar u$ as in
Eq.~(\ref{Kruskal}). In the new coordinates $(\bar u, v)$ the
expression for the bounce core takes the form
\be
\label{Ucore1}
\vf_{\rm b}\Big|_{\rm core}=
\ln\left[\frac{4\l^2\bar b_{U2}\e^{-\l
      v_2}}{\kappa\left(-\l^2(v-v_2)(\bar u-\bar u_2)+\bar
      b_{U2}\e^{-\l v_2}\right)^2}\right]-\l v\;,
\ee
where $\bar u_2=-\l^{-1}\e^{-\l u_2}$. We observe that, apart from the
last linear-in-$v$ term, this has the same form as for the bounce in
flat spacetime. The expression (\ref{G_U_near}) for the Green's 
function at close
separation is valid as long as all points are in the near-horizon
region and $|v-v_2|\ll 1/\sqrt{\l m}$ (note that there are no
restrictions on $|u-u_2|$). Thus, the matching region 
in the two-dimensional complex space of variables $(\bar u,v)$
is determined by
the conditions
\be
\big|\l^2 (v-v_2)(\bar u-\bar u_2)\big|\gg\bar b_{U2}\e^{-\l v_2}~,~~~~~
|v-v_2|\ll 1/\sqrt{\l m}~,~~~~~\Re\big(\l v+\ln(-\l\bar u)\big)<0\;.
\label{2domain}
\ee
The last condition here ensures that in the matching region one can use the
near-horizon form of the metric.
We additionally require that there must be a continuous deformation of
the contour ${\cal C}$, such that on this deformation the matching
region surrounds the core. 

The combination of the first and last conditions in (\ref{2domain})
turns out to provide the strongest restrictions. 
Any deformation of the contour must cross the
wedge where both $\bar u$ and $v$ are real and $\bar u<\bar u_2$, $v>v_2$
(i.e., the right wedge with respect to the point $(u_2,v_2)$ in the
original coordinates). The possibility to have an overlap on this
intersection implies that there is $v_*>v_2$, such that 
\be
\l(v_*-v_2)\big(\e^{-\l(v_*-v_2)}+\l\bar u_2\e^{\l v_2}\big)\gg\bar
b_{U2}. 
\ee
Given that $\bar u_2<0$, the l.h.s. of the above inequality does not exceed a value of order
one. Hence, we conclude that the necessary condition for matching is 
\be
\label{bU2cond}
\bar b_{U2}\ll 1\;.
\ee
One can verify that this condition is also sufficient. It restricts
the validity of the bounce solutions (\ref{Inst_U}) to 
$\l\lesssim \Lambda_{U2}$, where
\begin{equation}
\label{TempCritU}
\Lambda_{U2}=\frac{3\pi m}{16}
\bigg(\ln\frac{m}{\sqrt{\kappa}}+\gamma_E+\frac{3}{4}\bigg) \;.
\end{equation}
Note that this critical temperature is higher than its counterpart
(\ref{TempCritHH}) for 
tunneling from the Hartle--Hawking vacuum.

Similarly to the Hartle--Hawking case, we have obtained a family of
bounces parameterized by the position of the center. Their shape
depends on $x_2$. Still, their suppression is $x_2$-independent and
reads (see Appendix~\ref{app:Bcalc} for the calculation)
\begin{equation}\label{B_U_near}
B_{U2}=\frac{16\pi}{\gc^2}\left(\ln\sqrt{\frac{\l m}{\kappa}}
-\dfrac{8\l}{3\pi m}+\frac{\ln 2+\gamma_E}{2}-\frac{5}{8}\right)
 \; , ~~~~~ \l\lesssim\Lambda_{U2} \;.
\end{equation}
This implies less suppression than for tunneling far away from
the BH, Eq.~(\ref{B_U_far}). The enhancement of the transition rate 
is due to the
excitation of modes localized on the BH by the potential barrier
$U_{\rm eff}(x)=m^2\O(x)$
(see Fig.~\ref{fig:PotForModes1}). The existence of such localized
states is, in turn, a consequence of the nontrivial geometry. On the
other hand, as one could conjecture \cite{Arnold:1989cq}, 
the suppression is stronger
than for the decay of the Hartle--Hawking vacuum,
Eq.~(\ref{B_BH_HH}). 
Our result
(\ref{B_U_near}) provides the first confirmation of this conjecture
from first principles. 

As in the Hartle--Hawking case, we expect the degeneracy between
the bounces with different $x_2$ to be lifted if we take into account
deviation of the near-horizon metric from the Rindler form. What is
then the least suppressed solution? A natural candidate is the ``Unruh
sphaleron'' obtained by taking the limit\footnote{This corresponds to
  the limit $u_2\to\infty$ at fixed $v_2$ in the original expressions
  (\ref{Inst_U}).} $\bar u_2\to 0^-$ in
Eq.~(\ref{Ucore1}). Setting for simplicity $v_2=0$, we have in the
original coordinates
\be
\label{Usphaleron}
\vf_{\rm sph}\Big|_{\rm core}=\ln\left[\frac{4\l^2 \bar
    b_{U2}}{\kappa(\l v\,\e^{-\l u}+\bar b_{U2})^2}\right]-\l v\;.
\ee 
This solution is regular at $t>0$, thus it does not exhibit run-away
towards the true vacuum. We conjecture that it may instead describe at
late times formation of the Hartle--Hawking sphaleron of
Sec.~\ref{Ssec:BH_HH} plus excitations flying away from the BH. Our
analytic method does not allow us to trace this late-time evolution,
which would require a numerical study. Notice that, unlike any other
solutions considered so far, the sphaleron profile
(\ref{Usphaleron}) diverges linearly when $x\to -\infty$ at fixed
$t$. We do not know if this is a serious drawback. The
above limit corresponds to approaching the bifurcation point between
the past and future horizons on the extended Penrose diagram of the BH
spacetime, see Fig.~\ref{fig:Penrose}. So, the divergence is consistent
with the fact that the Unruh vacuum is
singular on the past horizon. On the other hand, it violates an
implicit assumption made throughout our analysis that the field is
bounded at $x\to-\infty$. In any case, even if the sphaleron
(\ref{Usphaleron}) should be excluded from the space of admissible
solutions, it is likely to reappear as the limiting configuration of
regular bounces (\ref{Inst_U}) that in the full treatment will arise
as constrained instantons (cf.~\cite{Affleck:1980mp}).  

\begin{figure}[t]
	\centering 
	\includegraphics[width=0.5\textwidth]{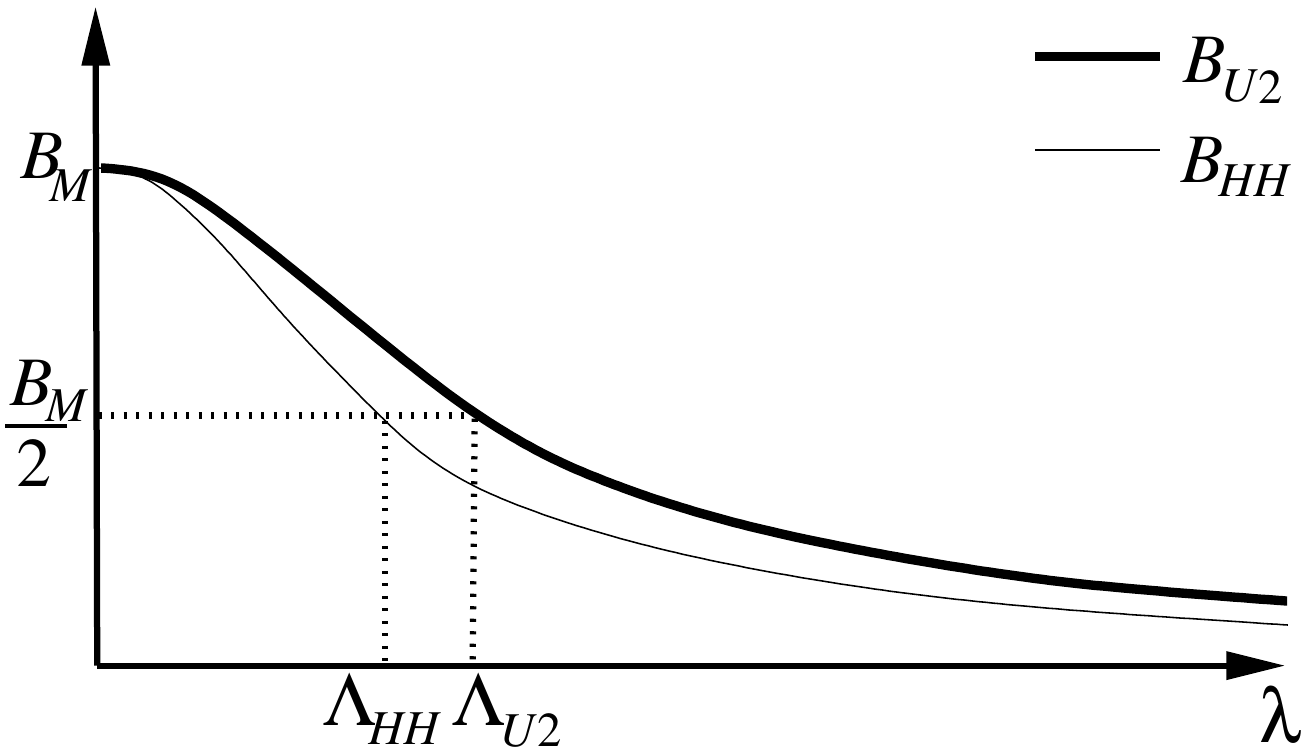}
	\caption{Suppression of the Unruh vacuum decay as a function
          of BH temperature $T_{BH}=\l/(2\pi)$ (thick
          line). The critical temperature $\Lambda_{U2}$
          (Eq.~(\ref{TempCritU})) marks the transition from the
          low-temperature regime dominated by bounce solutions in the
          near-horizon region to high-temperature stochastic jumps
          over the potential barrier. The suppression of the
          Hartle--Hawking vacuum decay is shown by the thin line for comparison, with
          the critical temperature $\Lambda_{HH}$ given by
          Eq.~(\ref{TempCritHH}). $B_M$ is the value of the
          suppression in the absence of BH, Eq.~(\ref{B0}).
}
	\label{fig:MainPlotMain}
\end{figure}

At BH temperature above $\Lambda_{U2}$, the bounce core stops fitting
into the near-horizon region, and our analytic method fails to produce
the solution. However, we can again estimate the tunneling rate using
the stochastic picture. From the expression (\ref{G_U_far}) (upper
line) 
for the Green's function
outside, but not very far from the BH 
we read out the amplitude of the field
fluctuations,
\be
\label{U2fluc}
\delta\vf_{U2}^2\approx \frac{4{\rm g}^2\l}{3\pi^2 m}\;.
\ee    
Substituting this into the estimate of the decay rate gives
\be
\label{GammaU2stoch}
\varGamma_{U2,\,\text{high-}\l}\sim\exp\left(-\frac{\vf_{\rm
      max}^2}{2\delta\vf_{U2}^2}\right)
\sim \exp\left[-\frac{3\pi^2 m}{2{\rm
      g}^2\l}\left(\ln\frac{m}{\sqrt\kappa}\right)^2\right]\;. 
\ee
We observe that in the leading-log approximation the suppression is
continuous across the transition from the low- to high-temperature
regime. 

The suppression of the false vacuum decay from the Unruh vacuum is plotted in Fig.~\ref{fig:MainPlotMain}, where it is
compared to the suppression of tunneling from the Hartle--Hawking
state. We see that the former is always larger than the latter, though
it also goes to zero with the increase of the BH temperature. The
situation is qualitatively similar to the case of tunneling far away
from the BH, cf. Fig.~\ref{fig:MainPlotU1}. We are going to argue,
however, that vanishing of the Unruh suppression at high BH
temperature is not expected to be generic.

\section{Discussion and outlook}
\label{Sec:disc}

In this paper we have developed an approach for the analysis of false
vacuum decay in BH background. This approach is rooted in the in-in
formalism and treats various vacua in the BH geometry as mixed
states. It reduces the task of finding the vacuum decay rate to the
solution of classical field equations on a contour in complex
Schwarzschild time. The solution --- bounce --- must interpolate
between the basins of attraction of the false and true vacua and is
subject to boundary conditions in the asymptotic past that encode the
details of the initial quantum state. These are the same boundary
conditions as for the time-ordered Green's function in the respective
state. They allow one to discriminate between different vacua in the
BH background: Boulware, Hartle--Hawking and Unruh.

Our method is general and can be applied to many situations other than
an isolated BH. For example, realistic BHs in the early Universe are
immersed in thermal plasma whose temperature may be different from
that of the BH \cite{1706.01364}. 
Another example of nontrivial environment is the de
Sitter spacetime relevant for inflation, which also possesses its own
ambient temperature. These cases present an interesting arena for the
use of our approach. Our method can also be of interest beyond the BH
physics, for the study of tunneling in various non-equilibrium quantum
systems. 

It is worth comparing our method to the approach of
\cite{Braden:2018tky,Hertzberg:2019wgx} proposing to describe
tunneling as classical evolution with stochastic initial
conditions. Despite some resemblance, there are important
differences. Refs.~\cite{Braden:2018tky,Hertzberg:2019wgx} assume that
the
fields are purely real and evolve in real time. Whereas in our
method the tunneling solutions are essentially complex and live on a
contour in the complex time that in general cannot be continuously
deformed into the real axis because of branch-cut
singularities.\footnote{Though this can be possible in some special
  cases \cite{Levkov:2004ij,Demidov:2015bua}.} Further, in the case of
false vacuum decay in flat space the stochastic approach tends to
significantly overestimate the decay rate \cite{Hertzberg:2020tqa},
whereas our method recovers exactly the results of the standard
Euclidean description. On the other hand, we expect the stochastic
approach to work well for tunneling from states with large occupation
numbers \cite{Grigoriev:1988bd,Grigoriev:1989je,Grigoriev:1989ub,
Khlebnikov:1998sz}, such as the Hartle--Hawking and Unruh vacua in the
background of a two-dimensional BH with high temperature. It will be
interesting to apply the method 
of~\cite{Braden:2018tky,Hertzberg:2019wgx,Hertzberg:2020tqa} to this
case and compare it with our results.

We have demonstrated the efficiency of our method on an example of a
two-dimensional scalar toy model with self-interaction given by the 
inverted Liouville
potential. Due to the specific properties of the model, the bounce
solutions could be found analytically in most regimes. Their structure
is in a sense opposite to the widely used thin-wall picture where the
size of the bounce solution is much larger than the Compton wavelength
of the field. By contrast, the bounces in our model have a tiny
nonlinear core with the size much smaller than the Compton
wavelength, and a wide tail of size $1/m$ where the field is linear.

Using this example, we connected our method to the
standard Euclidean approach to tunneling from equilibrium states, such
as the flat-space vacuum, thermal bath and the 
Hartle--Hawking state. We clarified several
details of these processes along the way. 
We further found the bounce solutions
describing decay of the Unruh vacuum where the Euclidean formalism is
not available. 

In the Hartle--Hawking case the tunneling regime changes from what can
be called ``direct tunneling'' at low BH temperature, where the bounce
describes run-away to the true vacuum at late times, to the
``sphaleron-driven'' transitions at high temperature that proceed via
formation of the sphaleron. The latter can be viewed as jumps on the
saddle point of the potential separating the two vacua, driven by
stochastic thermal fluctuations. We saw indications of a similar
change of the tunneling regime in the case of Unruh vacuum, but were
unable to find the high-temperature sphaleron-forming solutions
analytically. Still, we could estimate the tunneling suppression using the
stochastic picture and showed that it continuously connects in the leading order
to the rate found from the analytic bounces at low temperature. It
will be interesting to perform a numerical study of the
high-temperature Unruh bounces in order to rigorously establish the
transition between different tunneling regimes and refine the
computation of the transition rate at high BH temperatures.  

To our knowledge, the present work gives the first calculation of
the catalyzing effect of BH on vacuum decay fully taking into account
the structure of the Unruh state. This catalyzing effect is twofold.
Part of it is due to the change in geometry. Another part is due to
the excitation of the field modes by the BH. We found that
the two effects are of the same order and are closely
intertwined. Namely, the gravitational redshift near the BH horizon
gives rise to extra states localized on the BH, in addition to the
modes radiated away to infinity. These bound states get excited and
significantly enhance the field fluctuations in the BH vicinity,
thereby facilitating the vacuum decay.

Our results confirm the conjecture \cite{Arnold:1989cq} that the decay
rate of the Unruh vacuum is exponentially smaller than that of the
Hartle--Hawking state. Still, we found that the suppression disappears
at high BH temperature. As discussed in the Introduction, if this
result were to hold for realistic four-dimensional BHs, it would have
dramatic consequences for phenomenological model building. There are,
however, important properties of the four-dimensional setup that are
not captured by the two-dimensional model and may alter the
conclusion. 

\begin{figure}[t]
	\centering 
	\includegraphics[width=0.45\textwidth]{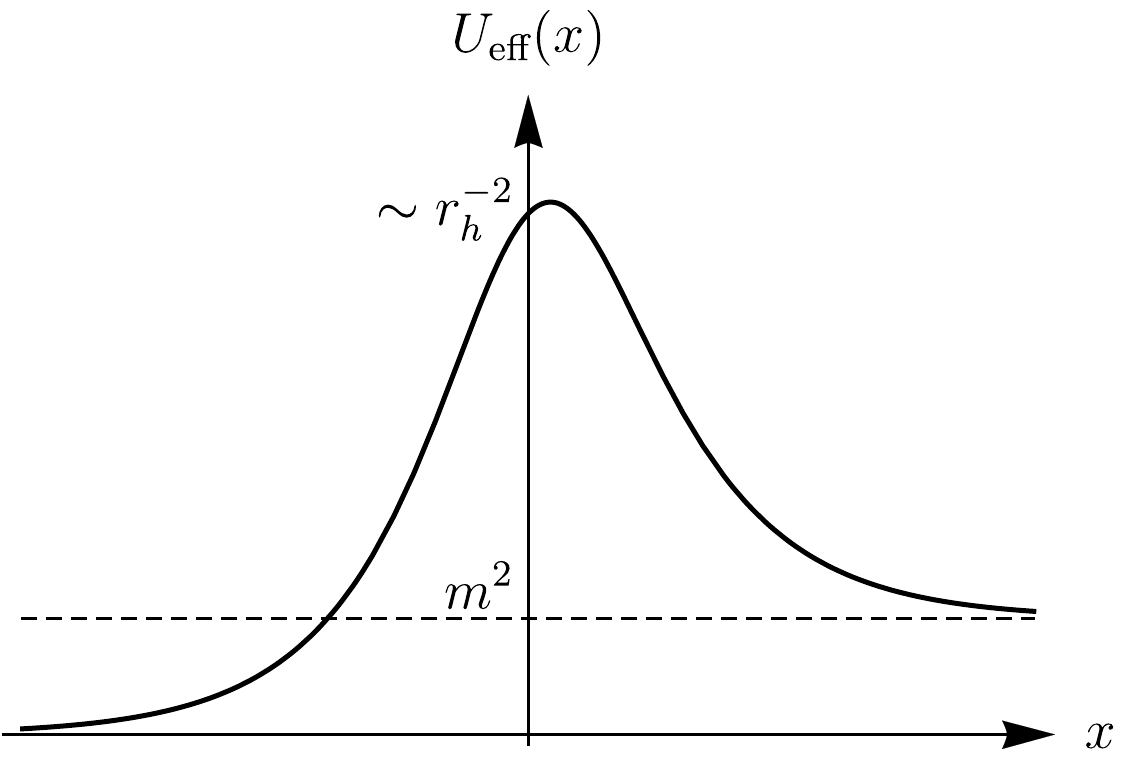}
	\caption{Effective potential for spherically symmetric 
 linear massive modes in the four-dimensional Schwarzschild
 geometry.}
	\label{fig:Potential4}
\end{figure}

One of them is the structure of the effective potential for the field
modes. In the two-dimensional case, it is a monotonic function of
the tortoise coordinate $x$, varying between $0$ and $m^2$, see
Fig.~\ref{fig:PotForModes1}. This allows all modes with frequencies
below and of
order $m$ to contribute into the field
fluctuations within the distance $\sim 1/m$ from
the horizon. Due to large occupation numbers,
these modes dominate tunneling at high temperature. On the other hand,
the effective potential in the Schwarzschild metric, obtained upon
spherical reduction, has the form depicted in
Fig.~\ref{fig:Potential4}. The important new feature is the presence
of a potential barrier with the height scaling as 
$\sim r_h^{-2}$, where $r_h$ is the Schwarzschild radius. It
suppresses the escape of the low-frequency modes, the effect being
encapsulated by the well-known fall-off of the BH grey-body factors at
small $\omega$. This is expected to suppress the field
fluctuations outside the BH resulting in a qualitatively different
behavior of the vacuum decay rate at high temperature
\cite{1704.05399}. 
Another effect that is expected to further reduce particle number
density outside the BH is the geometric $1/r^2$ spreading of the
particle flux. The general formalism developed in Sec.~\ref{Sec:gen}
of the paper in principle allows one to take both these effects into
account, and we plan to return to this topic in future work.

One more direction for future research is calculation of the
pre-exponential factor in the decay rate formula
(\ref{DecayRateLO}). Of course, its effect is subleading to that of
the exponential suppression which was the focus of this
paper. Still, it is important to set the overall dimensionful scale of
the rate. A related question is inclusion of the thermal corrections
to the scalar field potential. Again, these are generally expected to
be small~\cite{Hayashi:2020ocn}, but may be enhanced in theories with
large number of 
particle species and not-so-small couplings
\cite{1706.01364,1708.02138}. 
The latter class includes
the Standard Model --- admittedly, the most interesting case from the
phenomenological perspective. 

Finally, one would like to include dynamical gravity. This would open
the way to address such questions as the relation between the
tunneling probability and the BH entropy, or the possibility for a
complete BH disappearance as a result of tunneling~\cite{1401.0017}. Several
obstacles must be overcome to achieve this goal. The most important
one is the generalization of the boundary conditions on the bounce
solution formulated in this work to the case when the position of the
BH horizon is not fixed, but is itself a dynamical variable.

\section*{Acknowledgments}

We thank Matthew Johnson, Dmitry Levkov and Valery Rubakov for useful
discussions. The work of A.S. was in part supported by the Department of Energy Grant DE-SC0011842. The work of S.S. was partially supported by the
Natural Sciences and Engineering Research Council (NSERC) of Canada and
by the Russian
Foundation for Basic Research grant 20-02-00297.
Research at Perimeter Institute is supported in part by the Government
of Canada through the Department of Innovation, Science and Economic
Development Canada and by the Province of Ontario through the Ministry
of Colleges and Universities. 

\appendix
\renewcommand{\theequation}{\Alph{section}.\arabic{equation}}

\section{Dilaton black holes}
\label{app:Dil_grav}

Two-dimensional dilaton gravity includes the metric $g_{\mu\nu}$ and
the dilaton $\phi$ with the action \cite{Callan:1992rs} 
\begin{equation}
S_{\rm DG}=\int d^2x \sqrt{-g}
\;\e^{-2\phi}\big[R+4(\nabla_\mu\phi)^2+4\l^2\big]\;.
\end{equation}
Here $R$ is the scalar curvature and $\l$ is a constant parameter. The
strength of gravitational interactions is governed by the
field-dependent coupling $\e^{2\phi}$. The theory admits a
one-parameter family of solutions with the metric of the form
(\ref{metral}) and 
\begin{equation}
\O=1-\frac{M}{2\l} \e^{-2\l r}~,~~~~~\phi=-\l r \;.
\end{equation}
They describe BHs with the mass $M$ and horizon radius 
\begin{equation}\label{rh}
r_h=\frac{1}{2\l}\ln \frac{M}{2\l}\;.
\end{equation}
Introducing the tortoise coordinate 
\begin{equation}
x=\frac{1}{2\l}
\ln\bigg(\e^{2\l r}-\frac{M}{2\l}\bigg) -\frac{1}{2\l}\ln\frac{M}{2\l}
\end{equation}
and re-expressing $\Omega$ as a function of $x$, we obtain
Eq.~(\ref{O_BH}) from the main text.

Throughout the paper we neglect the back-reaction of the tunneling
field $\vf$ on the geometry. This is justified if the gravitational
coupling is sufficiently small. The maximal value of the coupling in
the BH exterior, which is the only region relevant for the false vacuum
decay, is achieved at the BH horizon and equals $2\l/M$. Thus, the
back-reaction can indeed be neglected for heavy enough BHs. Note that
due to peculiarities of the two-dimensional theory, the BH temperature
does not depend on its mass, being set by the parameter $\l$ in the
dilaton gravity action.

\section{More on modes and Green's functions}
\label{app:Green}

\subsection{Solutions to the Shr\"odinger equation}
\label{app:Green_modes}

Here we study some general properties of the massive linear modes $f_{R,\o}$,
$f_{L,\o}$ introduced in Sec.~\ref{Ssec:gen_vacuum} and give their
explicit expressions for a number of cases.

\subsubsection{General properties}

We assume that the effective
potential for the modes $U_{\rm eff}(x)$ behaves as $\propto\e^{2\l
  x}$ at large negative $x$ and goes to $1$ at large positive $x$,
see Fig.~\ref{fig:PotForModes1}. 
This is the case for the dilaton BH where
$U_{\rm eff}(x)=m^2\O(x)$ and $\O(x)$ is given in Eq.~\eqref{O_BH}. In
this case the potential grows monotonically as $x$ increases. However, 
the latter 
property is not used in this subsection which also applies to more
general nonmonotonic potentials
like the one in Fig.~\ref{fig:Potential4}. 

The asymptotic expansions of the modes $f_{R,\o}$, $f_{L,\o}$ are
given in Eqs.~(\ref{fRoas}), (\ref{fLoas}). 
We focus on $\o>m$
as only in this case two linearly-independent solutions exist and the
relation between their expansions is nontrivial. 
Note that $\gamma_\o$ and $\b_\o$ are
the transmission and reflection amplitudes of 
the potential $U_{\rm eff}$.
We will now show that all other
coefficients in the asymptotic expansions 
(\ref{fRoas}), (\ref{fLoas})
are expressed through these amplitudes.

The eigenfunctions $f_{R,\o}$, $f_{L,\o}$ belong to the continuous
spectrum of a Schr\"o\-din\-ger equation. Thus
their norms and orthogonality are determined by their
behavior at infinity. Each plane-wave integral should be treated as
contributing half of the $\delta$-function,
\[
\int_{-\infty}dx \,\e^{i(\o-\o')x}\sim
\frac{1}{2}\delta(\omega-\omega')~,~~~~~
\int^{+\infty}dx \,\e^{i(k-k')x}\sim
\frac{1}{2}\delta(k-k')=\frac{k}{2\o}\delta(\o-\o')\;.
\]
Then the orthogonality and normalization conditions \eqref{ModeOrth},
\eqref{ModeNorm}
imply
\bseq
\label{AppModes123}
\begin{align}
\label{AppModes3}
&\b\tilde\b^*+\frac{k}{\o}\gamma \tilde\gamma^*=0\;,\\
&
\frac{1}{2}|\a|^2+\frac{1}{2}|\b|^2+\frac{k}{2\o}|\gamma|^2=1
\;, \label{AppModes1}\\ 
&
\frac{1}{2}|\tilde\b|^2+\dfrac{k}{2\o}|\tilde\gamma|^2
+\frac{k}{2\o}|\tilde\delta|^2=1 
\;, \label{AppModes2} 
\end{align}
\eseq
where we have omitted the subscript ``$\o$'' to avoid clutter of
notations. 
Next, from the viewpoint of the effective Schr\"odinger equation, the
modes $f_{R,\o}$, $f_{L,\o}$ represent wavefunctions of a particle
scattering off the potential barrier. These wavefunctions preserve
the probability flux which gives us two more conditions,
\bseq
\label{AppModes45}
\begin{align}
&|\a|^2  =|\b|^2+\dfrac{k}{\o}|\gamma|^2 \;, \label{AppModes4}\\
&\dfrac{k}{\o}|\tilde\delta|^2=|\tilde\b|^2+\dfrac{k}{\o}|\tilde\gamma|^2 \;. 
\label{AppModes5}
\end{align}
\eseq
Finally, we note that the complex conjugated functions $f_{R,\o}^*$,
$f_{L,\o}^*$ must be linear combinations of $f_{R,\o}$,
$f_{L,\o}$ as the latter form a complete set of linearly-independent
solutions,  
\bseq
\label{conjff}
\begin{equation}
f^*_{R,\o}=A f_{R,\o}+B f_{L,\o} \;, ~~~~
f^*_{L,\o}=C f_{R,\o}+D f_{L,\o} \;. 
\end{equation}
We write these relations at $x\to\pm\infty$ using the expressions 
(\ref{fRoas}), (\ref{fLoas})
and equate terms with the same exponential
factors to obtain eight conditions on six amplitudes and four complex
coefficients $A,..,D$. Solving them with respect to the latter gives 
\begin{equation}
A=\dfrac{\b^*}{\a} \;, ~~~ B=\dfrac{\gamma^*}{\tilde\delta} \;, ~~~
C=\dfrac{\tilde\beta^*}{\a} \;, ~~~ D=\dfrac{\tilde\gamma^*}{\tilde\delta} \;. 
\end{equation}
\eseq
The remaining conditions become
\bseq
\label{AppModes67}
\begin{align}
& \a^*=\dfrac{|\b|^2}{\a}+\dfrac{\gamma^*\tilde\beta}{\tilde\delta}  \;, 
& \tilde\delta^*=\dfrac{\tilde\beta^*\gamma}{\a}
+\dfrac{|\tilde\gamma|^2}{\tilde\delta} \;, \label{AppModes6}\\
& \dfrac{\tilde\beta^*\b}{\a}
+\dfrac{\tilde\gamma^*\tilde\beta}{\tilde\delta}=0 \;, 
& \dfrac{\b^*\gamma}{\a}+\dfrac{\gamma^*\tilde\gamma}{\tilde\delta}=0
\;. 
\label{AppModes7}
\end{align}
\eseq
Note that not all of the relations (\ref{AppModes123}),
(\ref{AppModes45}), (\ref{AppModes67}) are independent.  

From Eqs.~\eqref{AppModes1}, \eqref{AppModes4} we get $|\a|=1$ and 
the relation between the absolute
values of $\beta$ and $\gamma$
\bseq
\label{modecoeff}
\be
\label{gammabeta}
|\gamma_\o|^2=\frac{\o}{k}\big(1-|\beta_\o|^2\big)~,\qquad\qquad \o>m\;,
\ee
where we have restored the index $\o$ and have explicitly emphasized
that this relation is valid 
at $\o>m$.
We use
the freedom in choosing the overall phase to set 
\be
\label{modecoeff1}
\a_\o=1\;.
\ee  
Next, combining Eqs.~\eqref{AppModes2}, \eqref{AppModes5} gives
$|\tilde\delta_\omega|=\sqrt{\o/k}$.
Again, we fix the phase so that $\tilde\delta_\omega$ is positive real.
Finally, the last of Eqs.~(\ref{AppModes7}) and Eq.~(\ref{AppModes3})
yield the expressions for $\tilde\gamma_\omega$ and
$\tilde\beta_\omega$. Collecting 
them together, we get
\be
\label{modecoeff2}
\tilde \beta_\o=\sqrt{\frac{k}{\o}}\,\gamma_\o~,~~~~
\tilde
\gamma_\o=-\sqrt{\frac{\o}{k}}\,\frac{\gamma_\o\beta^*_\o}{\gamma^*_\o}~,~~~~
\tilde\delta_\o=\sqrt{\frac{\o}{k}}
~,\qquad\qquad \o>m\;.
\ee 
\eseq
Thus, we have expressed all coefficients in terms of the reflection
and transmission amplitudes $\beta_\o$, $\gamma_\o$ of the potential $U_{\rm
  eff}$. It is straightforward to check that the 
remaining relations in (\ref{AppModes67}) are satisfied
and do not yield any further restrictions.

\subsubsection{Special cases}

\paragraph{Dilaton black hole background.} In this case $\O$ is given
by Eq.~\eqref{O_BH} and the eigenmode equation can be solved
exactly,
\bseq
\label{Modes_BH}
	\begin{align}
\label{Modes_BHR}
	& f_{R,\o}=\N_\o \e^{ikx}\: _2 F_1\left({\frac{i}{2\l}(\o-k) \: , \: -\frac{i}{2\l}(\o+k) \: ,\: -\frac{ik}{\l}+1 \: ; \: -\e^{-2\l x} }\right) \;, \\
	& f_{L,\o}=\N_\o\sqrt{\frac{k}{\o}} \e^{-i\o x}\: _2
        F_1\left({ \frac{i}{2\l}(k-\o) \: , \:
            -\frac{i}{2\l}(\o+k) \: , \: -\frac{i\o}{\l}+1  \:; \:
            \e^{2\l x} } \right) \theta(\o-m) \;,
\label{Modes_BHL}
	\end{align}
\eseq
where $_2F_1$ is the hypergeometric function. The momentum $k$ 
is given by $\sqrt{\o^2-m^2}$ when $\o>m$ and by the analytic continuation
$k=i\sqrt{m^2-\omega^2}$ when $\o<m$. Only modes $f_{R,\o}$ exist for
$\o<m$, whereas $f_{L,\o}$ vanish, which is ensured by the
$\theta$-function in Eq.~(\ref{Modes_BHL}). The normalization factor 
$\N_\o$ reads
\begin{equation}
\N_\o=\frac{\Gamma\left( -\frac{i}{2\l}(\o+k) \right)\Gamma\left( 1-\frac{i}{2\l}(\o+k) \right)}{\Gamma\left( -\frac{i\o}{\l}\right)\Gamma\left( 1-\frac{ik}{\l}\right)} \;.
\end{equation}
To calculate the Green's functions, we need the explicit form of
the amplitudes $\b_\o$, $\gamma_\o$ at $\o\ll\l$. 
Expanding
Eqs.~\eqref{Modes_BH} in the limits $x\to\pm\infty$
and comparing with Eqs.~(\ref{fRoas}), (\ref{fLoas}), we obtain
\begin{equation}\label{ab_BH}
\b_\o=\frac{\o-k}{\o+k}~,~~~~ \gamma_\o=\frac{2\o}{\o+k} \;,~~~~~~~~ \o\ll\l \;.
\end{equation}
In this derivation we  
have used
the asymptotics of the hypergeometric function following from 
Eq.~\eqref{hypergrel}. Note that these amplitudes are the
same as for a step-like potential of height $m^2$. They are valid both
for $\o>m$ and for $\o<m$ with the analytic continuation of momentum
(\ref{kappa}). 

\paragraph{Rindler metric.}
In Sec.~\ref{Sec:R} we consider dynamics of the scalar field in
Rindler spacetime. In this case the eigenmodes are still solutions to the 
Schr\"odinger equation (\ref{EqModes}), but now with the effective
potential $U_{\rm eff}=m^2\e^{2\l x}$ that grows to infinity at large positive
$x$. Therefore, there is no separation into left- and right-moving
modes: all modes must decay at $x\to+\infty$.
The complete set of eigenfunctions is
given by
\begin{equation}\label{Modes_R} 
f_{\o}=\sqrt{\frac{4\o}{\pi\l}\sh\frac{\pi\o}{\l}}\,K_{\frac{i\o}{\l}}
\left(\frac{m}{\l}\e^{\l  x}\right)\;,
\end{equation}
where $K$ is the modified Bessel function of the second kind. Note
that these modes are real. They
decay faster than exponentially at $x\to+\infty$ and become a sum of massless
plane waves at $x\to-\infty$.

\subsection{Green's functions at close separation}
\label{Ssec:Green_as}

In the main text we use the expressions for various Green's functions
in the dilaton BH background when the two points in their arguments
are close to 
each other. In this Appendix we derive these expressions. Given that
the Green's functions are time-translation invariant, we can set
$t'=0$. Then we consider the limit
\begin{equation}
\label{As_cond1}
|x-x'|\ll m^{-1}\;, ~~~~~ |t|\ll m^{-1}\;,~~t'=0\;,
\end{equation}
which we will denote with the superscript ``close''. On top of this, to
derive analytic formulae, we will assume high BH temperature, $\l\gg
m$, and will place the two points in the Green's function either to
the right of the transition region where the BH potential changes
significantly, 
\be
\label{rightlim}
x,x'>0\;,~~~~~x,x'\gg \l^{-1}\qquad\qquad\text{(``right'')},
\ee
or to the left of it in the near-horizon region,
\be
\label{leftlim}
x,x'<0\;,~~~~~|x|,|x'|\gg \l^{-1}\qquad\qquad\text{(``left'')}.
\ee 
In these regions we can use the asymptotics for the mode functions in
terms of plane waves (\ref{fRoas}), (\ref{fLoas}). Note that in both
cases $|x+x'|$, can be smaller or larger than~$m^{-1}$.

\subsubsection{Boulware Green's function}

\paragraph{Right region $x,x'>0$.}
Using the expression (\ref{G_B}) and
the asymptotics of the mode functions (\ref{fRoas}), (\ref{fLoas})
with the relations between the coefficients
(\ref{modecoeff}), 
we obtain
\be
\label{GBright1}
\begin{split}
&\GG_B\big|_{\rm right}=\int_m^\infty \frac{d\o}{4\pi k}
\big[\e^{ik(x-x')}+\e^{-ik(x-x')}\big]\e^{-i\o|t|}\\
&-\int_m^\infty \frac{d\o}{4\pi k}
\bigg[\frac{\g_\o\b_\o^*}{\g_\o^*}\e^{ik(x+x')}
+\frac{\g_\o^*\b_\o}{\g_\o}\e^{-ik(x+x')}\bigg]\e^{-i\o|t|}
+\int_0^m \frac{d\o}{4\pi \o} |\g_\o|^2\e^{-\vk(x+x')-i\o|t|},
\end{split}
\ee
where $k$ and $\vk$ are defined in Eqs.~(\ref{k}) and (\ref{kappa}).
In the first line we recognize the expression for the Feynman Green's
function in flat spacetime given by Eq.~(\ref{G_F_flat}). At close
separation it reduces to 
\be
\label{G_F_close}
\GG_F\big|^{\rm close}=-\frac{1}{4\pi}
\ln\big[m^2(x-x'-t)(x-x'+t)+i\epsilon\big]+\frac{\ln 2-\g_E}{2\pi}\;,
\ee
where $\g_E$ is the Euler constant. Notice that away from the
coincident points it splits into a sum of right- and left-moving
contributions depending only on $(x-x'-t)$ and $(x-x'+t)$,
respectively. 

\begin{figure}[t]
\centering{
\includegraphics[width=0.4\textwidth]{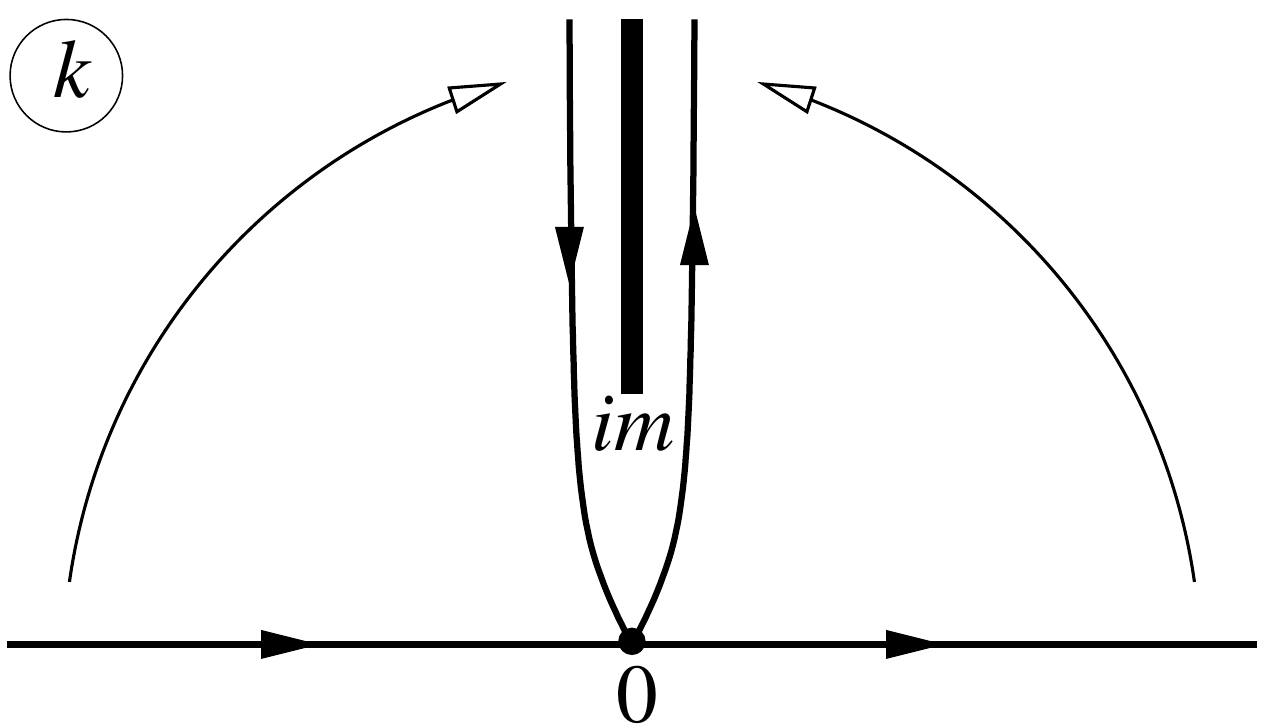}}
\caption{Contours in the $k$-plane used in the calculation of the Boulware
  Green's function at positive $x$, $x'$.}
\label{fig:GBright}
\end{figure}

The terms in the second line of Eq.~(\ref{GBright1}), which we denote by
$\GG_B^{(2)}\big|_{\rm right}$, arise due to the
nontrivial BH potential. To compute them, we make the following
transformations. First, we use the expressions (\ref{ab_BH}) for the
reflection and transmission amplitudes. Second, we notice that the
integrals in $\GG_B^{(2)}\big|_{\rm right}$ are saturated by $\o\sim m$,
implying that we can use $\o|t|\ll 1$ to drop the time dependence in
the integrands. Third, we change the integration variable from $\o$ to
$k$ in the first integral and to $\vk$ in the second one. Finally, we
flip the sign of the variable $k$ in the part of the integral
involving $\e^{-ik(x+x')}$. This yields
\be
\label{GBright2}
\GG_B^{(2)}\big|_{\rm right}^{\rm close}
=-\!\int_{-\infty}^0 \frac{dk\,(\o+k)}{4\pi\o(\o-k)}
\e^{ik(x+x')}
\!-\!\int^{\infty}_0 \!\frac{dk\,(\o-k)}{4\pi\o(\o+k)}\e^{ik(x+x')}
\!+\!\int_0^m \frac{d\vk\,\vk}{\pi m^2} \e^{-\vk(x+x')},
\ee
where now $\o=\sqrt{m^2+k^2}$. Next, using that $(x+x')>0$, we rotate
the integration contours in the first two integrals as shown in
Fig.~\ref{fig:GBright}. The parts of the integrals from $0$ to $im$
cancel with the third term, whereas the parts from $im$ to $+i\infty$
combine into
\be
\label{GBright3}
\GG_B^{(2)}\big|_{\rm right}^{\rm close}
=-\int_{m}^\infty \frac{d\vk}{2\pi\sqrt{\vk^2-m^2}}\cdot
\frac{\sqrt{\vk^2-m^2}-\vk}{\sqrt{\vk^2-m^2}+\vk}\cdot\e^{-\vk(x+x')}.
\ee
This contribution exponentially decays as $\e^{-m(x+x')}$
at large $x+x'$. On the other hand, when $m(x+x')\ll 1$, it
evaluates to $(4\pi)^{-1}$. Combining it with Eq.~(\ref{G_F_close}), we
arrive at
\be
\label{GBcloseright}
\GG_B\big|_{\rm right}^{\rm close}=
-\frac{1}{4\pi}\ln\big[m^2(x-x'-t)(x-x'+t)+i\epsilon\big]
+\begin{cases}
\frac{\ln2-\g_E}{2\pi}+\frac{1}{4\pi},& x+x'\ll m^{-1}\\
\frac{\ln2-\g_E}{2\pi}, & x+x'\gg m^{-1}
\end{cases}
\ee

\paragraph{Left region $x,x'<0$.} Here we
have
\be
\label{GBleft1}
\GG_B\big|_{\rm left}=
\int_0^\infty \frac{d\o}{4\pi\o} \big(\e^{i\o(x-x')}+\e^{-i\o(x-x')}
+\b_\o^* \e^{i\o(x+x')}+\b_\o \e^{-i\o(x+x')}\big) \e^{-i\o|t|},
\ee 
where we have again used the asymptotics (\ref{fRoas}), (\ref{fLoas})
and the relations
(\ref{modecoeff}). Notice that the presence of reflected waves
(contributions proportional to $\b_\o$ and $\b_\o^*$) ensures
convergence of the integral at $\o=0$. We split the integral (\ref{GBleft1})
into two parts by introducing an
arbitrary separation point $\o_1$, such that 
\be
\label{o1ineq}
m\ll\o_1\ll |t|^{-1},~|x-x'|^{-1}\;.
\ee  
In the integral over $\o>\o_1$ the reflections
amplitudes can be neglected. Then it reads,
\begin{align}
\label{GBleft2}
\GG_B^{(1)}\big|_{\rm left}^{\rm close}&=
\int_{\o_1}^\infty\frac{d\o}{4\pi\o}
\big(\e^{i\o(x-x')}+\e^{-i\o(x-x')}\big)\e^{-i\o|t|}\notag\\
&=-\frac{1}{4\pi}\big\{\ln\big[\o_1^2(x-x'-t)(x-x'+t)+i\epsilon\big]+2\g_E\big\}.
\end{align}
We recognize here the characteristic logarithmic singularity at
coincident points. For the remaining integral we write
\be
\label{GBleft3}
\begin{split}
\GG_B^{(2)}\big|_{\rm left}^{\rm close}=&
\int_{0}^m\frac{d\o}{4\pi\o}\bigg(2+\frac{\o+i\vk}{\o-i\vk}
\e^{i\o(x+x')}+\frac{\o-i\vk}{\o+i\vk}\e^{-i\o(x+x')}\bigg)\\
&+\int_{m}^{\o_1}\frac{d\o}{2\pi\o}
\bigg(1+\frac{\o-k}{\o+k}\cos{\o(x+x')}\bigg)\;.
\end{split}
\ee
It can be easily evaluated in two limits. If $|x+x'|\ll m^{-1}$, we can
set the exponents and the cosine to 1 and get elementary integrals
that evaluate to 
\be 
\label{GBleft4}
\GG_B^{(2)}\big|_{\rm left}^{\rm
  close}=\frac{1}{2\pi}\ln\frac{\o_1}{m} +\frac{\ln
  2}{2\pi}+\frac{1}{4\pi}~,
\qquad\qquad |x+x'|\ll m^{-1}.
\ee
On the other hand, if $|x+x'|\gg m^{-1}$, the exponents and the cosine
quickly oscillate and their contribution vanishes outside a small
vicinity of $\o=0$. Thus, we have 
\begin{align}
\GG_B^{(2)}\big|_{\rm left}^{\rm close}
&=\int_0^m\frac{d\o}{2\pi\o}\big(1-\cos\o(x+x')\big)
+\int_m^{\o_1}\frac{d\o}{2\pi\o} \notag\\
&=\frac{1}{2\pi} \big[\ln(\o_1|x+x'|)+\gamma_E\big]~,\qquad\qquad |x+x'|\gg m^{-1}.
\label{GBleft5}
\end{align}
Combining Eqs.~(\ref{GBleft2}), (\ref{GBleft4}), (\ref{GBleft5}), we
arrive at the final result,
\be
\label{G_B_near}
\GG_B\big|_{\rm left}^{\rm close}=\begin{cases}
-\frac{1}{4\pi} \ln\big[m^2(x-x'-t)(x-x'+t)+i\epsilon\big]
+\frac{\ln 2-\gamma_E}{2\pi}+\frac{1}{4\pi},& |x+x'|\ll m^{-1}\\
-\frac{1}{4\pi}
\ln\bigg[\frac{(x-x'-t)(x-x'+t)}{(x+x')^2}+i\epsilon\bigg]~,
&|x+x'|\gg m^{-1}
\end{cases}
\ee
Notice that the upper expression here coincides exactly with the
expansion of $\GG_B$ on the right not-so-far from BH, see the upper
expression in Eq.~(\ref{GBcloseright}). In other words, the Boulware
Green's function appears to be continuous through the region where the
BH potential rapidly changes. We will see that this property is
shared by other Green's functions. 

\subsubsection{Hartle--Hawking Green's function}

\paragraph{Right region $x,x'>0$.} Using the mode asymptotics, we find 
\be
\label{GHHright1}
\begin{split}
&\GG_{HH}\big|_{\rm right}=\int_m^\infty \frac{d\o}{2\pi
  k}\cos[k(x-x')]\,Q(\o)\\
&-\int_m^\infty \frac{d\o}{4\pi
  k}\bigg[\frac{\g_\o\b^*_\o}{\g_\o^*}\e^{ik(x+x')}
+\frac{\g_\o^*\b_\o}{\g_\o}\e^{-ik(x+x')}\bigg]\,Q(\o)
+\int_0^m\frac{d\o}{4\pi\o} |\gamma_\o|^2 \e^{-\vk (x+x')}\,Q(\o)\;,
\end{split}
\ee
where 
\be
\label{Qfactor}
Q(\o)=\frac{\e^{-i\o|t|}}{1-\e^{-2\pi\o/\l}}
+\frac{\e^{i\o|t|}}{\e^{2\pi\o/\l}-1}\;.
\ee
The term in the first line is just the thermal Green's function with
temperature $\l/(2\pi)$ in flat spacetime. Let us focus on it first.

\begin{figure}[t]
\centering{
\includegraphics[width=0.4\textwidth]{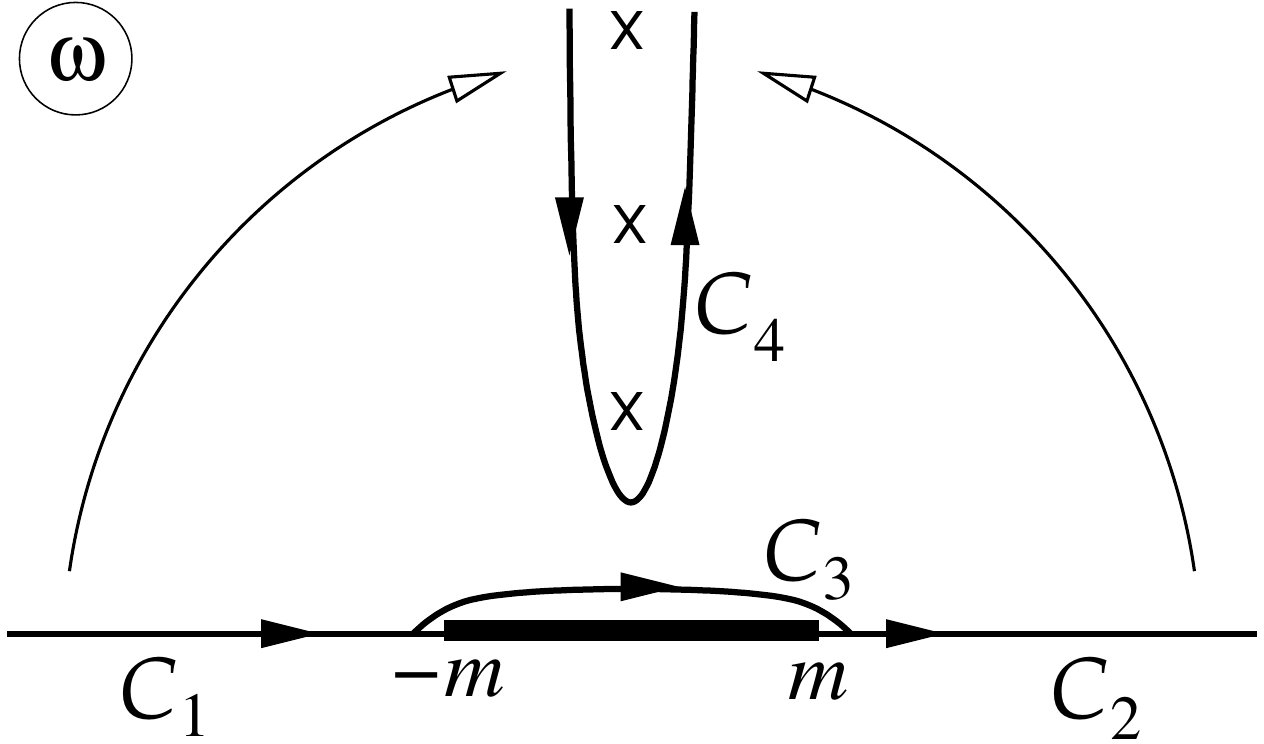}}
\caption{Contours in the $\o$-plane used in the calculation of the
  thermal, Hartle--Hawking and Unruh Green's functions.}
\label{fig:GHHright}
\end{figure}

We will perform the computation assuming $|x-x'|>|t|$ and then
analytically continue to the remaining portion of spacetime. Writing
cosine as the sum of exponents $\e^{\pm ik|x-x'|}$ and changing the
sign of $\o$ in the part containing the second exponent, we obtain
\be
\label{GGth1}
\GG_{th}=-\!\int_{-\infty}^{-m} \!\!\frac{d\o}{4\pi\sqrt{\o^2\!-\!m^2}}
\e^{-i\sqrt{\o^2-m^2}|x-x'|}\,Q(\o)
+\!\int_m^{\infty}\!\! \frac{d\o}{4\pi\sqrt{\o^2\!-\!m^2}}
\e^{i\sqrt{\o^2-m^2}|x-x'|}\,Q(\o)\;,
\ee
where we have used that $Q(-\o)=-Q(\o)$. We now notice that the
analytic continuation of $k$ from 
$\o>m$ to $\o<-m$ results in a minus sign,\footnote{We use the
  convention that the square root of a positive real number is positive.} 
$k=-\sqrt{\o^2-m^2}$ at $\o<-m$.
Therefore, the
expression (\ref{GGth1}) can be written as the integral of a single
analytic function with a branch cut from $-m$ to $m$ along the sum of
contours ${\cal C}_1$ and ${\cal C}_2$ shown in
Fig.~\ref{fig:GHHright}. We complete this contour by adding and
subtracting the integral along the path ${\cal C}_3$,
\be
\label{GGth2}
\GG_{th}=\int\limits_{{\cal C}_1+{\cal C}_3+{\cal C}_2}
\frac{d\o}{4\pi k}\e^{ik|x-x'|}\,Q(\o)
-\int\limits_{{\cal C}_3}\frac{d\o}{4\pi i\vk}
\e^{-\vk|x-x'|}\,Q(\o)\;.
\ee
The union of ${\cal C}_1+{\cal C}_3+{\cal C}_2$ can be now deformed 
into the contour ${\cal C}_4$ encircling the
poles of the function $Q(\o)$ at $\o_n=i n\l$, $n>0$. 
Summing the residues at
the poles, we obtain for the first term in Eq.~(\ref{GGth2}),
\begin{align}
\GG_{th}^{(1)}\big|^{\rm close}&=\sum_{n=1}^\infty\frac{1}{4\pi n}
\big(\e^{-n\l(|x-x'|+|t|)}+\e^{-n\l(|x-x'|-|t|)}\big)\notag\\
&=-\frac{1}{4\pi}\ln\Big[1-2\e^{-\l|x-x'|}\ch{\l t}
+\e^{-2\l|x-x'|}\Big]\;.
\label{GGth3}
\end{align}
In deriving this formula, we have used the approximation $\l\gg m$ to
write $k(\o_n)\approx \o_n$.

In the second term in Eq.~(\ref{GGth2}), due to the antisymmetry of
$Q(\o)$, the only contribution comes from the half-residue at $\o=0$
and is equal to
\be
\GG_{th}^{(2)}=\frac{\l}{4\pi m}\e^{-m|x-x'|}\approx
\frac{\l}{4\pi}
\bigg(\frac{1}{m}-|x-x'|\bigg)\;. 
\ee
Combining with Eq.~(\ref{GGth3}), we obtain
\be
\GG_{th}\big|^{\rm close}
=-\frac{1}{4\pi}\ln\big[2\ch{\l(x-x')}-2\ch{\l t}\big]
+\frac{\l}{4\pi m}\;.
\ee
Recall that this result has been derived under the assumption of
spacelike separation, ${|x-x'|>|t|}$. The continuation inside the future
and past light-cones is straightforward and is implemented by adding
$+i\epsilon$ to the argument of the logarithm. Finally, representing
the difference of the hyperbolic cosines as the product of sines, 
we arrive at
our final result for the thermal Green's function,
\begin{equation}\label{G_th_as}
\GG_{th}\big|^{\rm close}=-\frac{1}{4\pi}\ln\left[ 
4\sh\bigg(\frac{\l}{2}(x-x'-t)\bigg)\sh\bigg(\frac{\l}{2}(x-x'+t)\bigg)+i\epsilon\right] 
+\frac{\l}{4\pi m} \;.
\end{equation}
Notice the large constant piece inversely proportional to the mass in
this expression. Its appearance is a peculiarity of two dimensions
where the integrals for the Green's function are infrared divergent in
the massless limit. Finite mass regulates this divergence.

We return to the remaining contributions in the Hartle--Hawking
Green's function (\ref{GHHright1}). Using the expressions
(\ref{ab_BH}) for $\beta_\o$ and $\g_\o$, we notice that the integrals
are saturated at $\o\sim m\ll \l$. Expanding $Q(\o)$ in this limit, we
obtain
\[
\GG_{HH}^{(2)}\big|^{\rm close}_{\rm right}=
-\l\int_m^\infty \frac{d\o}{4\pi^2 k\o}\cdot\frac{\o-k}{\o+k}
\big(\e^{ik(x+x')}+\e^{-ik(x+x')}\big)
+\l \int_0^m \frac{d\o}{\pi^2 m^2} \e^{-\vk(x+x')}\;.
\]
We proceed similarly to the case of the thermal Green's function
above. Namely, we flip the sign of $\o$ in the integral containing 
$\e^{-ik(x+x')}$; notice that we obtain the integrals along the
contours ${\cal C}_1$ and ${\cal C}_2$ of the same analytic function,
with $k(\o)$ analytically continued through the upper half-plane; add
and subtract the integral along ${\cal C}_3$; deform the union of the
contours 
${\cal C}_1+{\cal C}_3+{\cal C}_2$ into the upper half-plane (this is
allowed because $x+x'$ is positive). The latter contour now does not
encounter any singularities, so the integral along it vanishes. We are
left with 
\begin{align}
\GG_{HH}^{(2)}\big|^{\rm close}_{\rm right}&=
\l\int_{{\cal C}_3} \frac{d\o}{4\pi^2 i \o\vk}\cdot\frac{\o-i\vk }{\o+i\vk}
\e^{-\vk(x+x')}
+\l \int_0^m \frac{d\o}{\pi^2 m^2} \e^{-\vk(x+x')}\notag\\
&=\frac{\l}{4\pi m}\e^{-m(x+x')}\;,
\label{GHHright3}
\end{align}
where again only the half-residue at $\o=0$ contributes.

Combining Eqs.~(\ref{G_th_as}) and (\ref{GHHright3}), we obtain the
Hartle--Hawking Green's function in the right region. It is convenient
to write it in two limits,
\be
\label{GHHcloseright}
\begin{split}
\GG_{HH}\big|^{\rm close}_{\rm right}=
&-\frac{1}{4\pi}\ln\left[ 
4\sh\bigg(\frac{\l}{2}(x-x'-t)\bigg)
\sh\bigg(\frac{\l}{2}(x-x'+t)\bigg)+i\epsilon\right] \\
&+\begin{cases}
-\frac{\l}{4\pi}(x+x')+\frac{\l}{2\pi m}~,&x+x'\ll m^{-1}\\
\frac{\l}{4\pi m}~,& x+x'\gg m^{-1}
\end{cases} 
\end{split}
\ee

\paragraph{Left region $x,x'<0$.} 
Using the asymptotics of the modes, we obtain
\be
\label{GHHleft1}
\GG_{HH}\big|_{\rm left}=\int_0^\infty \frac{d\o}{4\pi\o}
\big[2\cos\o(x-x')+\beta_\o\e^{-i\o(x+x')}+\beta^*_\o\e^{i\o(x+x')}\big]
Q(\o)\;.
\ee
We again assume $|x-x'|>|t|$, keeping in mind that we can always
analytically continue to $|t|<|x-x'|$ by the $i\epsilon$-prescription. The calculation is very similar to the one described in
the previous paragraph, so we only briefly outline it here, without
going into details. One uses the expression (\ref{ab_BH}) for
$\beta_\o$ and separates 
Eq.~(\ref{GHHleft1}) into the integrals over $0<\o<m$ and $m<\o$. One
further rewrites the second integral as the sum of integrals over the
contours ${\cal C}_1$ and ${\cal C}_2$ (see Fig.~\ref{fig:GHHright})
of a single analytic function decreasing into the upper
half-plane. Upon adding and subtracting the integral over ${\cal C}_3$,
one deforms the contour into ${\cal C}_4$ to pick up the residues at
the poles of the thermal factor $Q(\o)$. In this way one arrives at
\[
\begin{split}
\GG_{HH}\big|_{\rm left}\!=&-\frac{1}{4\pi}\ln\Big[1\!-\!2\e^{-\l|x-x'|}\ch{\l t}
\!+\!\e^{-2\l|x-x'|}\Big]\!-\!\!\int_{{\cal C}_3}\!\frac{d\o}{4\pi\o}
\bigg[\e^{i\o|x-x'|}+\frac{\o\!-\!i\vk}{\o\!+\!i\vk}\e^{-i\o(x+x')}\bigg]
Q(\o)\\
&+\int_0^m \frac{d\o}{4\pi\o} 
\bigg[2\cos{\o|x-x'|}+\frac{\o-i\vk}{\o+i\vk}\e^{-i\o(x+x')}
+\frac{\o+i\vk}{\o-i\vk}\e^{i\o(x+x')}
\bigg]Q(\o)\;.
\end{split}
\]
The two integrals in the last
formula almost
cancel each other, up to a half-residue at $\o=0$. Evaluating this
residue, one obtains the
final result
\be
\label{G_HH_near}
\GG_{HH}\big|_{\rm left}= 
-\frac{1}{4\pi}\ln\left[4\sh\!\bigg(\frac{\l}{2}(x-x'-t)\!\bigg)
\sh\!\bigg(\frac{\l}{2}
(x-x'+t)\!\bigg)\!+\!i\epsilon\right]-\frac{\l}{4\pi}(x+x')+\frac{\l}{2\pi m}\,, 
\ee
where $i\epsilon$ has been inserted to ensure the analytic
continuation. Note that this expression is the same as the upper
case in Eq.~(\ref{GHHcloseright}). Note also that it is valid without
any restrictions on $|x+x'|$, $|x-x'|$, $|t|$ provided $x$ and $x'$
are in the near-horizon region (i.e., they satisfy the condition
(\ref{leftlim})). 
If
we keep $x'$ fixed and send $x$ to $-\infty$, the Green's function
goes to a constant,
\be
\GG_{HH}\big|_{\rm left}\to \frac{\l}{2\pi}\big(m^{-1}-x'\big)~,\qquad\qquad
x\to-\infty. 
\ee
This reflects the regularity of the Hartle--Hawking state at the BH horizon.

\subsubsection{Unruh Green's function}

\paragraph{Right region $x,x'>0$.} A convenient way to compute
the Unruh Green's function on the right is to relate it to the
Boulware Green's function. Using Eq.~(\ref{G_U}) and the modes'
asymptotics (\ref{fRoas}), (\ref{fLoas}), we obtain
\begin{equation}\label{G_U_far2}
\begin{split}
\GG_{U}\big|_{\rm right}\!=\GG_{B}\big|_{\rm right} \!+ \!
\int_m^\infty\!\!\dfrac{d\o}{2\pi\o}|\gamma_\o|^2
\dfrac{\cos[k(x\!-\!x')\!-\!\o t]}{\e^{2\pi\o/\l}-1} 
+ \!\int_0^m\!\!\dfrac{d\o}{2\pi\o}|\gamma_\o|^2\dfrac{\cos{\o
    t}}{\e^{2\pi\o/\l}\!-\!1} 
\e^{-\vk(x+x')}.
\end{split}
\end{equation} 
Let us assume $x-x'-t>0$ and compute the second term. Using the
expression (\ref{ab_BH}) for $\gamma_\o$ and performing analytic
continuation in $\o$, we can write it in the form
\be
\GG_{U}^{(2)}\big|_{\rm right}=\int\limits_{{\cal C}_1+{\cal C}_2}
\frac{d\o\,\o}{\pi(\o+k)^2}\cdot \frac{\e^{ik(x-x')-i\o t}}{1-\e^{-2\pi\o/\l}}
-\int_m^\infty \frac{d\o\,\o}{\pi(\o+k)^2} \e^{ik(x-x')-i\o t},
\ee
where the contours ${\cal C}_1$ and ${\cal C}_2$ are shown in
Fig.~\ref{fig:GHHright}. Next, we complete the first integral with the
contour ${\cal C}_3$ and deform it into the upper half-plane, picking
up the poles at $\o=i\l n$. In the second integral we split the
integration domain by introducing a separation scale 
$m\ll \o_1\ll |x-x'|^{-1},|t|^{-1}$. This yields 
\be
\label{GUright1}
\begin{split}
\GG_{U}^{(2)}\big|_{\rm right}^{\rm close}=
&-\frac{1}{4\pi}\ln\big[1-\e^{-\l(x-x'-t)}\big]
-\int_{-m}^{m} \frac{d\o\,\o}{\pi(\o+i\vk)^2}
\cdot \frac{\e^{-\vk(x-x')-i\o t}}{1-\e^{-2\pi\o/\l}}\\
&-\int_m^{\o_1}\frac{d\o\,\o}{\pi(\o+k)^2}\, \e^{ik(x-x')-i\o t}
-\int_{\o_1}^\infty\frac{d\o\,\o}{\pi(\o+k)^2}\, \e^{ik(x-x')-i\o t}\;.
\end{split}
\ee
Now we can simplify the integrands.
In the integral in the first line we expand the exponents at small
$\o$. As the leading term is enhanced by the large ratio $\l/\o$, 
we keep subleading terms to retain $\mathcal{O}(1)$-contributions. In the first
integral in the second line we set the exponent to $1$. Whereas in the
last integral we use the approximation $k\approx \o$. After these
simplifications the evaluation of the integrals is straightforward
and we arrive at
\be
\label{GUright2}
\GG_{U}^{(2)}\big|_{\rm right}^{\rm close}=
-\frac{1}{4\pi}
\ln\bigg[\frac{2\sh\frac{\l}{2}(x-x'-t)}{m(x-x'-t)}\bigg]
+\frac{\l}{3\pi^2 m}-\frac{\ln 2-\gamma_E}{4\pi}+\frac{1}{16\pi}\;.
\ee 
The calculation at $x-x'-t<0$ proceeds in a similar way and gives the
same result.

We still have to evaluate the last term in Eq.~(\ref{G_U_far2}). Clearly,
this term vanishes if $x+x'\gg m^{-1}$. In the opposite limit,
$x+x'\ll m^{-1}$, we expand the integrand to subleading order and
after an elementary integration obtain
\be
\label{GUright3}
\GG_U^{(3)}\big|_{\rm right}=\begin{cases}
\frac{\l}{\pi^2 m}-\frac{\l(x+x')}{4\pi}-\frac{1}{2\pi}~,& x+x'\ll
m^{-1}\\
0~,& x+x'\gg m^{-1}
\end{cases}
\ee 
Combining together the expression (\ref{GBcloseright}) for the
Boulware Green's function and Eqs.~(\ref{GUright2}), (\ref{GUright3}),
we end up with
\be
\label{G_U_far}
\begin{split}
\GG_{U}\big|_{\rm right}^{\rm close}= &-\frac{1}{4\pi}
\ln \left[ 2\sh\left(\frac{\l}{2} (x-x'-t)\right) m (x-x'+t)+i\epsilon\right]\\
&+\begin{cases}
-\frac{\l}{4\pi}(x+x')+\frac{4\l}{3\pi^2 m}+\frac{\ln 2-\g_E}{4\pi}
-\frac{3}{16\pi}~,&x+x'\ll m^{-1}\\
\frac{\l}{3\pi^2 m}+\frac{\ln 2-\g_E}{4\pi}
+\frac{1}{16\pi}~,&x+x'\gg m^{-1}
\end{cases}
\end{split}
\ee
We observe that the Unruh Green's function is a mixture of the thermal
right-moving and vacuum left-moving contributions. It is important to
note, however, that this mixture is not a simple linear superposition:
the nonsingular part in $\GG_U$ is not an arithmetic mean of the
nonsingular parts of $\GG_{HH}$ and $\GG_B$. In particular, the large
terms $\propto \l/m$ produced by soft modes with $\o\sim m$ have
different coefficients in Eqs.~(\ref{G_U_far}) and
(\ref{GHHcloseright}). These terms play
the key role in determining the tunneling solution and the vacuum
decay probability in the model studied in the main text.

Note that an enhanced $\mathcal{O}(\l/m)$-term is present in $\GG_U$ even far
away from the BH implying that the soft modes are still highly populated
in the asymptotic region. This is a feature of two dimensions: the
radiation emitted by the BH does not spread out and persists as a constant
flux of particles at arbitrarily large  distance from the horizon.

Finally, the leading corrections to the expression (\ref{G_U_far})
come from the last term in (\ref{G_U_far2}) and the second term in
(\ref{GUright1}) and are of order 
${{\cal O}\big(\l m(x\!-\!x')^2,\l mt^2,\l m(x\!-\!x')t\big)}$.
Requiring that these corrections are smaller than the ${\cal O}(1)$
terms kept in Eq.~(\ref{G_U_far}) restricts the domain of validity
of this formula to 
\be
|x-x'|,~|t|\ll 1/\sqrt{\l m}\;.
\ee

\paragraph{Left region $x,x'<0$.} Here it is convenient to relate the
Unruh Green's function to the Hartle--Hawking Green's function. Upon
using the mode asymptotics, we get
\begin{equation}\label{G_U_near2}
\GG_{U}\big|_{\rm left}=\GG_{HH}\big|_{\rm left}
-\int_m^\infty\dfrac{d\o\,k}{2\pi\o^2}\,|\g_\o|^2\,
\dfrac{\cos[\o(x-x'+t)]}{\e^{2\pi\o/\l}-1} \; .
\end{equation}
Evaluation of the additional integral proceeds in complete analogy
with the calculation of $\GG_U^{(2)}\big|_{\rm right}$ above. We do
not repeat this calculation, and just give the result,
\be
\GG_{U}^{(2)}\big|_{\rm left}^{\rm close}=
\frac{1}{4\pi}\ln\bigg[\frac{2\sh\frac{\l}{2}(x-x'+t)}{m(x-x'+t)}\bigg]
+\frac{\l}{m}\bigg(-\frac{1}{2\pi}+\frac{4}{3\pi^2}\bigg)
+\frac{\ln 2-\g_E}{4\pi}-\frac{3}{16\pi}\;. 
\ee 
Combining with Eq.~(\ref{G_HH_near}) we obtain the final answer,
\be
\label{G_U_near}
\begin{split}
\GG_{U}\big|_{\rm left}^{\rm close}= 
-\frac{1}{4\pi}\ln \left[ 2\sh\bigg(\!\frac{\l}{2}(x\!-\!x'\!-\!t)\!\bigg) m
  (x\!-\!x'\!+\!t)\!+\!i\epsilon\right]  
\!-\!\frac{\l(x\!+\!x')}{4\pi}\!+\!\frac{4\l}{3\pi^2 m}\!
+\!\frac{\ln 2\!-\!\g_E}{4\pi} 
\!-\!\frac{3}{16\pi} \;.
\end{split}
\ee
The leading corrections to this formula are of order ${\cal O}\big(\l m
(x-x'+t)^2\big)$. On the other hand, there are no corrections in
$(x-x'-t)$.  
This is clear
from the representation (\ref{G_U_near2}) and the fact that the form
(\ref{G_HH_near}) of the
Hartle--Hawking Green's function is valid in the entire near-horizon
region. Thus, the expressions (\ref{G_U_near}) can be used as long as  
\[
|x-x'+t|\ll 1/\sqrt{\l m}\;.
\]
As expected, Eq.~(\ref{G_U_near}) coincides with the expression for
$\GG_{U}\big|_{\rm right}^{\rm close}$ at $x+x'\ll m^{-1}$ (upper case in
Eq.~(\ref{G_U_far})).

\section{Boundary conditions for the bounce}
\label{app:Vac}

Here we derive the boundary conditions for the bounce solution
formulated in Sec.~\ref{Ssec:gen_amplitude}. The path integral for
the false vacuum decay probability (\ref{Pdecay}) contains the
elements of the density matrix in the configuration-space basis,
$\bra{\vf_i,t_i^{\rm up}}\varrho\ket{\vf_i',t_i^{\rm low}}$. 
It is convenient to switch to the (over-complete) 
basis of coherent states
\be
\label{adef}
\ket{a}=\exp\Big\{\int_0^\infty d\o\sum_{I=R,L} 
\Big[-\tfrac{1}{2}|a_{I,\o}|^2+a_{I,\o}\hat a_{I,\o}^\dagger\Big]\Big\}
\ket{0}_B\;.
\ee
They are eigenstates of the annihilation operator,
\be
\hat a_{I,\o}\ket{a}= a_{I,\o}\ket{a}~,~~~~~~I=R,L\;,
\ee
and provide a decomposition of unity,
\be
\int D[a]D[a^*]\;\ket{a}\bra{a}=\mathbbm{1}\;.
\ee
Therefore, the matrix element of interest takes the form
\be
\bra{\vf_i,t_i}\varrho\ket{\vf_i',t_i'}=\int D[a]D[a^*]D[a']D[a'^*]
\braket{\vf_i,t_i}{a}\bra{a}\varrho\ket{a'}\braket{a'}{\vf_i',t_i'}\;.
\ee
We now compute the elements entering into this formula.

In Boulware, Hartle--Hawking and Unruh vacua different modes are
populated incoherently, according to the thermal distribution. Their
temperature $T_I$ is either equal to the BH temperature
$\l/(2\pi)$ or is zero (the mode is in vacuum). The single-mode
thermal density matrix is
\be
\varrho_{I,\o}=\sum_{n} \frac{(\hat a^\dagger_{I,\o})^n}{\sqrt{n!}}
\ket{0}\bra{0}\frac{(\hat a_{I,\o})^n}{\sqrt{n!}} \e^{-n \o/T_I}
\ee
and its elements in the coherent-state representation are easily
calculated, 
\be
\label{rhosingle}
\bra{a}\varrho_{I,\o}\ket{a'}=\exp\Big\{-\tfrac{1}{2}|a_{I,\o}|^2
-\tfrac{1}{2}|a'_{I,\o}|^2+\e^{-\o/T_I}a^*_{I,\o}a'_{I,\o}\Big\}.
\ee
The total density matrix is obtained as the product of single-mode
density matrices over all modes. For all states of interest the result
has the general form
\be
\label{rhotot}
\bra{a}\varrho\ket{a'}=\exp\Big\{\int_0^\infty d\o\sum_{I=R,L}
\Big[-\tfrac{1}{2}|a_{I,\o}|^2
-\tfrac{1}{2}|a'_{I,\o}|^2+r_I(\o)a^*_{I,\o}a'_{I,\o}\Big]\Big\}\;,
\ee
and the difference between the Boulware, Hartle--Hawking and Unruh
states is encapsulated by the coefficients $r_I(\o)$, whose values are
given in Eqs.~(\ref{rBHHU}).

Next, we need the wavefunction of the coherent state,
$\braket{\vf_i,t_i}{a}$. To this aim, we derive a set of differential
equations that this wavefunction obeys. 
We notice that the annihilation and creation
operators can be represented as
\bseq
\label{aavfpi}
\begin{align}
&\hat a_{I,\o}=\e^{i\o t_i}\int\frac{dx}{\sqrt{4\pi}}f^*_{I,\o}(x)
\bigg(\frac{\sqrt{\o}}{{\rm g}}\hat\vf(t_i,x)
+\frac{i{\rm g}}{\sqrt{\o}}\hat\pi(t_i,x)\bigg),\\
&\hat a_{I,\o}^\dagger=\e^{-i\o t_i}\int\frac{dx}{\sqrt{4\pi}}f_{I,\o}(x)
\bigg(\frac{\sqrt{\o}}{{\rm g}}\hat\vf(t_i,x)
-\frac{i{\rm g}}{\sqrt{\o}}\hat\pi(t_i,x)\bigg),
\end{align}
\eseq
where 
\be
\label{pihat}
\hat\pi(t,x)={\rm g}^{-2}\d_t\vf(t,x)
\ee
is the canonical momentum operator. It acts by derivative on the
configuration-space wavefunctions,
\be
\label{pihatact}
\bra{\vf_i,t_i}\hat\pi(t_i,x)\ket{\psi}
=-i\frac{\delta}{\delta\vf_i(x)} \braket{\vf_i,t_i}{\psi}
\ee
for any quantum state $\ket{\psi}$. Hence, we can write
\bseq
\label{eqsfora}
\begin{align}
&a_{I,\o}\braket{\vf_i,t_i}{a}=\bra{\vf_i,t_i}\hat a_{I,\o}\ket{a}
=\e^{i\o t_i}\int\frac{dx}{\sqrt{4\pi}} f_{I,\o}^*(x)
\bigg[\frac{\sqrt{\o}}{{\rm g}} \vf_i(x)+
\frac{{\rm g}}{\sqrt{\o}}\frac{\delta}{\delta\vf_i(x)}\bigg]
\braket{\vf_i,t_i}{a}\;,\\
&\bigg[\frac{\delta}{\delta a_{I,\o}}\!+\!\frac{1}{2}a^*_{I,\o}\bigg]
\!\braket{\vf_i,t_i}{a}
\!=\!\bra{\vf_i,t_i}\hat a_{I,\o}^\dagger\ket{a}
\!=\!\e^{-i\o t_i}\!\!\int\!\!\frac{dx}{\sqrt{4\pi}} f_{I,\o}(x)
\!\bigg[\frac{\sqrt{\o}}{{\rm g}} \vf_i(x)\!-\!
\frac{{\rm g}}{\sqrt{\o}}\frac{\delta}{\delta\vf_i(x)}\bigg]
\!\braket{\vf_i,t_i}{a},\\
&\bigg[\frac{\delta}{\delta a^*_{I,\o}}+\frac{1}{2}a_{I,\o}\bigg]
\braket{\vf_i,t_i}{a}=0\;.
\end{align}
\eseq
The solution of this system is straightforward and yields the following
result: 
\be
\label{wavea}
\begin{split}
\braket{\vf_i,t_i}{a}\propto\exp\Big\{
&-\tfrac{1}{2{\rm g}^2}\int dxdx'\,A(x,x')\vf_i(x)\vf_i(x')\\
&+\tfrac{1}{{\rm g}} \int d\o dx\,\sqrt{\o/\pi}\sum_I f_{I,\o}(x)\vf_i(x)
a_{I,\o}\e^{-i\o t_i}\\
&-\tfrac{1}{2}\int d\o\Big[\sum_I|a_{I,\o}|^2
+\sum_{I,J} B_{IJ} a_{I,\o}a_{J,\o} \e^{-2i\o t_i}\Big]\Big\},
\end{split}
\ee
where 
\bseq
\begin{align}
&A(x,x')=\int\frac{d\o}{2\pi}\, \o\sum_I f_{I,\o}(x)
f^*_{I,\o}(x')\;,\\
&2\pi B_{IJ}\,\delta(\o-\o')=\int dx\, f_{I,\o}(x) f_{J,\o}(x)\;.
\end{align}
\eseq
Note the absence of complex conjugation in the expression for
$B_{IJ}$. 
Though Eq.~(\ref{wavea}) looks lengthy, its 
structure is simple: it is just an exponent of a quadratic form in the field
$\vf_i$ and the mode amplitudes $a_{I,\o}$.

We are now ready to combine Eqs.~(\ref{rhotot}), (\ref{wavea}) and
substitute them into the path integral for the decay probability. The
integrals over $\vf_i$, $a$, $a^*$ and their primed counterparts are
Gaussian and hence are saturated by the saddle point. The saddle-point
condition obtained by variation with respect to $\vf_i(x)$, $a_{I,\o}$,
$a_{I,\o}^*$ are
\bseq
\label{bbc*}
\begin{align}
\label{bbc1}
&i\frac{\delta S[\vf_{\rm b}]}{\delta\vf_i(x)}
\!-\!\frac{1}{{\rm g}^2}\int dx'\,A(x,x')\vf_i(x')
\!+\!\frac{1}{{\rm g}}\int d\o\sqrt{\frac{\o}{\pi}}\sum_I 
f_{I,\o}(x)a_{I,\o}\e^{-i\o t_i^{\rm up}}=0\;,\\
\label{bbc2}
&\frac{1}{{\rm g}}\sqrt{\frac{\o}{\pi}}\int dx\,
f_{I,\o}(x)\vf_i(x)\e^{-i\o t_i^{\rm up}}-\sum_J
B_{IJ}a_{J,\o}\e^{-2i\o t_i^{\rm up}}
-a^*_{I,\o}=0\;,\\
\label{bbc3}
&-a_{I,\o}+r_I(\o)\,a'_{I,\o}=0\;.
\end{align} 
\eseq
Next, we notice that the variation of the action evaluated on
the bounce solution with respect to the initial value of the field is
the initial momentum, taken with minus sign,
\be
\label{piclas}
\frac{\delta S[\vf_{\rm b}]}{\delta\vf_i(x)}=-\pi(t_i^{\rm up},x)
=-{\rm g}^{-2} \d_t\vf_{\rm b}(t_i^{\rm up},x)\;.
\ee
Substituting this into Eq.~(\ref{bbc1}) and performing a series of
straightforward manipulations, one arrives to a very simple relation
\bseq
\label{bcsimp}
\be
\label{bcsimp1}
c_{I,\o}^{\rm up}=a_{I,\o}\;,
\ee
where $c_{I,\o}^{\rm up}$ are the positive-frequency amplitudes of the bounce
solution on the upper side of the contour ${\cal C}$ (see
Eq.~(\ref{vful}) and Fig.~\ref{fig:Contour}). 
Inserting this result into Eq.~(\ref{bbc2}) leads
once more to a massive simplification and yields\footnote{Note that at
  the saddle point the value of $a^*_{I,\o}$ is {\em not} complex conjugate to
  $a_{I,\o}$.}
\be
\label{bcsimp2}
\bar c_{I,\o}^{\rm up}=a_{I,\o}^*\;,
\ee
\eseq
where $\bar c_{I,\o}^{\rm up}$ are the negative-frequency amplitudes of the bounce
on the upper side of the contour. The third equation (\ref{bbc3}) is
left as it is.

The saddle-point conditions following from variation with respect to
the primed variables are handled in exactly the same manner. They read
\be
\label{bbcprim}
c_{I,\o}^{\rm low}=a'_{I,\o}~,~~~~\bar c_{I,\o}^{\rm low}=a'^*_{I,\o}~,~~~~
r_I(\o)\,a^*_{I,\o}=a'^*_{I,\o}\;,
\ee
where $c_{I,\o}^{\rm low}$, $\bar c_{I,\o}^{\rm low}$ 
are the positive- and negative-frequency amplitudes of the bounce on the lower part of the contour
${\cal C}$. Eliminating $a$, $a^*$, $a'$, $a'^*$ from
Eqs.~(\ref{bbc3}), (\ref{bcsimp}), (\ref{bbcprim}), we arrive at the
boundary conditions (\ref{cpmrel}) from the main text.

Finally, let us work out the initial state contribution into the
tunneling exponent. This is given by the saddle-point value of the
density matrix. As discussed above, both the matrix elements
$\bra{a}\varrho\ket{a'}$ and the wavefunction $\braket{\vf_i,t_i}{a}$ are
exponents of homogeneous quadratic forms in $\vf_i$ and $a$. On the
other hand, the bounce action $S[\vf_{\rm b}]$ is linear in $\vf_i$,
$\vf_i'$. Hence, the evaluation of all Gaussian integrals in the
initial state variables leaves behind a simple expression,
\be
\bra{\vf_i,t_i^{\rm up}}\varrho\ket{\vf_i',t_i^{\rm low}}\big|_{\rm saddle}\propto
\exp\bigg\{\int dx\bigg[-\frac{i}{2}\vf_{i}(x)
\frac{\delta S[\vf_{\rm b}]}{\delta\vf_i(x)}
-\frac{i}{2}\vf'_{i}(x)
\frac{\delta S[\vf_{\rm b}]}{\delta\vf'_i(x)}\bigg]\bigg\}
\ee
This gives Eq.~(\ref{bterms}) upon using the relation (\ref{piclas}).

\section{Calculation of the bounce suppression}
\label{app:Bcalc}

\paragraph{Periodic instanton in flat space.}
The suppression of the periodic instanton in flat space is given by
Eq.~(\ref{B}) where one should substitute the inner core of the
instanton (\ref{PIin2}). 
As discussed in the main text, the integration can
be performed over the contour ${\cal C}'$ consisting of two
semi-infinite rays with 
$-\infty<\Re t<0$, $\Im t=\pm \pi/\l$ and a piece of imaginary time
axis $\Re t=0$, $-\pi/\l<\Im t<\pi/\l$ connecting them, see
Fig.~\ref{fig:bounceth}a. 
The integrals
over the rays cancel each other due to the periodicity of the
solution in imaginary time. Thus, we are left with an integral over
one period of the 
instanton in Euclidean time: 
\be
\label{SPI1}
B_{th}=\frac{1}{{\rm \gc}^2}\!\!\int\limits_{-\pi/\l}^{\pi/\l}\!\!d\tau\!
\int\limits_{-\infty}^\infty\!\! dx \;
\frac{\l^2 b_{th}}{\big(\ch\l x\!-\!\sqrt{1\!-\!b_{th}}\cos\l\tau\big)^2}
\Bigg\{\!\ln \left[\frac{\l^2b_{th}\e^{-2\l x}}
{\kappa\big(\ch\l x\!-\!\sqrt{1\!-\!b_{th}}\cos\l\tau\big)^2}\right]
\!-\!2\!+\!2\l x\Bigg\},
\end{equation}
where we have added and subtracted a linear piece $2\l x$ in the square
brackets. Because of the symmetry of  the first factor under $x\mapsto
-x$, the integral of this linear piece vanishes, so we omit it in what
follows. Next, we make a change of variables 
\begin{equation}\label{toflat}
\TT=\e^{\l x} \sin{\l \tau}~,~~~~X=\e^{\l x}\cos\l\tau\;.
\end{equation}
Note that this is the same transformation that connects the inertial
frame coordinates $(\TT,X)$ with
the frame of a uniformly accelerating observer $(\tau,x)$ in the
Euclidean signature. Of course, there is no physical accelerating
frame in the present calculation and we are using the change of
variables (\ref{toflat}) simply as a mathematical tool.
The domain of integration is mapped  
to $-\infty< \TT, X<\infty$, and the integral becomes
\begin{equation}\label{SPI2}
\begin{split}
B_{th}=\frac{1}{{\rm \gc}^2}\!\!\int\limits_{-\infty}^{\infty}\!d\TT\!\!
\int\limits_{-\infty}^\infty \!dX  \;
\frac{4b_{th}}{\big(\TT^2\!+\!(X\!-\!\sqrt{1\!-\!b_{th}})^2\!+\!b_{th}\big)^2} 
\Bigg\{
\!\ln\left[\frac{4\l^2b_{th}}{\kappa\big(\TT^2\!+\!(X\!-\!\sqrt{1-b_{th}})^2
\!+\!b_{th}\big)^2}\right]\!-\!2\Bigg\}.
\end{split}
\end{equation}
This is easily evaluated in polar coordinates
centered at $\TT=0$, $X=\sqrt{1-b_{th}}$ and yields
\be
\label{SPI3}
B_{th}=\frac{4\pi}{{\rm g}^2}\bigg[\ln\bigg(\frac{4\l^2}{\kappa
  b_{th}}\bigg)-4\bigg]\;. 
\ee
Upon substitution of
$b_{th}$ from Eq.~(\ref{aTh}),
we arrive at Eq.~\eqref{SPI}. 

\paragraph{Hartle--Hawking bounce.}
For the periodic instantons (bounces) in the vicinity of the BH we use
the expression (\ref{HHcore}). Then the bounce action integrated by
parts~is
\begin{equation}
\begin{split}
B_{HH}=&\frac{1}{{\rm g}^2}\int_{-\pi/\l}^{\pi/\l}d\tau
\int_{-\infty}^\infty dx\,\O(x)\kappa(\vf_{\rm b}-2)\e^{\vf_{\rm b}}\\
=&\frac{1}{\gc^2}\int_{-\pi/\l}^{\pi/\l}d\tau \int_{-\infty}^\infty dx\;
\frac{\l^2b_{HH}}{\big(\ch\l(x-x_{HH})-\sqrt{1-b_{HH}}\cos\l\tau\big)^2}\\
&\qquad\times \Bigg\{
\ln\left[\frac{\l^2b_{HH}\e^{-2\l (x-x_{HH})}}{
\kappa \big(\ch\l (x-x_{HH})-\sqrt{1-b_{HH}}\cos\l\tau\big)^2}
\right]-2-2\l x_{HH}\Bigg\}.
\end{split}
\end{equation}
This integral is analogous to Eq.~(\ref{SPI1}) and is computed using the
same change of variables (\ref{toflat}). Upon using the relation
(\ref{HH_a}) between $b_{HH}$ and $x_{HH}$, we obtain the result
(\ref{B_BH_HH}) 
given in the main text.

\begin{figure}[t]
\centering
\begin{picture}(210,80)
\put(0,0){\includegraphics[width=3in]{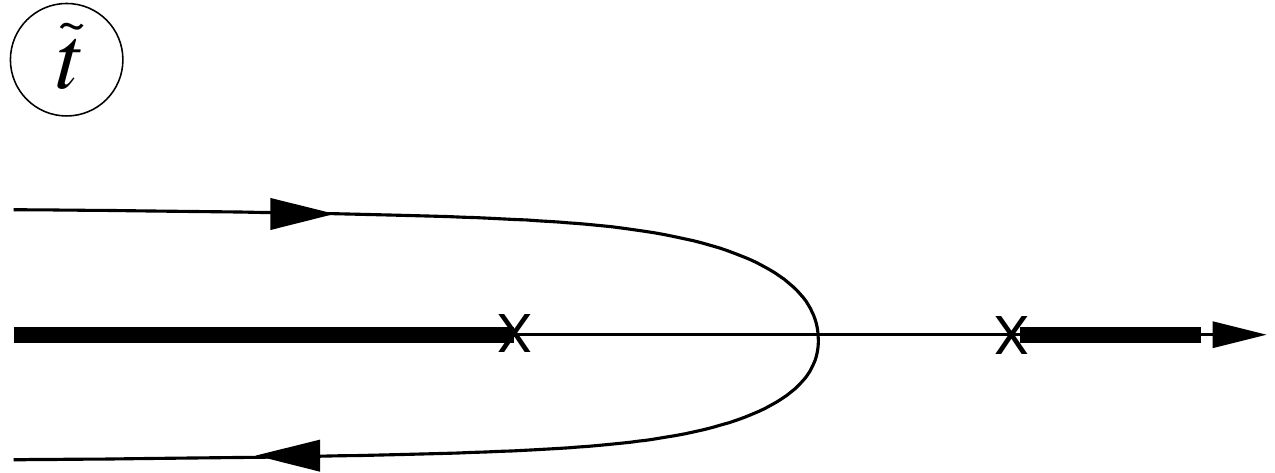}}
\put(50,9){$-\sqrt{\tilde x^2+b_{U1}}$}
\put(160,9){$\tilde x+1$}
\put(110,47){$\tilde{\cal C}_{\tilde x}$}
\end{picture}
\caption{Time integration contour in Eq.~(\ref{BU1_2}).}
\label{fig:tildeC}
\end{figure}
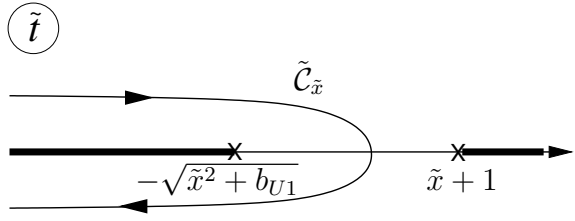

\paragraph{Unruh bounce far from horizon.} 
Substituting the expression (\ref{U_core_far}) in the general formula
(\ref{B}), we get
\be
\label{BU1_1}
\begin{split}
B_{U1}=\frac{i}{{\rm g}^2}
&\int_{-\infty}^{\infty}dx\int_{\cal C} dt
\frac{4\l^2 b_{U1}}{\left(-2\l(v-v_1)\sh\left(\frac{\l}{2}(u-u_1)\right)
+b_{U1}\e^{\frac{\l}{2}(u-u_1)}\right)^2}\\
&\times\left\{\ln\left[
\frac{4\l^2 b_{U1}}{\kappa\left(-2\l(v-v_1)\sh\left(\frac{\l}{2}(u-u_1)\right)
+b_{U1}\e^{\frac{\l}{2}(u-u_1)}\right)^2}\right]-2\right\}.
\end{split}
\ee
Recall that the time integral is performed over a contour ${\cal
  C}$ running along the real axis and encircling the singularity of
the integrand. This singularity is quite complicated: it is a
second-order pole superimposed on a logarithmic branch cut. 
The calculation is greatly simplified by a change of variables. First,
we introduce a new advanced coordinate
\be
\tilde u=\l^{-1}\big(1-\e^{-\l (u-u_1)}\big)\;.
\ee
Notice that when $u$ varies from $-\infty$ to $+\infty$, $\tilde u$
varies from $-\infty$ to $1/\l$. Second, we introduce new time and
space variables
\be
\tilde t=\tfrac{\l}{2}(v-v_1+\tilde u)~,~~~~~
\tilde x=\tfrac{\l}{2}(v-v_1-\tilde u)\;.
\ee
With this replacement the integral becomes
\be
\label{BU1_2}
\begin{split}
B_{U1}=\frac{i}{{\rm g}^2}
\!\int\limits_{-\infty}^{\infty}\!d\tilde x\!\int\limits_{\tilde{\cal C}_{\tilde
    x}}\! d\tilde t\; 
\frac{4b_{U1}}{\left(-\tilde t^2\!+\!\tilde x^2\!+\!b_{U1}\right)^2}
\left\{\ln\left[
\frac{4\l^2 b_{U1}}{\kappa\left(-\tilde t^2\!+\!\tilde
    x^2\!+\!b_{U1}\right)^2}\right]\!-\!2
\!+\!\ln(1\!-\!\tilde t\!+\!\tilde x)\right\}.
\end{split}
\ee
The time contour $\tilde{\cal C}_{\tilde x}$, which is now $\tilde
x$-dependent, runs along the real axis and encircles the singularity at
$\tilde t=-\sqrt{\tilde x^2+b_{U1}}$, but avoids the logarithmic cut
of the last term in braces at $\tilde t=\tilde x+1$, see
Fig.~\ref{fig:tildeC}. We make a crucial observation: if we restrict
the integrand to the first two terms in braces, the contour can be
freely deformed into the imaginary time axis. The resulting integral
has the same form as in Eq.~(\ref{SPI2}) and can be easily evaluated in polar
coordinates. It leads to the same result as Eq.~(\ref{SPI3}) with the
replacement $b_{th}\mapsto b_{U1}$.

The remaining contribution is computed by residues and
turns out to vanish,
\be
\begin{split}
&\frac{2\pi}{{\rm g}^2}\int_{-\infty}^{\infty}d\tilde x\;
\frac{\d}{\d\tilde t}\bigg[\frac{4b_{U1}
\ln(1-\tilde t+\tilde x)}{(\tilde t-\sqrt{\tilde x^2+b_{U1}})^2}\bigg]
\bigg|_{\tilde t=-\sqrt{\tilde x^2+b_{U1}}}\\
&=\frac{2\pi}{{\rm g}^2}\int_{-\infty}^{\infty}d\tilde x\;
\bigg[\frac{b_{U1}\ln(1+\tilde x+\sqrt{\tilde x^2+b_{U1}})}{(\tilde
  x^2+b_{U1})^{3/2}}
-\frac{b_{U1}}{(\tilde x^2+b_{U1})(1+\tilde x+\sqrt{\tilde
    x^2+b_{U1}})}\bigg]
=0\;.
\end{split}
\ee 
Thus, we conclude that 
\be
\label{BU1_3}
B_{U1}=\frac{4\pi}{{\rm g}^2}\bigg[\ln\bigg(\frac{4\l^2}{\kappa
  b_{U1}}\bigg)-4\bigg]\;,
\ee
which upon substitution of $b_{U1}$ from Eq.~(\ref{U_a_far}) yields
Eq.~(\ref{B_U_far}). 

\paragraph{Unruh  bounce near horizon.} 
Here the starting expression is
\be
\label{BU2_1}
\begin{split}
B_{U2}=\frac{i}{{\rm g}^2}
&\int_{-\infty}^{\infty}dx\int_{\cal C} dt
\frac{4\l^2 b_{U2}}{\left(-2\l(v-v_2)\sh\left(\frac{\l}{2}(u-u_2)\right)
+b_{U2}\e^{\frac{\l}{2}(u-u_2)}\right)^2}\\
&\times\left\{\ln\left[
\frac{4\l^2 b_{U2}}{\kappa\left(-2\l(v-v_2)\sh\left(\frac{\l}{2}(u-u_2)\right)
+b_{U2}\e^{\frac{\l}{2}(u-u_2)}\right)^2}\right]-2\l x-2\right\}.
\end{split}
\ee
Upon the change of variables analogous to the previous paragraph, we obtain
\be
\label{BU2_2}
\begin{split}
B_{U2}=\frac{i}{{\rm g}^2}
\!\int\limits_{-\infty}^{\infty}\!d\tilde x\!\int\limits_{\tilde{\cal C}_{\tilde
    x}}\! d\tilde t\; 
\frac{4b_{U2}}{\left(-\tilde t^2\!+\!\tilde x^2\!+\!b_{U2}\right)^2}
\left\{\ln\left[
\frac{4\l^2 b_{U2}}{\kappa\left(-\tilde t^2\!+\!\tilde
    x^2\!+\!b_{U2}\right)^2}\right]\!-\!\tilde t\!-\!\tilde x\!
-\!2\l x_2\!-\!2\right\}.
\end{split}
\ee
We can now deform the time integration contour into the imaginary
axis and take the integral in polar coordinates. In this way we
obtain
\be
\label{BU2_3}
B_{U2}=\frac{4\pi}{{\rm g}^2}\bigg[\ln\bigg(\frac{4\l^2}{\kappa
  b_{U2}}\bigg)-2\l x_2-4\bigg]\;,
\ee
which yields Eq.~(\ref{B_U_near}) after substituting $b_{U2}$ from
Eq.~(\ref{U_a}).

\section{A failed attempt: Nonminimal coupling}
\label{app:Dil_nonmin}

Throughout the main text we consider a minimally coupled scalar
field. As a result, the field equation for our toy model
(\ref{phieqOmega}) is not exactly solvable, even in the massless
limit, unless the metric function $\O$ satisfies $(\ln \O)''=0$. On
the other hand, it is well-known that the classical Liouville theory
remains exactly solvable in an arbitrary metric if one adds a
nonminimal coupling of the scalar field to curvature. In this
Appendix we consider such coupling and explain why it is
not suitable for our purposes. 

The nonminimal coupling in question has the form
\begin{equation}\label{DeltaS}
S_{\rm nm}=-\frac{1}{\gc^2}\int d^2x\sqrt{-g}\, R\vf \;,
\end{equation}
where $R$ is the Ricci scalar given by Eq.~(\ref{OmegaInt}) for an
arbitrary metric written in the conformally flat form. Notice that
this coupling is linear 
in the field $\vf$ and thus leads to a source term $-\Box \ln\O$ on
the r.h.s. of the field equation (\ref{phieqOmega}). It is
straightforward to see that in the presence of this source the
expression (\ref{GenSolCurv}) provides the general solution for $\vf$
in the massless limit, for an arbitrary metric function $\O$. 

However, the presence of the source in the field equation has an
unwanted consequence. It modifies the {\em classical} false vacuum by
shifting it away from $\vf=0$. To see this, we use the 
metric of dilaton BH \eqref{O_BH}. The result does not
qualitatively depend on this
particular choice, as long as the metric has a horizon with the
temperature $\l\gg m$. Due to the space-dependent source
the false vacuum configuration
$\vf_{\rm fv}(x)$ becomes
inhomogeneous. To find it,
we omit the nonlinear term in the potential and
obtain the equation
\begin{equation}\label{gseq}
\vf_{\rm fv}''-m^2\Omega \vf_{\rm fv}+(\ln \Omega)''=0\;.
\end{equation}
In the region $x\ll1/m$ the mass term can be neglected and the solution is
\begin{equation}\label{gsin}
\vf_{\rm fv}=-\ln\Omega+2\l x+ C_{\rm fv}\;.
\end{equation}
The linear term has been fixed by the regularity of the field at the
horizon, $x\to -\infty$, whereas the constant term $C_{\rm fv}$ must be
determined from the behavior at $x\gg 1/\l$.
In that region the field has the usual exponential form,
$\vf_{\rm fv}=A_{\rm fv}\e^{-mx}$. Comparing to Eq.~(\ref{gsin}), we get
$C_{\rm fv}=A_{\rm fv}=-2\l/m$. 
We see that in the near-horizon region the field
acquires large negative values:
$\vf_{\rm fv}=-2\l/m$ at $x\to-\infty$. 
In other words, it is driven away from the tunneling region located at
large positive $\vf$. 

It is instructive to compute the false vacuum
energy. A straightforward calculation gives
\begin{equation}\label{Egs}
E_{\rm fv}=-\frac{2\l^2}{{\gc}^2m}\bigg(1-\frac{m}{2\l}\bigg)\;,
\end{equation}
where we have taken into account the potential energy coming from the
nonminimal coupling term (\ref{DeltaS}).
We see
that $E_{\rm fv}$ is negative and its absolute value grows with $\l$
faster than the BH temperature $T_{BH}\propto \l$.

To get a sense of how this property affects false vacuum decay,
let us find the sphaleron energy separating the false and true
vacua in the vicinity of the BH. We add the Liouville term to
Eq.~(\ref{gseq}) and, as usual, solve it separately for the core and
tail. 
At
$x\ll 1/m$ we have
\begin{equation}\label{sphinNM}
\vf_{\rm sph}\Big|_{x\ll
  m^{-1}}=\ln\left[\frac{\l^2}{\Omega(x)\kappa\ch^2\l(x-x_{\rm sph})}\right]
\end{equation} 
with $x_{\rm sph}$ an arbitrary constant. By matching to the solution
of the free massive equation at $x\gg 1/\l$,
\be
\vf_{\rm sph}\Big|_{x\gg \l^{-1}}=A_{\rm sph} \e^{-mx}\;,
\ee
we find 
\begin{equation}
x_{\rm
  sph}=\frac{1}{m}-\frac{1}{\l}\ln\frac{2\l}{\sqrt{\kappa}}~,~~~~~
A_{\rm sph}=\frac{2\l}{m}\;.
\end{equation}
Note that $x_{\rm sph}$ is always smaller than $1/m$ and the matching
region always exists. 
The sphaleron energy then reads
\begin{equation}
E_{\rm sph}=\frac{\l}{{\gc}^2}\bigg[8\ln\frac{2\l}{\sqrt\kappa}-\frac{2\l}{m}
-7\bigg]\;.
\end{equation}
Thus, the energy difference between the sphaleron and the false vacuum,
\begin{equation}\label{EnDiff}
E_{\rm sph}-E_{\rm
  fv}=\frac{8\l}{{\gc}^2}\bigg(\ln\frac{\l}{\sqrt\kappa}+\ln 2-1\bigg)\;, 
\end{equation}
increases with 
the growth of
$\l$ and for $\l>\Lambda_0/2$ becomes bigger than the sphaleron energy in
flat space, Eq.~(\ref{Esphfin1}). The corresponding suppression of
jumps over the sphaleron in 
the Hartle--Hawking state is
(cf. Sec.~\ref{Ssec:BH_HH}) 
\be
B_{HH}=\frac{2\pi (E_{\rm sph}-E_{\rm fv})}{\l}=\frac{16\pi}{{\rm
    g}^2}
\left(\ln\frac{\l}{\sqrt\kappa}+\ln 2-1\right)\;,
\ee
which is larger than the suppression in flat spacetime, Eq.~(\ref{B0})
(recall that we assume $\l\gg m$). 
This means that in the theory with the nonminimal coupling transitions
from the false to true vacuum in the BH vicinity are 
suppressed, instead of being catalyzed. In fact, the vacuum decay will
be dominated by tunneling far away from the BH, in the asymptotically
flat region. 

In realistic situations, such as, e.g., a scalar field in the
background of a Schwarz\-schild BH, one does not expect any modification
of the {\em classical} false vacuum. Thus, we do not 
want this property to be present in the toy model. This is why we focus
on the study of a minimally coupled scalar in this paper.

\section{Decay of the Boulware vacuum}
\label{Ssec:BH_B}

In this Appendix we construct bounce solutions describing tunneling
from the Boulware vacuum in the BH background. This vacuum is empty from
the viewpoint of an observer at asymptotic infinity. However, the
corresponding vacuum energy-momentum tensor diverges at the horizon
\cite{Birrell:1982ix} rendering this state unphysical. The problem of
decay of this state is still of academic interest for comparison with
the realistic cases of Hartle--Hawking and Unruh vacua. Also, the
Boulware vacuum may provide an adequate description for the quantum
field state in the metric of BH mimickers: horizonless compact
objects with the size only slightly exceeding the gravitational
radius \cite{Oshita:2018ptr} (see \cite{Cardoso:2019rvt} for review).

We observe that the singular part of the Boulware Green's function
in Eqs.~(\ref{GBcloseright}), (\ref{G_B_near}) is similar to
that of the Feynman function in flat spacetime. This suggests to use a
linear Ansatz for the functions $F$ and $G$ in the general solution
(\ref{GenSolCurv}) for the bounce core. Working in the Euclidean time,
we write
\be
\label{FGB}
F(z)=C_B\,(z-x_B)~,~~~~~G(\bar z)=C_B\,(\bar z-x_B)\;,
\ee
where $z,\bar z$ are defined in Eq.~(\ref{zzbar}) and $x_B$ is the
coordinate of the bounce center. To determine the constant $C_B$, we
need to match the core to the Green's function, including the
subleading nonsingular terms. This matching works
differently when the bounce center is in one of the following four regions: 
\begin{align}
&(a)~~ x_B>0~,~~ x_B\gg 1/m\;, &&(b)~~ x_B>0~,~~ 1/\l\ll x_B\ll 1/m\;,\notag\\
&(c)~~ x_B<0~,~~ 1/\l\ll |x_B|\ll 1/m\;, &&(d)~~ x_B<0~,~~ |x_B|\gg 1/m\;.\notag
\end{align}
We consider these possibilities one after the other. 

{\it (a)} In this case, the Green's function is the same as in flat
spacetime. Correspondingly, one obtains $C_B=C_M$, with $C_M$ given by
Eq.~(\ref{Aabounce}). The tunneling suppression is also
the same as in flat space, Eq.~(\ref{B0}).  

{\it (b)} Since $\O\approx 1$ in this region, the bounce core 
still has the same form as in flat space,
\be
\label{Bcoreout}
\vf_{\rm b}\Big|_{\text{core},\;x_B>0}=\ln\left[
\frac{4C_B^2}{(1+\kappa C_B^2 |z-x_B|^2)^2}\right]\;.
\ee
On the other hand, the constant in the Green's function differs from
that in flat space, see the upper line in
Eq.~(\ref{GBcloseright}). This can be interpreted as a manifestation of
the vacuum polarization by the geometry.\footnote{Though the space is
  close to flat in this region, the field modes feel the gradients of
  the metric within the distance of order $1/m$.} The matching gives
\be
\label{CBout}
C_B=\frac{m^2}{2\kappa}\e^{2\gamma_E-1}\;,
\ee 
which is smaller than $C_M$ by a factor $\e$. The suppression is
calculated in the same way as in flat space\footnote{The suppression 
is saturated by the nonlinear core which is $O(2)$
symmetric in the Euclidean spacetime, even though the metric is
not. Thus, we can easily evaluate the suppression in polar coordinates
centered at $(\tau=0,x=x_B)$.} and we obtain
\be
\label{BB}
B_B=\frac{16\pi}{{\rm
    g}^2}\left(\ln\frac{m}{\sqrt\kappa}+\gamma_E-\frac{3}{2}\right)\;. 
\ee
This is slightly below the flat-space suppression (\ref{B0}) due to
the difference in the last term. Notice, however, that the leading
logarithmic part does not change. 

{\it (c)} One might think that tunneling could be further enhanced in
the near-horizon region. However, this does not happen, as we now
show. Here the metric function is $\O\approx\e^{2\l x}$ and the bounce
core gets modified,
\be
\label{Bcorein}
\vf_{\rm b}\Big|_{\text{core},\;x_B<0}=\ln\left[
\frac{4C_B^2}{(1+\kappa C_B^2 |z-x_B|^2)^2}\right]-2\l x\;.
\ee
This brings a problem: we cannot match the linear term in this
expression to the expansion of the Green's function, see the upper
line in Eq.~(\ref{G_B_near}). It is straightforward to check that modifying the
Ansatz (\ref{FGB}) will not help. In particular, quadratic corrections
added to the functions $F$ and $G$ will cancel in the long-distance
asymptotics of the core, whereas the cubic ones will produce quadratic
contributions in $\vf_{\rm b}$, instead of the required linear term. 
We conclude that precise matching is
impossible, excluding bounce solutions with the core localized in the
near-horizon region. 
 
To see this in more detail, let us perform a `partial' matching
assuming that the gradients of the first term in Eq.~(\ref{Bcorein}) are
much larger than $\l$. This will hold in the matching region if the
following inequalities can be simultaneously satisfied:
\be
\label{Bineq}
(\sqrt\kappa C_B)^{-1}\ll|z-x_B|\ll\l^{-1}\;.
\ee
Then we can set $x\approx x_B$ in the second term in
Eq.~(\ref{Bcorein}) and, comparing with the expansion of the Green's
function, we obtain
\be
\label{CBin}
C_B=\frac{m^2}{2\kappa}\e^{2\gamma_E-1+\l |x_B|}\;.
\ee
Since $|x_B|\gg 1/\l$, this is much larger than the value of $C_B$ in
the outer region, Eq.~(\ref{CBout}). Thus, our assumption
(\ref{Bineq}) is a posteriori justified. Using the found value of $C_B$, we
compute the bounce suppression as a function of the core position,
\begin{equation}\label{S_B}
B_B=\frac{16\pi}{\gc^2}\left(\ln\dfrac{m}{\sqrt{\vk}}
+\g_E-\frac{3}{2}+\l |x_B|\right) \;. 
\end{equation}
This grows for bounces that are deeper in the near-horizon region, and
decreases towards the value (\ref{BB}) when the bounce core approaches
the outer boundary of this region.

{\it (d)} One can check that in this region the suppression of
`partially matched' bounces further increases with $|x_B|$. Therefore,
these bounces are even further suppressed than in the region {\it
  (c)}.

To sum up, we have found that the Boulware bounces are `pushed out' from
the near-horizon region and the optimal tunneling rate is achieved
when the bounce center is in the region ${\it (b)}$. The corresponding
suppression (\ref{BB}) is only marginally weaker than in flat
spacetime. 

A comment is in order. With our analytic approach we cannot exclude
existence of a bounce solution with the core at $|x_B|\lesssim 1/\l$,
i.e., right at the boundary between the near-horizon and the outer
regions. Such solution, if it exists, can have a lower suppression
than (\ref{BB}). Still, we do not expect it to differ in
the first logarithmically enhanced term. A detailed investigation of
this issue requires numerical analysis, which is beyond the scope of
this paper.

\section{Hartle--Hawking sphaleron at low temperatures}
\label{app:Sph}

Here we study the static sphaleron solution which provides a tunneling channel
from the Hartle--Hawking vacuum. We focus on the BH temperatures
below the critical value \eqref{TempCritHH}.  
Unlike in the main text, we do not assume $\l\gg m$.
The sphaleron satisfies
the equation (\ref{HH_sph_eom}).
As usual, we will look for the solution separately in two regions. In
the core (tail) we will neglect the mass (Liouville)
term. Then we will glue the two solutions in the overlap.  

In the inner region the form of the solution is fixed by the
requirement of regularity at the horizon and is given by
Eq.~(\ref{HH_sph}). It is straightforward to see that the derivation of
this expression does not assume any hierarchy between $\l$ and $m$. 

In the outer region, the sphaleron 
is given by the zero-frequency solution of the Schr\"o\-din\-ger equation
(\ref{EqModes}) vanishing at $x\to +\infty$. This is expressed in
terms of the hypergeometric function,
\begin{equation}\label{BH_sph_out}
\vf_{\rm sph}=A_{HH,{\rm sph}}\, \e^{-mx}\, {}_2F_1\left(\frac{m}{2\l},\frac{m}{2\l},1+\frac{m}{\l} \:; -\e^{-2\l x}\right) \;.
\end{equation} 
Its asymptotics at large negative $x$ is inferred from Eq.~\eqref{hypergasymp},
\begin{equation}\label{BHsphout1}
\vf_{\rm sph}\approx A_{HH,{\rm sph}}\,
\frac{\Gamma\left(1+\frac{m}{\l}\right)}{
\left[\Gamma\left(1+\frac{m}{2\l}\right)\right]^2}
\bigg[1-mx
+\frac{m}{\l}\bigg(\psi(1)-\psi\Big(1+\frac{m}{2\l}\Big)\bigg)
\bigg]\;.
\end{equation}
where $\psi(s)=\Gamma'(s)/\Gamma(s)$. Matching this to the expansion
of Eq.~(\ref{HH_sph}) at $(x-x_{HH,{\rm sph}})\gg 1/\l$, we find 
\be
x_{HH,{\rm sph}}=\frac{2}{m}-\frac{1}{\l}\ln\dfrac{2\l}{\sqrt{\kappa}}
+4\bigg(\psi(1)-\psi\Big(1+\frac{m}{2\l}\Big)\bigg)~,~~~
 A_{HH,{\rm
     sph}}=\frac{4\l\left[\Gamma\left(1+\frac{m}{2\l}\right)\right]^2}{m\,\Gamma\left(1+\frac{m}{\l}\right)}\;.  
\label{BHshpx0}
\ee
Note that the expression for $x_{HH,{\rm sph}}$
reduces to Eq.~(\ref{xHHsph}) in the limit
$\l\gg m$.

Let us compute the sphaleron energy and the associated tunneling
suppression. As usual, it is convenient to integrate by
parts in the expression for the energy. 
The boundary terms vanish, because $\vf_{\rm sph}\d_x\vf_{\rm
  sph}\to 0$ both at the horizon (due to the vanishing of $\d_x\vf_{\rm
  sph}$) and at infinity. Then for the energy we obtain 
\begin{equation}\label{BHEsph}
\begin{split}
E_{\rm sph}=\frac{1}{\gc^2}\! \int\limits_{-\infty}^\infty \!dx\;
\Omega(x)\kappa (\vf_{\rm sph}\!-\!2)\e^{\vf_{\rm sph}}
=\frac{8\l}{\gc^2}\bigg[\!\ln\frac{\l}{\sqrt\kappa}
\!-\!\frac{\l}{m}\!+\!\ln 2\!-\!1\!+\!\psi\bigg(\!1\!+\!\frac{m}{2\l}\bigg)\!-\!\psi(1)\bigg].
\end{split}
\end{equation}
Thus, at $\l\lesssim\Lambda_{HH}$ 
the sphaleron provides a tunneling channel with the suppression
\begin{equation}\label{SBHsph}
B_{\rm sph}=\frac{16\pi}{\gc^2}\left[\ln\frac{\l}{\sqrt\kappa}
-\frac{\l}{m}+\ln2-1+\psi\bigg(1+\frac{m}{2\l}\bigg)-\psi(1)\right] \;.
\end{equation}
Using $\psi(1)=-\gamma_E$ and 
Eq.~(\ref{psiasymp}) for the asymptotics of $\psi(s)$ at large values of
its argument,  
one can check that at $\l\to 0$ the expression (\ref{SBHsph})
tends to the vacuum suppression \eqref{B0} with the corrections
starting at the quadratic order ${\cal O}(\l^2/m^2)$. 
In the regime $\l\gg m$ it
coincides with the 
periodic instanton action \eqref{B_BH_HH} computed in
Sec.~\ref{Ssec:BH_HH}. 

\section{Some useful formulas}
\label{app:int}
Asymptotic expansion of the $\psi$-function at infinity 
(Eq.~(5.11.2) from~\cite{NIST}):
\be
\label{psiasymp}
\psi(z)\sim \ln{z}-\frac{1}{2z}-\sum_{k=1}^\infty
\frac{B_{2k}}{2k\,z^{2k}}\;, 
~~~~~~ z\to\infty,~ |\arg z|<\pi\;,
\ee
where $B_{2k}$ are the Bernoulli numbers.

Transformation of variables in the 
hypergeometric function (from Eqs.~(15.8.2), (15.1.2) of \cite{NIST}):
\begin{equation}
\begin{split}
\label{hypergrel}
\frac{\sin(\pi(b\!-\!a))}{\pi \Gamma(c)} ~_2F_1 (a,b,c;z) 
&=\frac{(-z)^{-a}}{\Gamma(b)\Gamma(c\!-\!a)\Gamma(a\!-\!b\!+\!1)}
~_2F_1 \bigg(a,a\!-\!c\!+\!1,a\!-\!b\!+\!1;\frac{1}{z}\bigg) 
\\
&-\frac{(-z)^{-b}}{\Gamma(a)\Gamma(c\!-\!b)\Gamma(b\!-\!a\!+\!1)}
~_2F_1 \bigg(b,b\!-\!c\!+\!1,b\!-\!a\!+\!1;\frac{1}{z}\bigg).
\end{split}
\end{equation}
Asymptotic expansion of the hypergeometric function when $b-a$ is a
nonnegative integer (from Eqs.~(15.8.8), (15.1.2) of \cite{NIST}):
\begin{equation}
\begin{split}
\label{hypergasymp}
_2F_1 & (a, a+m,c;z) =\frac{(-z)^{-a}\Gamma(c)}{\Gamma(a+m)}\sum_{k=0}^{m-1}
\frac{(a)_k(m-k-1)!}{k!\, \Gamma(c-a-k)}z^{-k}\\
&+\frac{(-z)^{-a}\Gamma(c)}{\Gamma(a)}\sum_{k=0}^\infty
\frac{(a+m)_k}{k!\,(k+m)!\, \Gamma(c-a-k-m)}(-1)^k z^{-k-m}\\
&\times
\big(\ln(-z)+\psi(k+1)+\psi(k+m+1)-\psi(a+k+m)-\psi(c-a-k-m)\big)\;.
\end{split}
\end{equation}
In both above equations it is assumed that $|\arg(-z)|<\pi$. In addition, in
the second equation $|z|>1$, $m$ is a nonnegative integer, 
$(a)_k=a(a+1)\ldots(a+k-1)$ and $\psi(s)=\Gamma'(s)/\Gamma(s)$.

\bibliographystyle{JHEP}
\bibliography{Refs}

\end{document}